\documentclass[varenna,nocopyright]{cimento}
\renewcommand{\i}{\sy{i}}
\newcommand{\g}{\sy{g}}
\newcommand{\B}{\sy{B}}
\newcommand{\e}{\sy{e}}
\newcommand{\N}{\sy{N}}

\usepackage{amssymb, amsmath} 

\usepackage[showonlyrefs]{mathtools}		
\mathtoolsset{showonlyrefs=true}

\usepackage{array}
\newcolumntype{L}[1]{>{\raggedright\arraybackslash}m{#1}} 
\newcolumntype{C}[1]{>{\centering\arraybackslash}m{#1}} 
\newcolumntype{R}[1]{>{\raggedleft\arraybackslash}m{#1}} 
\usepackage{graphicx}
\usepackage{grffile} 
\usepackage{bbm}
\DeclareMathAlphabet{\mathscr}{OT1}{pzc}{m}{it} 
\usepackage{arcs}

\renewcommand{\i}{\mathrm{i}}

\renewcommand{\vec}[1]{\boldsymbol{#1}}

\newcommand{\ket}[1]{| #1 \rangle}
\newcommand{\bra}[1]{\langle  #1 |}
\newcommand{\braket}[1]{\langle #1  \rangle}
\newcommand{\ketbra}[2]{\ket{ #1} \bra{#2} }
\newcommand{\sandwich}[3]{\langle  #1 |\, #2 \,| #3 \rangle}

\title{The interface of gravity and quantum mechanics\newline illuminated by Wigner phase space}

\shorttitle{Interface of gravity and quantum mechanics in Wigner phase space}

 \author{E.~Giese, W.~Zeller, S.~Kleinert, M.~Meister, V.~Tamma, A.~Roura}
\institute{Institut f{\"u}r Quantenphysik and Center for Integrated Quantum Science and Technology ($\text{IQ}^\text{ST}$), 
Universit{\"a}t Ulm, D-89069 Ulm, Germany}
\author{\atque W.~P.~Schleich}
\institute{Institut f{\"u}r Quantenphysik and Center for Integrated Quantum Science and Technology ($\text{IQ}^\text{ST}$), 
Universit{\"a}t Ulm, D-89069 Ulm, Germany\\
Texas A\&M University Institute for Advanced Study, Institute for Quantum Science and Engineering (IQSE) and Department of Physics and Astronomy, Texas A\&M University, College Station, Texas 77843-4242, USA}

\shortauthor{Giese, Zeller, Kleinert, Meister, Tamma, Roura, Schleich}

\PACSes{
\PACSit{03.75.Dg}{Atom and neutron interferometry}
\PACSit{37.25.+k}{Atom interferometry techniques}
\PACSit{42.50.-p}{Quantum optics}
\PACSit{04.80.Cc}{Experimental tests of gravitational theories}
}

\begin{document}

\maketitle

\begin{abstract}
We provide an introduction into the formulation of non-relativistic quantum mechanics using the Wigner phase-space distribution function and apply this concept to two physical situations at the interface of quantum theory and general relativity: (\textit{i}) the motion of an ensemble of cold atoms relevant to tests of the weak equivalence principle, and (\textit{ii}) the Kasevich-Chu interferometer. In order to lay the foundations for this analysis we first present a representation-free description of the Kasevich-Chu interferometer based on unitary operators.
\end{abstract}

%
%
\section{Introduction}
%
%
Although the physics of the twentieth century was dominated by relativity~\cite{Misner} and quantum mechanics~\cite{Bohm} the two theories could not be merged in a completely satisfactory way. To bring them together remains one of the great challenges~\cite{DeWitt,Kiefer} for the twenty-first century. Today we witness impressive experiments at the interface of these theories using tools of atom optics~\cite{Cronin09}. Indeed, the recent creation of~\cite{van-Zoest10} and interferometry with~\cite{Muetinga12} a Bose-Einstein condensate in micro-gravity~\cite{Arimondo:Atomic-and-Space-Physics} in the drop tower of Bremen together with equivalent experiments employing laser-cooled atoms on a parabolic flight on a plane~\cite{Stern2009,Geiger11,Menoret11} or large atomic fountains~\cite{Dickerson2013} bring new precision tests of the weak equivalence principle with different atomic species~\cite{Fray04,Fray09,Bonnin13,Schlippert14} within reach. Moreover, the Kasevich-Chu atom interferometer~\cite{Kasevich91,Kasevich92} has become an important instrument~\cite{Graham2013,Dimopoulos2008} for testing general relativity using cold atoms. In the present lectures we analyze experiments of 
this type from quantum phase space using the Wigner function~\cite{Wigner1932,Schleich:Quantum-optics,Zachos:quantum-mechanics-phase-space}.

\subsection{How to include weak gravity in quantum mechanics}
%
Since there is no unified theory for gravity and quantum mechanics one might wonder how to describe atom optics in the presence of gravity. However, since we are only interested in experiments performed on earth, or in microgravity, it suffices to consider the weak field and low curvature limit of gravity. In this case, the Newtonian potential enters into the Schr\"odinger equation as a regular potential energy. Throughout these lectures we restrict ourselves to this description but emphasize that there is a rigorous way of deriving this result when the effects of quantum fluctuations of the metric are neglected.

Indeed, in the weak-field limit there is a strong analogy~\cite{DeWitt66,Schleich:New-Trends-Atomic-Physics} between classical electrodynamics and gravity. The role of the scalar potential of electrostatics is played by the Newtonian potential and the off-diagonal elements of the metric associated with rotation act as a type of vector potential. The two potentials are sometimes referred to as gravito-electric and gravito-magnetic potentials. It is, therefore, not surprising that we arrive at a Schr\"odinger equation with a minimal coupling to these gravito-electromagnetic potentials. We emphasize that a closely related result has been obtained in~\cite{Parker80} starting from the Dirac equation in curved space-time and considering the non-relativistic limit in Fermi-normal coordinates~\cite{Kajari09}.
\subsection{Outline of the lectures}
%
Our lecture notes are organized as follows: In section~\ref{sec:Wigner_functions} we briefly recall the formulation 
of non-relativistic quantum mechanics~\cite{Bohm} in terms of the Wigner function~\cite{Wigner1932,Schleich:Quantum-optics,Zachos:quantum-mechanics-phase-space}. 
This quantity is most suited to compare and contrast classical 
and quantum dynamics. For this purpose, we first review the key features of a classical phase-space distribution 
function and then turn to the Wigner function. In particular, we derive the quantum Liouville equation and the 
energy eigenvalue equation in phase space. We show that for the case of a linear gravitational potential 
with gravity gradients the quantum Liouville equation reduces to the classical Liouville equation, 
which is free of Planck's constant. Hence, classical and quantum dynamics are identical. 
In contrast, the corresponding energy eigenvalue equation for the Wigner function does contain Planck's constant. 
This feature is a clear indication that even for potentials whose dynamics is independent of Planck's constant 
the energy eigenstates do depend on it.

These considerations are most relevant~\cite{Kajari10} for the discussion of tests of the weak equivalence principle with 
cold atoms such as a Bose-Einstein condensate. For a single atom its gravitational and inertial mass enter 
into the Newton equation only through their ratio. Therefore, the acceleration does not depend on the kind of particle provided the weak equivalence principle holds true. However, in the case of a thermal cloud of atoms we do not deal with a single particle but an ensemble of particles, and the 
Wigner function is the perfect tool to analyze the dynamics of the ensemble.  Moreover, it can also describe the pure quantum state of the atoms in a condensate. In section~\ref{sec:Equiv_of_inertial&gravitational} 
we discuss with the help of the Wigner function the dynamics of cold atoms in a linear gravitational potential and show that in 
the dynamics the inertial and gravitational mass only appear through their ratio. However, they can enter independently from 
each other through the initial distribution. Moreover, the energy eigenstate, either expressed in 
Wigner phase space, or in position space contains the gravitational and inertial mass not as a ratio 
but in products of powers which can even be different. We illustrate this feature using three different 
potentials.

Sections~\ref{sec:KC_interferometer} and~\ref{subsec:analysis_in_wigner_space}  are dedicated to yet another application of the Wigner function. 
Here we analyze the Kasevich-Chu interferometer~\cite{Kasevich91,Kasevich92}. In section~\ref{sec:KC_interferometer} we first recall a 
representation-free operator description~\cite{Kajari10,Schleich13,SchleichNJP13} of the Kasevich-Chu interferometer 
that clearly demonstrates that it is an accelerometer. Indeed, we show that the phase shift originates from 
(\textit{i}) a combination of the phases of the three laser pulses in the form of a discrete second time derivative, 
and from (\textit{ii}) the non-commutativity of two time evolutions reflecting the acceleration of the atom. This result holds true 
for any form of the gravitational potential. However, in the case of a linear potential the unitary phase-shift operator 
determining the number of atoms at the exit ports of the interferometer is a c-number phase factor given by the difference of 
these two phases.

In section~\ref{subsec:analysis_in_wigner_space} we then turn to the Wigner description of this interferometer. We show that the Wigner function of the center-of-mass motion of the atom in the state $\ket{g_1}$ at the exit of the interferometer consists of three contributions corresponding to the upper and the lower path in the interferometer, and an interference term. In all three cases the time evolution is a sequence of shifts in the momentum and motions along classical trajectories in phase space. The final Wigner functions at the end of the two paths are solely the result of a classical transport of the original Wigner function. In contrast, the interference term is also modulated in phase space by a cosine function. This modulation is the origin of the phase in the interferometer. In the presence of gravity gradients or an asymmetric pulse sequence the two paths do not close and the interference term is located symmetrically between the two corresponding Wigner functions. This behavior is reminiscent of that of a Schr\"odinger cat state~\cite{Schleich91}.

We conclude in section~\ref{sec:Summary} by summarizing our results and providing an outlook.

In order to focus on the main ideas we have moved more extensive calculations to appendices. For example, in appendices~\ref{app:Weyl-Wigner-displacement-operator} and~\ref{app.quantum_liouville} we discuss the Weyl-Wigner correspondences of the displacement operator, and of the product of an arbitrary operator and functions of the position and momentum operator. For this purpose we introduce the Bopp operators and derive expressions for the special case of the non-relativistic Hamiltonian of a particle. This approach allows us to derive quantum Liouville equations for the Wigner functions describing the motion of the atom in the two internal states while it propagates through the interferometer. Moreover, we translate the time-independent Schr\"odinger equation into phase space.

In appendix~\ref{app:Gradients} we generalize the representation-free description of the 
Kasevich-Chu interferometer to include gravity gradients. This expression allows us to make contact with the corresponding formulation in terms of Wigner  functions.

We dedicate appendix~\ref{app:phase-space-dynamics} to an analysis of the phase space trajectories corresponding to the upper, lower and inference paths in the interferometer. Here we derive closed-form expressions for the end points of these trajectories and show that the one representing the interference term lies between the ones of the upper and lower paths. Moreover, we derive an expression for the phase shift.

%
%
%
\section{Introduction to Wigner functions}
%
\label{sec:Wigner_functions}
In the present section we provide an introduction to the Wigner phase-space distribution function 
\cite{Wigner1932,Schleich:Quantum-optics,Zachos:quantum-mechanics-phase-space} 
which allows us to formulate non-relativistic quantum mechanics in complete analogy to classical statistical mechanics. 
For this purpose we first recall the essential ingredients of a classical phase-space distribution function and motivate 
its time evolution given by the 
uation. We then turn to its quantum analogue, motivate the form of 
the Wigner function using a heuristic argument~\cite{Suessmann91} based on quantum jumps and briefly discuss its properties. 
Most relevant in this context is the comparison of quantum and classical dynamics as expressed by the quantum Liouville equation. 
Moreover, we derive the phase-space analogue of the time-independent Schr{\"o}dinger equation, that is, of an energy eigenstate.

\subsection{Ensemble of particles: classical dynamics \`{a} la Liouville}
%
In a gas of atoms we cannot specify the coordinates and the momentum of every atom. Therefore, we have to resort to a description based on a distribution function which evolves in time. 
Since the dynamics of classical mechanics is governed by phase space, it is natural to consider a phase-space 
distribution function.
%
\subsubsection{Definition of phase-space distribution}
%
In classical statistical mechanics we describe an ensemble of particles by the distribution function 
$f = f(z,v;t)$ which depends on the coordinate $z$ and the velocity $v$ of the particle. Moreover, $f$ evolves 
in time as indicated by the argument $t$. In order to distinguish between the variables $(z,v)$ of phase space 
and the parameter $t$ of time we have separated them by a semicolon.

\begin{figure}
\centering
\includegraphics{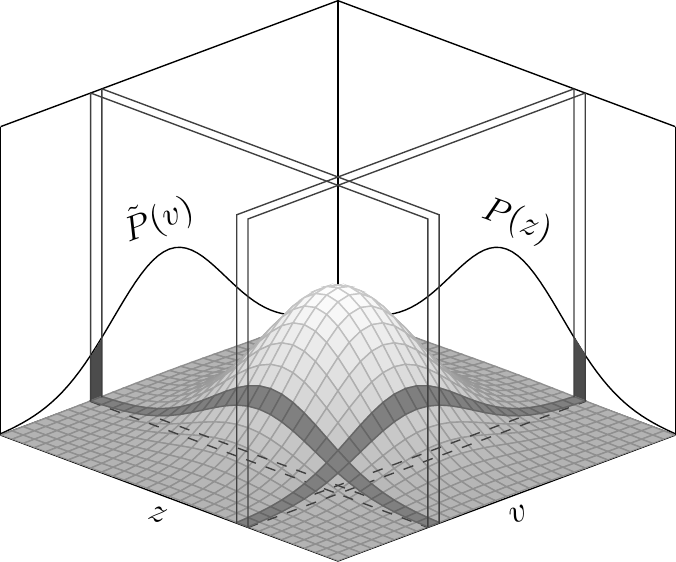}
\caption{Classical phase-space distribution function for the example of a Gaussian. Integration over position $z$ and velocity $v$ yields the corresponding probability distributions $\tilde{P}=\tilde{P}(v)$ and $P=P(z)$ for velocity and position, respectively.}
\label{fig.cl_phase-space}
\end{figure}

This distribution function allows the interpretation of a probability density where 
\begin{align}
      f(z,v;t)\, \drm z \drm v  	&\equiv	\left(\begin{matrix} \sy{probability~to~find~particles~at~time}~t~\sy{between} \\ 
							z~\sy{and}~z+\drm z,~\sy{and~between}~v~\sy{and}~v+\drm v \end{matrix}\right). \nonumber
\end{align}
Moreover, we obtain the probability densities $P=P(z;t)$ or $\tilde{P}=\tilde{P}(v;t)$ to find particles at time $t$ between the 
positions $z$ and $z+\drm z$, or in the velocity interval between $v$ and $v+\drm v$ by integrating over 
the conjugate variable, that is by integration over $v$, or over $z$. Indeed, we arrive at 
\begin{align}
      \int\limits_{-\infty}^{\infty}\!\! \drm v \ f(z,v;t) &=	P(z;t) 
\end{align}
and
\begin{align}
      \int\limits_{-\infty}^{\infty}\!\! \drm z \ f(z,v;t) &=	\tilde{P}(v;t) \,.
\end{align}
The  integration over all phase space yields the number $N$ of particles, that is
\begin{align}\label{eq:particle_number}
      \int\limits_{-\infty}^{\infty} \!\!\drm z \int\limits_{-\infty}^{\infty} \!\!\drm v \ f(z,v;t) &=	N\, .
\end{align}
In fig.~\ref{fig.cl_phase-space} we illustrate this property of the marginals for the case of a Gaussian. The integration 
over position provides us with the velocity distribution, whereas the integration over velocity leads to the position 
distribution.
%
%
%
\subsubsection{Dynamics: classical Liouville equation}
%
Next we have to find an equation describing the time evolution of $f$. Here we are guided by the conservation of particles which 
requires that the total time derivative of $f$ vanishes, that is
\begin{align}
      \frac{\drm}{\drm t} f = 0 \, .
\end{align}
This statement is made in the coordinate system moving with the flow.
In the non-moving frame the total time derivative reads with the help of the chain rule
\begin{align}
    \left(\frac{\partial}{\partial t} + \dot{z} \frac{\partial}{\partial z} +\dot{v} \frac{\partial}{\partial v} \right) f = 0\,.
\end{align}
Throughout this article dots indicate derivatives with respect to time. 

Hence, we arrive at the familiar Liouville equation
\begin{align}\label{equ:liouville_equ}
\hat{\mathcal{L}}f \equiv  \left(\frac{\partial}{\partial t} + v \frac{\partial}{\partial z}  + a \frac{\partial}{\partial v}\right) f(z,v;t) = 0 \,,
\end{align}
where $a\equiv\dot{v}$ denotes the acceleration.
%
%
\subsubsection{Solution of the Liouville equation}
%
%
The next step in our analysis is to obtain the general solution of the Liouville equation~\eqref{equ:liouville_equ}. 
Since we deal with a partial differential equation of first order, the method of characteristics comes to mind. 
However, a more physical approach respects the requirement of conservation of particles and rests on the 
transport along classical trajectories. Indeed, we can only reach at $t$ the phase-space point ($z,v$) provided we have 
started at the appropriate initial phase-space point ($z_0,v_0$) at time $t=0$ such that the classical 
trajectories $\overline{z}=\overline{z}(z_0,v_0;t)$ and $\dot{\overline{z}}=\dot{\overline{z}}(z_0,v_0;t)$ in 
position and velocity space connect $(z,v)$ with $(z_0,v_0)$. This idea leads to the solution 
\begin{align}\label{equ:f}
  f(z,v;t) 	&= 	\int\limits_{-\infty}^{\infty} \!\!\drm z_0 \int\limits_{-\infty}^{\infty} \!\!\drm v_0  
			\ \delta[z-\overline{z}(z_0,v_0;t)] \ \delta[v-\dot{\overline{z}}(z_0,v_0;t)]
			\ f_0 (z_0,v_0)
\end{align}
of the Liouville equation. The delta function in $z$ and $v$ ensures that the particles follow along the classical 
trajectories and $f_0 (z_0,v_0) \equiv f (z_0,v_0;t=0)$ denotes the initial distribution. It is straightforward 
to verify that the so-defined distribution function satisfies the Liouville equation~\eqref{equ:liouville_equ}.

\subsection{Definition of Wigner function}
%
In the formulation of classical statistical mechanics in terms of phase-space distribution functions we 
associate with every point in phase space a probability. At first sight such an approach might not work 
in quantum mechanics since the uncertainty principle does not allow us to attribute a quantum state to a single point in phase space. Nevertheless, there exists an infinite amount of quantum mechanical distribution functions 
which are equivalent to standard quantum mechanics. We now focus on the Wigner function since it is particularly useful in the context of the dynamics of cold atoms in a gravitational field.
\begin{figure}
\centering
\includegraphics{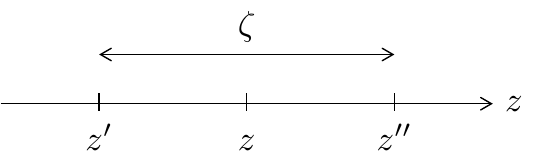}
\caption{Heuristic argument~\cite{Suessmann91} 
motivating the definition of the Wigner function (\ref{equ:Wigner}) 
as a jump 
process from $z'$ to $z''$ around the center $z$ of the jump of length $\zeta \equiv z'-z''$.}
\label{fig:quantumjump}
\end{figure}

One key ingredient of quantum mechanics are quantum jumps. Indeed, Werner Heisenberg in his very first paper~\cite{Heisenberg1925} 
on quantum mechanics recognized that it is not the individual Bohr orbits that matter but the jumps between them. 
In the one-dimensional motion of a particle along the z-axis we should therefore consider a jump from the 
position $z'$ to $z''$ as shown in fig.~\ref{fig:quantumjump} and represented by the operator 
\begin{align}
  \hat{J}	&\equiv	\ket{z''} \bra{z'}  
\end{align}
in terms of position eigenstates $\ket{z'}$ and $\ket{z''}$.

It is useful to express these coordinates in terms of the length $\zeta \equiv z' - z'' $ of the jump and 
its center $ z \equiv \frac{1}{2} (z'+z'')$ which yields
\begin{align}
  \hat{J}	&=	\ket{z - \frac{1}{2}\, \zeta} \bra{z + \frac{1}{2}\, \zeta}.
\end{align}
This jump is associated with a momentum $p$ due to the Fourier transform 
\begin{align}
\quad  \hat{W}(z,p)	&\equiv	\frac{1}{2\pi\hbar} \int\limits_{-\infty}^{\infty} \!\! \drm \zeta \
				      \e^{-\i p \zeta /\hbar} \ket{z - \frac{1}{2}\, \zeta} \bra{z + \frac{1}{2}\, \zeta}
\end{align}
of the jump. Here we have introduced the appropriate normalization factor $1/(2\pi\hbar)$.

So far, we have only considered the jump operator $\hat{W}(z,p)$ connecting all jumps with momentum $p$ 
symmetrically around the position $z$. The quantum state $\ket{\psi}$ of the center-of-mass motion of the particle 
has not entered yet. Since we aim for a c-number phase-space distribution function, it is natural to take the 
expectation value of $\hat{W}(z,p)$ which leads us to the familiar expression 
\cite{Wigner1932,Schleich:Quantum-optics,Zachos:quantum-mechanics-phase-space}
\begin{align}
  W_\psi (z,p)	&\equiv	\bra{\psi} \hat{W}(z,p) \ket{\psi} =
				\frac{1}{2\pi\hbar} \int\limits_{-\infty}^{\infty} \!\! \drm \zeta\
				      \e^{-\i p \zeta /\hbar} \braket{\psi|\,z - \frac{1}{2}\, \zeta} 
						    \braket{z + \frac{1}{2}\, \zeta \,| \psi}  
\end{align}
for the Wigner function $W_{\psi}$ of the state $\ket{\psi}$. When we introduce the wave function 
$\psi(z) \equiv \braket{z|\psi}$ in position representation we arrive at
\begin{align}\label{equ:Wigner}
  W_\psi (z,p)	&=		\frac{1}{2\pi\hbar} \int\limits_{-\infty}^{\infty} \!\! \drm \zeta \
				      \e^{-\i p \zeta /\hbar} \  \psi^*(z - \frac{1}{2}\  \zeta) \  \psi(z + \frac{1}{2}\  \zeta)
\end{align}
which in terms of states reads 
\begin{align}\label{equ:Wigner_2}
  W_\psi(z,p)	&=		\frac{1}{2\pi\hbar} \int\limits\limits_{-\infty}^{\infty} \!\! \drm \zeta \
				      \e^{-\i p \zeta /\hbar} \ \braket{z + \frac{1}{2}\, \zeta | \psi} \braket{\psi | z- \frac{1}{2}\, \zeta }\,.
\end{align}

So far we have only considered pure states. However, it is also possible to define Wigner functions of quantum states
 described by a density operator $\hat{\rho}$. Indeed, from the definition \eqref{equ:Wigner_2} of the Wigner function in 
terms of states, and the fact that the density operator of a pure state 
reads $\hat{\rho} \equiv \ket{\psi}\bra{\psi}$, we find the definition 
\begin{align}\label {equ:Wigner_3}
  W_\rho(z,p)	&\equiv		\frac{1}{2\pi\hbar} \int\limits_{-\infty}^{\infty} \!\! \drm \zeta \
				      \e^{-\i p \zeta /\hbar} \ \bra{z + \frac{1}{2}\, \zeta} \hat{\rho} \ket{z- \frac{1}{2}\, \zeta }
\end{align}
of the Wigner function $W_{\rho}$ of the density operator $\hat\rho$.
%
%
%
\subsection{Properties}
%
%
\begin{figure}
\centering
\includegraphics{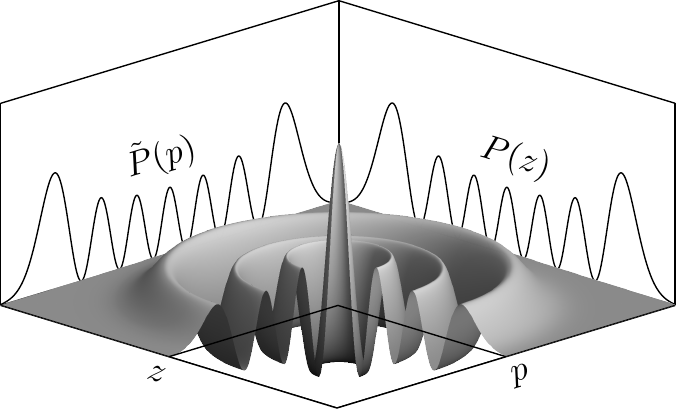}
\caption{Wigner function $W_6 = W_6 (z,p)$ of the 6-th excited state of a harmonic oscillator. Phase space is made up of annular 
domains where $W_6$ takes on consecutively positive and negative values. These regions are separated by circular lines where 
$W_6$ vanishes. The last positive domain is in the neighborhood of the classical trajectory corresponding to the energy of this 
state. These fringes in phase space are a manifestation of the bilinearity of the Wigner function in the wave function and thus 
reflect the interference of quantum waves.}
\label{fig.wignerfunction}
\end{figure}
From the definition~\eqref{equ:Wigner} of the Wigner function $W_\psi$ in terms of the wave function it is straightforward 
to show that $W_\psi$ is real. Indeed, by taking the complex conjugate of $W_\psi$ and introducing the integration 
variable $\bar\zeta \equiv - \zeta$, we can verify this statement.

Moreover, the marginals of the Wigner function, that is, the integrals over 
momentum $p$, or position $z$, yield the correct quantum mechanical probability densities $P=P(z)$, or $\tilde{P}=\tilde{P}(p)$, 
in the conjugate variables, that 
is, in position or momentum. Indeed, from the definition~\eqref{equ:Wigner} we find 
\begin{align}
  \int\limits_{-\infty}^{\infty} \!\! \drm p \ W_\rho(z,p) &= \bra{z}\hat\rho\ket{z}	 \equiv P(z)\, ,  
\end{align}
and analogously
\begin{align}
  \int\limits_{-\infty}^{\infty} \!\! \drm z \ W_\rho(z,p)	&= \bra{p}\hat\rho\ket{p} \equiv \tilde{P}(p)\,.
\end{align}
Here we have introduced the momentum eigenstates $\ket{p}$.

For normalized states we obtain
\begin{align}
  \int\limits_{-\infty}^{\infty} \!\! \drm z \int\limits_{-\infty}^{\infty} \!\! \drm p \  W_\rho (z,p) = 1. 
\end{align}

From the definition~\eqref{equ:Wigner_3} of the Wigner 
function of a density operator $\hat{\rho}$ we can easily establish the product formula 
\begin{align}\label{equ:trace_formula}
  \sy{Tr} (\hat{\rho}_1 \, \hat{\rho}_2) 	&=	2 \pi \hbar 
						  \int\limits_{-\infty}^{\infty} \!\! \drm z
						  \int\limits_{-\infty}^{\infty} \!\! \drm p \
							W_{\rho_1}(z,p)\  W_{\rho_2} (z,p)
\end{align}
for two density operators $\hat{\rho}_1$ and $\hat{\rho}_2$, 
which for the special case of pure states $\hat{\rho}_1 \equiv \ket{\psi_1}\bra{\psi_1}$ and $\hat{\rho}_2 \equiv \ket{\psi_2}\bra{\psi_2}$ 
reduces to
\begin{align}
  \sy{Tr} (\hat{\rho}_1 \, \hat{\rho}_2) &= 	
					      |\braket{\psi_1|\psi_2}|^2	
					    =	2 \pi \hbar 
						  \int\limits_{-\infty}^{\infty} \!\! \drm z
						  \int\limits_{-\infty}^{\infty} \!\! \drm p \,
							W_{\psi_1}(z,p)\  W_{\psi_2} (z,p) \,.
\end{align}
Hence, the scalar product $\braket{\psi_1|\psi_2}$ of two quantum states is determined by the area of 
overlap in phase space of the two corresponding Wigner functions. For orthogonal states $\ket{\psi_1}$ and $\ket{\psi_2}$ with 
$\braket{\psi_1|\psi_2} = 0$ this overlap must vanish. Hence, there must be domains in phase space where either 
$W_{\psi_1}$ or $W_{\psi_2}$, or both must assume negative values. As a result, the Wigner function cannot be considered a 
proper probability distribution. 

In fig.~\ref{fig.wignerfunction} we show the Wigner function $W_6$ of the 6-th excited state of a harmonic oscillator and emphasize 
the ring-shaped regions where $W_6$ takes on negative values.
They are a consequence of the bilinearity of the Wigner function and reflect the interference property of quantum mechanics.
%
%
\subsection{Weyl-Wigner correspondence}
%
\label{sec:Weyl-Wigner-correspondence}
The Wigner function is more than just a representation of the quantum state. It also allows us to calculate quantum 
mechanical expectation values in the same way as we have performed classical averages. However, there is a 
subtlety. The operators have to be ordered appropriately before we replace them by c-numbers. 
In particular, for the Wigner function the order in which the position and the momentum 
operator appear has to be symmetric. This task is achieved by the so-called Weyl-Wigner correspondence~\cite{Weyl}
which associates with an arbitrary operator $\hat{\mathcal{O}}	\equiv \hat{\mathcal{O}}(\hat z, \hat p)$ the c-number representation 
\begin{align}\label{eq:WW-corresp}
  \mathcal{O}_W (z,p)	&\equiv	\int\limits_{-\infty}^{\infty} \!\! \drm \zeta \
				      \e^{-\i p \zeta/\hbar} \ \bra{z + \frac{1}{2}\, \zeta} \hat{\mathcal{O}} \ket{z- \frac{1}{2}\, \zeta}\, .
\end{align}
We emphasize that apart from the factor $1/(2\pi\hbar)$ this expression is the definition of the Wigner function of a density 
operator. 

Therefore, we can use the product formula eq.~\eqref{equ:trace_formula} to express the quantum mechanical 
expectation value
\begin{align}\label{eq:erwO}
  \braket{\hat{ \mathcal{O}}} 	&= \sy{Tr} (\hat{\mathcal{O}} \hat{\rho})
			=	\int\limits_{-\infty}^{\infty} \!\! \drm z
				\int\limits_{-\infty}^{\infty} \!\! \drm p ~
				      \mathcal{O}_W (z,p)\  W_{\rho} (z,p)
\end{align}
in a way analogous to a classical average. Here we average the Weyl-Wigner-ordered operator $\mathcal{O}_W (z,p)$
with the Wigner function $W_{\rho} (z,p)$ by integrating over phase space. Hence, quantum mechanics enters 
in two ways: (\textit{i}) in the operator ordering, and (\textit{ii}) the Wigner function. 

In this context the displacement operator
\begin{align}\label{eq:Doperator}
 \hat{D} \equiv \e^{\i(\xi \hat{p}+ q \hat{z})/\hbar} \, ,
\end{align}
which involves the symmetric combination of position and momentum operators in the exponent together with the c-numbers $\xi$ and $q$ with dimensions of position and momentum plays a special role. Indeed, we show in appendix~\ref{app:Weyl-Wigner-displacement-operator} that the Weyl-Wigner correspondence $D_W$ of $\hat{D}$ defined by eq.~\eqref{eq:WW-corresp} is identical to $\hat{D}$ with the operators replaced by the corresponding c-number quantities, that is
\begin{align}\label{eq:DWeyl}
 D_W=\e^{\i(\xi p + qz)/\hbar}\,.
\end{align}

As a result, the corresponding expectation value
\begin{align}\label{eq:Derwartungswert}
 \langle \hat{D} \rangle = \int\limits_{-\infty}^{\infty} \!\! \drm z
			      \int\limits_{-\infty}^{\infty} \!\! \drm p \,
			      \e^{\i(\xi p + qz)/\hbar}\  W_{\rho} (z,p)
\end{align}
of $\hat{D}$ given by eq.~\eqref{eq:erwO} constitutes the phase space average of the c-number representation $D_W$ defined by eq.~\eqref{eq:DWeyl} using the Wigner function $W_\rho$.

It is interesting to note that here quantum mechanics enters solely through the quantum state, that is the Wigner function $W_\rho$ and not through the operator ordering. Indeed, this property is valid not only for the displacement operator but for any symmetrically ordered operator.

We conclude by mentioning that the phase-space representation eq.~\eqref{eq:Derwartungswert} of the expectation value of $\hat{D}$ in terms of the Wigner function will become important in the context of the Kasevich-Chu interferometer analyzed in sections \ref{sec:KC_interferometer} and \ref{subsec:analysis_in_wigner_space}.
%
%
%
\subsection{Quantum Liouville equation}
\label{subsec:Q-Liouville-eq}
In classical statistical mechanics the time evolution of the phase-space distribution function is determined by the Liouville 
equation~\eqref{equ:liouville_equ}. In the Schr\"odinger formulation of quantum mechanics the time evolution of the quantum state $\ket{\psi}$ follows from the Schr{\"o}dinger equation
\begin{align}
  \i \hbar \frac{\drm}{\drm t} \ket{\psi(t)}	 &= 	\hat H \ket{\psi(t)}
\end{align}
where the Hamiltonian of the system
\begin{align}
  \hat H \equiv \frac{\hat{p}^2}{2m} + V(\hat z)
\end{align}
contains the arbitrary potential $V=V(z)$.

In the case of a density operator $\hat\rho$ its dynamics follows from the von Neumann equation
\begin{align}
    \dot{\hat\rho} &= -\frac{\i}{\hbar} [\hat H, \hat{\rho}]\, ,
\end{align}
which we now translate into Wigner phase space. For this purpose we have to find the Weyl-Wigner 
correspondence, as given by eq.~\eqref{eq:WW-corresp}, that is we have to find the c-number representation of the operator $\dot{\hat\rho}$, and of the 
operator products $\hat H \hat\rho$ and $\hat\rho \hat{H} $.

Obviously, the Weyl-Wigner correspondence of 
$\dot{\hat\rho}$ is the time derivative of the Wigner function. However, the operator products are more 
complicated since the Weyl-Wigner correspondence of the product of two operators is not~\cite{Dahl2003} the product of the two 
Weyl-Wigner correspondences. Indeed, in appendix~\ref{app.weyl-wigner} we derive the quantum Liouville equation
\begin{align}\label{eq:quant_Liouville}
  \mathcal{\hat{L}}\, W(z,p;t) 	&=	\hat{\mathcal{L}}_\text{odd}\, W(z,p;t)\, ,
\end{align}
where $\mathcal{\hat{L}}$ denotes the classical Liouville operator defined in eq.~\eqref{equ:liouville_equ} and the operator
\begin{align}\label{eq:Liouville_op}
\hat{\mathcal{L}}_\text{odd}\equiv \sum_{l=1}^\infty \frac{(-1)^l (\hbar/2)^{2l}}{(2l+1)!} \frac{\partial^{2l+1} V}{\partial z^{2l+1}} 			    \frac{\partial^{2l+1}}{\partial p^{2l+1}}
\end{align}
containing the quantum corrections involves only the odd derivatives of the potential, and the odd derivatives of the 
Wigner function with respect to the momentum.

Moreover, it is a power expansion in the square of Planck's constant. 
The number of powers depends on the potential, and the lowest power is determined by the third derivative of the potential 
with respect to the position. Hence, for potentials which are at most quadratic in position, $\hat{\mathcal{L}}_\text{odd}$ vanishes 
and the time evolution of the Wigner function is solely determined by the classical Liouville equation~\eqref{equ:liouville_equ}.
%
\subsection{Energy eigenvalue equation in phase space}
%
One might therefore wonder: How does quantum mechanics manifest itself for potentials that are at most quadratic? The answer 
to this question lies in the energy eigenstates. In appendix~\ref{app.quantum_liouville} we show that the Wigner function 
$W_E$ of an energy eigenstate follows from the Schr{\"o}dinger equation 
\begin{align}\label{eq:Schr_eq_W}
  \left[\frac{p^2}{2m} 	+ V - \frac{\hbar^2}{8m} \frac{\partial^2}{\partial z^2} 
			+ \hat{\mathcal{L}}_\text{even}\right] W_E
		  &= E\  W_E
\end{align}
in phase space where the operator
\begin{align}
\hat{\mathcal{L}}_\text{even} \equiv \sum_{l=1}^\infty \frac{(-1)^l (\hbar/2)^{2l}}{(2l)!} \frac{\partial^{2l} V}{\partial z^{2l}} \frac{\partial^{2l}}{\partial p^{2l}} 
\end{align}
only contains even derivatives of the potential.

In the quantum Liouville equation~\eqref{eq:quant_Liouville} Planck's constant enters only through the operator $\hat{\mathcal{L}}_\text{odd}$ 
given by eq.~\eqref{eq:Liouville_op} which depends on the form of the potential. In contrast, in the phase-space analogue of the energy eigenvalue equation, eq.~\eqref{eq:Schr_eq_W}, there is a term which involves $\hbar$ and is independent of the 
potential. Indeed, the coefficient in front of the second derivative with respect to the position is proportional to the 
square of $\hbar$ and results from the sum of the squares of the Bopp operators $\hat{p}_B^{(l)}$ and $\hat{p}_B^{(r)}$ discussed in 
appendix~\ref{app.quantum_liouville}. 
Hence, even for a constant potential we find an appearance of $\hbar$.
%
\subsection{Summary}
%
The Wigner function makes a clear distinction between kinematics and dynamics. In the phase-space differential equation for an 
energy eigenstate only the even derivatives of the potential enter. In contrast, in the quantum Liouville equation only the 
odd derivatives appear. These derivatives involve powers of Planck's constant but they appear differently in the two equations. 
Indeed, the dynamics of a particle in a constant, linear or quadratic potential is independent of $\hbar$. In this case, the 
dynamics of the Wigner function is classical. However, in the energy eigenvalue equation for the Wigner function $\hbar$ 
appears explicitly even for these potentials. Therefore, the energy eigenstates are quantum states. One might therefore 
conclude that there are two different types of $\hbar$, one associated with the dynamics and one with the initial state 
and there is a kind of ``equivalence principle'' of the two $\hbar$'s.

We conclude this brief introduction to the Wigner function by noting that the definition~\eqref{equ:Wigner} can easily be 
generalized~\cite{Wigner1932} from one to an arbitrary number of space dimensions. Moreover, we emphasize that due to the shifted 
arguments in eq.~\eqref{equ:Wigner} the Wigner function of a separable wave function does not separate but contains correlations not 
obvious from the Schr{\"o}dinger formulation of quantum mechanics. This fact stands out most clearly for the example of the Wigner function~\cite{Dahl1982,Praxmeyer2006} of the ground state of the hydrogen atom. Although the wave function only depends on the radial coordinate, the corresponding Wigner function depends not only on the absolute radius of the position $\vec{r}$ and the momentum $\vec{p}$ but also on the angle between $\vec{r}$ and $\vec{p}$. This feature holds true~\cite{Dahl2007} for all s-waves, that is all radial waves. In this way we can even gain deeper insight~\cite{Dahl2006} into the concept of entanglement, which unfortunately goes beyond the scope of the present lectures.

%
\section{Equivalence of inertial and gravitational mass}
%
\label{sec:Equiv_of_inertial&gravitational}
The identity of inertial and gravitational mass expressed by the weak equivalence principle is a cornerstone 
of general relativity~\cite{Misner}. Einstein himself~\cite{ono:45} describes the birth of this idea as follows: 
\begin{quote}
\it I was sitting on a chair in my patent office in Bern. Suddenly, a thought struck me: If a man falls freely, he would not feel his weight. I was taken aback. This simple thought experiment made a deep impression on me. This led to the theory of gravity.
\end{quote}
In the present section we apply the concepts of a classical and a quantum phase-space distribution function to investigate the motion of an ensemble of particles in a linear gravitational potential. Here we focus especially on the appearance of the inertial and gravitational mass in the corresponding expressions. In particular, we show that the classical and the quantum dynamics are identical, and the two masses only enter as their ratio. However, in the energy eigenfunctions they appear independently from each other.

\subsection{Single particle}

When we consider a single particle of inertial mass $m_\i$ moving in one space dimension along the $z$-axis in the presence of a linear gravitational potential 
\begin{align}\label{eq:gravpot}
  V_\g(z) \equiv m_\g g z, 
\end{align}
which contains the gravitational mass $m_\g$ and the gravitational acceleration $g$, the Newton equation of motion reads 
\begin{align}
    m_\i \ddot{z} 	&= -\frac{\drm V_\g}{\drm z} = -m_\g g,
\end{align}
or
\begin{align}\label{equ:zddot}
    \ddot{z}		&= -\frac{m_\g}{m_\i} g \equiv -g'.
\end{align}
Hence, the acceleration of the particle involves the ratio of its inertial and its gravitational mass. 
In principle, this ratio could depend on the specific type of particle, e.\,g. its composition or nature. The weak equivalence principle states that this ratio is identical for all kinds of particles, and therefore can defined to be unity. As a result, the acceleration is independent of the species and moreover, independent of the mass.
In case the weak equivalence principle were violated the acceleration $g'$ defined on the right-hand side of eq.~\eqref{equ:zddot} would be different for different types of particles.

We conclude by recalling the solutions of the Newton equation for the position
\begin{align}\label{eq:zbar-solution}
     \overline{z}(z_0,v_0;t)   	&\equiv 	z_0 + v_0 t - \frac{1}{2} \frac{m_\g}{m_\i} g t^2
				= 		z_0 + v_0 t - \frac{1}{2} g' t^2	
\end{align}
and the velocitiy
\begin{align}\label{eq:vbar-solution}
       \overline{v}(z_0,v_0;t) 	&\equiv 	\dot{\overline{z}}(z_0,v_0;t)
				=	 	v_0 - \frac{m_\g}{m_\i} g t  
				=	 	v_0 -g' t				
\end{align}
subjected to the initial condition $\overline{z}(z_0,v_0;t=0) \equiv z_0$ and 
 $\dot{\overline{z}}(z_0,v_0;t=0) \equiv v_0$.
 
Hence, the acceleration reads
\begin{align}
      \overline{a}(z_0,v_0;t) 	&\equiv	 \ddot{\overline{z}}(z_0,v_0;t) 		
				=	- \frac{m_\g}{m_\i} g 
				=	-g'  ,
\end{align}
in complete agreement with the Newton equation~\eqref{equ:zddot}.
%
\subsection{Ensemble of particles: classical dynamics \`{a} la Liouville}
%
In the preceding section we have recalled the weak equivalence principle as stated for a single particle with a 
well-defined initial position $z_0$ and a well-defined initial velocity $v_0$. However, 
when we deal with an ensemble of many atoms  it is in principle impossible to obtain 
identical initial conditions for all atoms. We therefore have to resort to a description based on a 
distribution function which evolves in time. 
Since the dynamics of classical mechanics is governed by phase space, it is natural to consider a phase-space 
distribution function. In order to compare and contrast classical versus quantum dynamics we first briefly 
review in the present section the classical aspects and then turn in the next section to the quantum ones. 
%
%
\subsubsection{Dynamics: classical Liouville equation}
%
For the case of the linear gravitational potential eq.~\eqref{eq:gravpot} the classical Liouville equation~\eqref{equ:liouville_equ} takes the form
\begin{align}\label{eq:grav-liouville-equ}
  \left(\frac{\partial}{\partial t} + v \frac{\partial}{\partial z}  - \frac{m_\g}{m_\i} g  \frac{\partial}{\partial v}\right) f(z,v;t) = 0\,.
\end{align}
Since only the ratio of inertial and gravitational mass enters into this equation the dynamics of the distribution 
function is independent of the mass of the particles provided the weak equivalence principle holds true. 
In order to verify this feature directly in the time-dependent distribution function we now perform the integration in the formal solution eq.~\eqref{equ:f} of the Liouville equation. 

From the delta function in $v$ we find with the expression~\eqref{eq:vbar-solution} for $\overline{v}$ the relation $\overline{v} =\dot{\overline{z}} =v_0 - g't $, or 
$v_0 =\overline{ v} + g' t$ which allows us to perform the integration over $v_0$ that is 
\begin{align}
  f(z,v;t) 	&= 	\int\limits_{-\infty}^{\infty} \!\!\drm z_0   
			\ \delta[z-\overline{z}(z_0,v + g' t;t)] \ f_0 (z_0,v + g' t)\,.
\end{align}
Moreover, with the help of the delta function in $z$ we obtain from eq.~\eqref{eq:zbar-solution} with the relation
\begin{align}
  z 	&= 	\overline{z}(z_0,v+g' t; t) = z_0 + vt + \frac{1}{2} g' t^2
\end{align}
the dynamics 
\begin{align}\label{eq:f-distribution-dynamics}
  f(z,v;t)	&=	f_0(z - vt - \frac{1}{2} g' t^2, v + g' t) 
\end{align}
of the initial distribution $f_0$ in the linear gravitational potential. It is straightforward to verify that 
this expression satisfies the Liouville equation~\eqref{eq:grav-liouville-equ}.
%
%
\subsubsection{Initial distribution}
%
Our solution eq.~\eqref{eq:f-distribution-dynamics} of the classical Liouville equation shows that the distribution function $f$ of an ensemble 
of particles evolves so that every point $(z,v)$ in phase space moves 
along a classical trajectory. The trajectory starting from $(z_0,v_0)$ has the weight $f_0(z_0,v_0)$.
The trajectories depend on the ratio $m_\g/m_\i$ of the gravitational and inertial mass.
If there is an equivalence principle for the single particle, then it must also hold for the ensemble in a linear potential.

Does this statement imply that the observed probability densities $P=P(z;t)$ and $\tilde{P}=\tilde{P}(v;t)$ are independent 
of the mass of the particles? The answer to this question depends crucially on the preparation mechanism, 
that is, on the way the mass enters into the initial distribution.

In order to bring this feature out most clearly we consider as an initial distribution $f_0$ an ensemble of atoms stored in 
a binding potential $V=V(z)$ in equilibrium with a thermal reservoir of temperature $T$. In this case the 
stationary solution of the Boltzmann equation reads
\begin{align}
  f_0(z_0,v_0)	&=	\mathcal{N} \ \sy{exp} \left[-\frac{1}{2} \frac{m_\i v_0^2}{k_\B T} -\frac{V(z_0)}{k_\B T} \right] 
\end{align}
where $\mathcal{N}$ denotes a constant given by the normalization condition eq.~\eqref{eq:particle_number}.

In the case of an eigenstate in the linear gravitational potential eq.~\eqref{eq:gravpot} with a hard wall at $z=0$ the two masses appear in separate places
in the initial distribution. Although the dynamics depends only on the ratio of the gravitational and inertial masses, they might be separately present 
in the initial distribution $f_0$. Apart from the propagation, which only depends on the ratio, the calculation of the observed time-dependent densities involves integrating over the initial distribution.
Hence, the evolved distribution may depend not only on the ratio of the two masses.
Due to this feature, one might be led to conclude that the equivalence principle is violated. However, this is only 
a remnant of the initial distribution.
%
%
%
\subsection{Ensemble of particles: quantum dynamics and quantum kinematics \`{a} la Wigner}
%
%
In the preceding section we have analyzed the classical dynamics of an ensemble of particles in terms of a classical
phase-space distribution function. In the present section we discuss the corresponding problem in the framework of quantum mechanics using the Wigner function discussed in section \ref{sec:Wigner_functions}.

We first recall that for a potential involving not higher than quadratic expressions of the coordinate the quantum Liouville equation reduces to the classical one, which for a linear gravitational potential reads
\begin{align}\label{eq:W-Liouville-lin-grav}
  \left(\frac{\partial}{\partial t} + v \frac{\partial}{\partial z} - 
			  \frac{m_\g}{m_\i} g \frac{\partial}{\partial v} \right) W 	
 	&=	0    \, .
\end{align}
Hence, classical and quantum dynamics are identical, in particular, the gravitational and the inertial mass only enter through their ratio.

In contrast to the quantum Liouville equation~\eqref{eq:W-Liouville-lin-grav}, which is independent of $\hbar$, the energy eigenvalue equation in phase space 
\begin{align}
  \left[\frac{p^2}{2m_\i} 	 + m_\g g z - \frac{\hbar^2}{8m_\i} \frac{\partial^2}{\partial z^2} 
			\right] W_{E}
		  &= E\  W_{E}
\end{align}
for the Wigner function $W_E$ of an energy eigenstate $\ket{E}$ contains
$\hbar$ explicitly. More importantly, inertial and gravitational mass enter at different places of the equation. As a result, we expect that the Wigner function $W_E$ or the wave function $u_E$ of an energy eigenstate do not only involve the two masses as a ratio, but in a more complicated way.

%
\subsection{Appearance of inertial and gravitational mass in energy eigenstates}
%
The preceding analysis has shown that quantum mechanics manifests itself most prominently in energy eigenstates. 
For this reason we now consider the time-independent Schr{\"o}dinger equation
\begin{align}
  \frac{\drm^2 u_E(z)}{\drm z^2}  + \frac{2m_\i}{\hbar^2} \left[ E - V(z) \right] u_E (z) &= 	0
\end{align}
for the energy wave function $u_E=u_E(z)$.
Here we consider three examples for the center-of-mass motion of an atom: 
(\textit{i}) in a linear gravitational potential, (\textit{ii}) in a linear gravitational potential with a hard wall 
and (\textit{iii}) in the gravitational Coulomb potential. We show that in all three cases the gravitational and 
inertial mass do not appear in the energy wave function as a ratio.    
%
%
%
\subsubsection{Linear potential}
\begin{figure}
\centering
\includegraphics{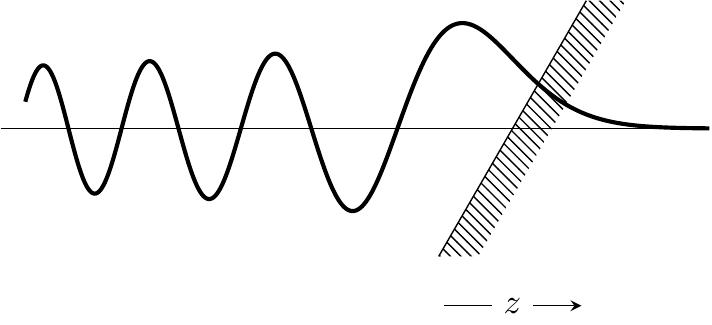}

\caption{The Schr\"odinger energy spectrum of a linear potential is continuous. For a given energy the wave function is given by an Airy function whose modulation is governed by the third root of the product of the inertial and gravitational mass.}
\label{fig.lin-potential}
\end{figure}

%
We start our discussion with the linear gravitational potential as defined in eq.~\eqref{eq:gravpot} and shown in fig.~\ref{fig.lin-potential}, for which the time-independent Schr{\"o}dinger equation reads
\begin{align}
   \frac{\drm^2 u_E(z)}{\drm z^2} - \frac{2m_\i}{\hbar^2} \left[ m_\g g z - E \right] u_E (z) &= 	0\,.
\end{align}
This equation has played a crucial role in the early days of quantum mechanics when Hendrik Kramers~\cite{Kramers26} in 1926 developed his approximation of the time-independent Schr\"odinger equation summarized by the Wentzel-Kramers-Brillouin (WKB) approach. Indeed, Kramers used as a solution the Bessel-function of order $1/3$. However, in 1928 Gregory Breit~\cite{Breit28} used the Airy function to express the solution. Here we follow his approach in order to see the scaling of the energy wave function with respect to the inertial and gravitational mass.

When we introduce the wave vector 
\begin{align}\label{eq:wave-vector}
  k 	&\equiv \left( \frac{2 m_\i m_\g g}{\hbar^2} \right)^{1/3}
\end{align}
together with the dimensionless position $\zeta \equiv k z$ and the energy 
\begin{align}\label{eq:energy-lin-potential}
	\varepsilon	&\equiv		 \left( \frac{2 m_\i}{\hbar^2 m_\g^2 g^2}  \right)^{1/3} E
\end{align}
we arrive at the differential equation 
\begin{align}
    \frac{\drm^2 u_\varepsilon }{\drm \zeta^2} - \left( \zeta - \varepsilon  \right) u_\varepsilon	&=	0 \,.
\end{align}

When we recall the differential equation
\begin{align}
   \frac{\drm^2  \sy{Ai}(\zeta)}{\drm \zeta^2}  - \zeta\, \sy{Ai}(\zeta) = 0
\end{align}
of the Airy function $\sy{Ai}$ we can identify the solution 
\begin{align}
  u_\varepsilon 	&= \mathcal{N} \sy{Ai}(kz-\varepsilon)
\end{align}
where $\mathcal{N}$ is an appropriate normalization constant.

Hence, the modulation of the wave function $u_\varepsilon$ in position space shown in fig.~\ref{fig.lin-potential} and expressed by the wave vector $k$ and defined by eq.~\eqref{eq:wave-vector} depends on the third root of the product of the inertial and gravitational mass. This feature might be experimentally relevant since this wave function manifests itself in the mode function of a Bose-Einstein condensate in the linear gravitational field being released from a trap, as observed for example in ref.~\cite{Koehl01}.
%
%
\subsubsection{Linear potential plus wall}
%
%
\begin{figure}
\centering
\includegraphics{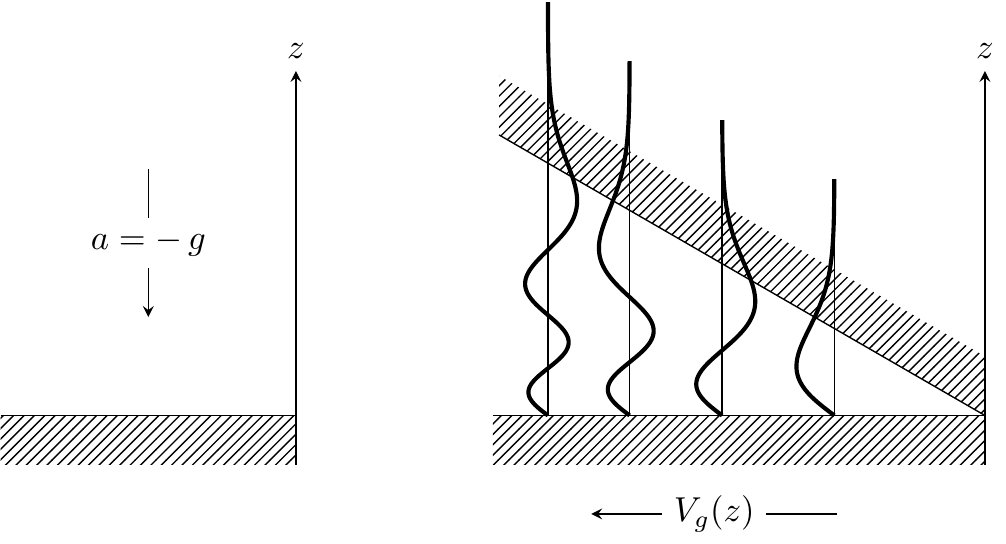}

\caption{The Schr\"odinger energy spectrum of a linear potential in the presence of a hard wall is discrete. The energy eigenvalues, determined by the fact that the Airy function has to vanish at the wall, depend on fractional powers of the inertial and gravitational mass.}
\label{fig.lin-potential+wall}
\end{figure}
In our second example we add to the linear gravitational potential an infinitely tall and infinitely steep 
wall at $z=0$ as shown in fig.~\ref{fig.lin-potential+wall}. This model is of great relevance for cold atoms~\cite{Cohen93,Grimm97} and neutrons~\cite{Abele02}. Indeed, in ref.~\cite{Cohen93,Grimm97} the evanescent wave from a laser beam impinging on a glass plate under total reflection has been used to create a hard wall and to bounce atoms in the gravitational potential. Similarly, neutrons have been used to observe the discrete energy spectrum of this potential. Most recently, even transitions between these states in the gravitational field have been observed~\cite{Jenke11}.

The associated boundary condition $u_\varepsilon(0) = N\cdot\sy{Ai}(-\varepsilon) = 0$ selects from the 
continuum of the preceding example a discrete set 
$\varepsilon_n 	\equiv	a_{n+1}$ where $a_{n+1}$ denotes the zeros of the Airy function. As a result of the scaling of the dimensionless energy $\varepsilon$, as defined in eq.~\eqref{eq:energy-lin-potential}, the 
discrete energies 
\begin{align}
    E_n 			&=		\left( \frac{m_\g^2}{2 m_\i} g^2 \hbar^2 \right)^{1/3} \!\! a_{n+1} 
				\sim  	m_\g^{2/3} m_\i^{-1/3}
\end{align}
involve fractional powers of the gravitational and inertial mass.
%
%
%
\subsubsection{Gravitational Coulomb potential}
%
One might argue that the strange combination of powers of the gravitational and inertial mass might be a consequence of the fact that the potential causing the atom to bounce, and in particular the hard wall, cannot be produced by gravitational fields only. In order to show that this argument does not eliminate the strange powers we now consider the Schr\"odinger equation of a two-body problem bound by gravitational forces only.

For this reason we consider the potential energy 
\begin{align}
          V_\N 	&\equiv 	-\frac{ m_\g M G }{r}
\end{align}
corresponding to the gravitational Coulomb potential originating from a mass $M$ with the gravitational constant 
$G$, where the radial energy eigenvalue equation reads
\begin{align}
 \frac{\drm^2 u_E}{\drm r^2}  + \frac{2m_\i}{\hbar^2} \left[E + \frac{m_\g M G}{r} - \frac{\hbar^2}{2 m_\i}\frac{l(l+1)}{r^2}\right] u_E 
			&= 	0 \,.
\end{align}
Here, $l$ denotes the angular momentum quantum number.

When we introduce the wave vector
\begin{align}
  k	 &\equiv 	\frac{m_\i m_\g M G}{\hbar^2}
\end{align}
and the dimensionless coordinate $\zeta \equiv k r $ we find
\begin{align}
  \frac{\drm^2 u_\varepsilon}{\drm \zeta^2} + \left[\varepsilon + \frac{2}{\zeta} - \frac{l(l+1)}{\zeta^2}\right] u_\varepsilon  
	      &= 	0
\end{align}
where we have introduced the dimensionless energy 
\begin{align}
  \varepsilon &\equiv	\frac{2m_\i E}{(\hbar k)^2} \,.
\end{align}
Hence, we arrive at the energy eigenvalues
\begin{align}
    \varepsilon_n 	&= 	- \frac{1}{n^2}
\end{align}
with the quantum number $n$ and therefore find
\begin{align}
    E_n 	&= 	\frac{m_\i m_\g^2 (M G)^2}{2\hbar^2} \varepsilon_n \, .
\end{align}
Again, the inertial and the gravitational mass do not enter in the form of a ratio but with completely different powers.

%
%
%
\section{Representation-free description of the Kasevich-Chu interferometer}
%
\label{sec:KC_interferometer}
In order to prepare the ground for our second application of the Wigner function formulation of quantum mechanics discussed in section \ref{subsec:analysis_in_wigner_space} we now revisit the Kasevich-Chu interferometer shown in table~\ref{tab.KC} and note that many extremely powerful descriptions of this interferometer exist. They range from a Feynman path-integral approach~\cite{Storey94}, via the ABCD formalism~\cite{Borde02,Borde04,Borde08} to the interpretation as a quantum clock~\cite{Mueller10,Hohensee12}.

In this section we introduce yet another approach in terms of state vectors in Hilbert space and non-commuting time-evolution operators. We first discuss the setup and, in particular, focus on the use of laser pulses to create a beam splitter and a mirror for atoms, and then address the motion of an atom in the gravitational field. Our representation-free approach~\cite{Kajari10,Schleich13,SchleichNJP13} describes the interferometer as a sequence of unitary operators. In this way we identify for a linear potential the non-commutativity of two time-evolutions as the origin of the phase shift and generalize our approach to include gravity gradients.

\subsection{Building blocks}
In the Kasevich-Chu interferometer~\cite{Kasevich91,Kasevich92} cold atoms interact with three appropriately 
tailored laser pulses formed by two counter-propagating running electromagnetic waves with wave 
vectors $\mathbf k_1$ and $\mathbf k_2$. The resulting two-photon Raman transition between the two atomic states $\ket{g_1}$ and $\ket{g_2}$ shown in table~\ref{tab.KC} is associated with an effective momentum transfer of $\hbar k \equiv \hbar|\mathbf k_1-\mathbf k_2|$. Whereas the transition from $\ket{g_1}$ to $\ket{g_2}$ 
increases the momentum of the center-of-mass motion of the atom by $\hbar k$, the reverse transition from 
$\ket{g_2}$ to $\ket{g_1}$ reduces it by the same amount. The parameters of the pulses are designed so that 
the first and the last at $t=0$ and $t=2T$, respectively, serve as perfect beam splitters, that is they create a coherent 
superposition of equal weight, whereas the second pulse at $t=T$ serves as a mirror. In this way the three 
Raman pulses coherently split and recombine the atomic beam and 
create the Kasevich-Chu interferometer.

In our description we carry out, for the sake of simplicity, a one-dimensional analysis with a gravitational 
potential $V=V(z)$   along the $z$-axis, which is also the propagation direction of the laser fields. 
Moreover, we assume that during the duration of the short pulses the atoms do not move significantly.

\subsubsection{Raman pulses}
%
%
In order to understand the action of the beam splitter made out of Raman pulses we consider the interaction of a 
three-level atom consisting of two ground states $\ket{g_1}$ and $\ket{g_2}$ and an excited state $\ket{e}$ as shown in table~\ref{tab.KC}  with 
two laser pulses of the form 
\begin{align}
  E_1(z,t)	&=\mathcal{E}_1(t) \e^{-\i\nu_1 t} \e^{\i(k_1 z+\phi_1)}+~\text{c.c.}
\end{align}
and
\begin{align}
  E_2(z,t)	&=\mathcal{E}_2(t) \e^{-\i\nu_2 t} \e^{-\i(k_2 z+\phi_2)}+~\text{c.c.}
\end{align}
Here $\mathcal{E}_j$, $\nu_j$, $k_j$ and $\phi_j$ denote the time-dependent envelope, the frequency, the 
wave number and the phase of the $j$th field.
 
We assume that the laser pulse is so short that the atom does not move significantly during this time.
Therefore, we can neglect for the time being the motion, that is the position $z$ of the atom is fixed during the laser pulse.

Within the rotating-wave approximation~\cite{Schleich:Quantum-optics} the elimination of the excited state~\cite{Kasevich92,SchleichNJP13} using second-order perturbation theory yields the unitary matrix
\begin{align}
 \hat{U}_\text{lp}(\theta)=\cos \theta \,\mathbbm{1} -\i \sin \theta \left[ \e^{\i[kz+\phi(t)]} \ketbra{g_2}{g_1} + \e^{-\i[kz+\phi(t)]} \ketbra{g_1}{g_2}\right]
\label{eq:KC:Ulp}
\end{align}
where $k\equiv k_1+k_2$, $\phi\equiv\phi_1+\phi_2$ and $\theta$ is the pulse area determined by the product of the two envelopes $\mathcal{E}_1$ and $\mathcal{E}_2$.
Moreover, we have introduced the identity operator 
\begin{align}
  \mathbbm{1} &\equiv  \ketbra{g_1}{g_1}+ \ketbra{g_2}{g_2}
\end{align}
of the atom confined in its dynamics to the two ground states only.

\subsubsection{Beam splitter}
%
%
Rather than deriving eq.~\eqref{eq:KC:Ulp} we now try to motivate its form. For a detailed derivation we refer to~\cite{SchleichNJP13}.

Indeed, it is instructive to see how $\hat{U}_\text{lp}(\theta)$ acts on a single ground state, for example $\ket{g_1}$. In this case, we find from eq.~\eqref{eq:KC:Ulp} the identity  
\begin{align}
 \hat{U}_\text{lp}(\theta)\ket {g_1}=\cos \theta \,\ket{g_1} -\i \sin \theta\, \e^{\i[kz+\phi(t)]} \ket{g_2} \label{e.U_lpket-g_1}\, ,
\end{align}
that is a superposition of the two ground states. The probability amplitude of the original state $\ket{g_1}$ is given by $\cos \theta$, whereas the one of $\ket{g_2}$, that is of the newly populated state consists of the product of  $(-\i) \sin \theta$ and the phase factor $\exp\left[\i(kz+\phi(t))\right]$. The factor $(-\i)$ is a consequence of the Schr\"odinger equation, and the appearance of the trigonometric functions is due to the effective Rabi oscillations between the two ground states. In order to make a transition a non-vanishing field is necessary. Therefore, the probability amplitude for the transition has to be at least linear in the Rabi frequency. This fact explains the appearance of the sine-function. In order to remain in the original state no field is required which gives rise to the cosine-function.

Most important in the present discussion of eq.~\eqref{e.U_lpket-g_1} is the position-dependent phase factor in front of $\ket{g_2}$, which is decisive when we now allow the atom to move after the pulse. In this case, the coordinate $z$ turns into a dynamical quantum mechanical variable, that is the position operator $\hat{z}$. With the unitary operators
\begin{align}
 \hat{U}^{(\pm)}(t)\equiv\e^{\pm\i[k \hat{z} + \phi(t)]} 
 \label{e.defUpm}
\end{align}
eq.~\eqref{e.U_lpket-g_1} takes the form 
\begin{align}
 \hat{U}_\text{lp}(\theta)\ket {g_1}=\cos \theta \,\ket{g_1} -\i \sin \theta\, \hat{U}^{(+)}(t) \ket{g_2}
\label{eq:KC:Ulpbs1}
\end{align}
and we recognize that apart from the phase $\phi(t)$ at the time $t$ of the pulse the operator $\hat{U}^{(+)}$ is a displacement operator in momentum. Hence, the transition from $\ket{g_1}$ to $\ket{g_2}$ is associated with a gain in momentum by an amount $\hbar k$. 

Similarly, we find from eq.~\eqref{eq:KC:Ulp} the relations
\begin{align}
 \hat{U}_\text{lp}(\theta)\ket {g_2}=\cos \theta \,\ket{g_2} -\i \sin \theta\, \hat{U}^{(-)}(t) \ket{g_1}
\label{eq:KC:Ulpbs2}
\end{align}
for a transition from $\ket{g_2}$ to $\ket{g_1}$ which implies a loss of the momentum by the same amount due to the appearance of the operator $\hat{U}^{(-)}$.

The pulse area $\theta$ depends on and can be tuned by the electric field envelopes $\mathcal{E}_1$ and $\mathcal{E}_2$. Indeed, for the special choice $\theta \equiv \pi/4 $ we find from eq.~\eqref{eq:KC:Ulpbs1} with 
$\cos(\pi/4)=\sin(\pi/4)={1}/{\sqrt{2}}$ the expression 
\begin{align}
\hat{U}_\text{lp}(\pi/4)\ket {g_1}=\frac{1}{\sqrt{2}} \left( \ket{g_1} -\i \, \hat{U}^{(+)}(t) \ket{g_2}\right) \label{e.beam-splitter}
\end{align}
indicating an equally weighted superposition of the ground state $\ket{g_1}$ and the state $\ket{g_2}$.

However, more importantly we note that there is also a superposition in the center-of-mass motion. Indeed, when we assume that the ground state is associated with a given wave function in momentum representation, the excited state is of the same form but displaced by $\hbar k$. As a result, the transformation eq.~\eqref{e.beam-splitter} corresponds to a beam splitter.

\subsubsection{Mirror}
%
%
Another interesting choice of $\theta$ is $\theta \equiv \pi/2$ where due to $\cos(\pi/2)=0$ and $ \sin(\pi/2)=1$ we find from eq.~\eqref{eq:KC:Ulpbs1} the expression
\begin{align}
 \hat{U}_\text{lp}(\pi/2)\ket {g_1}=-\i \, \hat{U}^{+}(t) \ket{g_2}\,.
\end{align}
In this case, we have not created a coherent superposition of $\ket{g_1}$ and $\ket{g_2}$, but have made a complete transition. The atom leaves the laser pulse in the state $\ket{g_2}$, has gained momentum, and obtained the additional phase $\phi(t)$. 

In contrast, an atom initially in $\ket{g_2}$ experiences a loss in momentum due to the transition to $\ket{g_1}$ as indicated by the relation
\begin{align}
 \hat{U}_\text{lp}(\pi/2)\ket {g_2}=-\i \, \hat{U}^{(-)}(t) \ket{g_1}
\end{align}
following from eq.~\eqref{eq:KC:Ulpbs2}. This discussion shows that the choice of $\theta=\pi/2$ corresponds to a mirror.

\subsubsection{Time evolution between the laser pulses}
%
So far, we have only concentrated on the influence of the laser pulse on the atom changing the internal states and the center-of-mass motion. We now analyze the motion of the atom between the laser pulses in the presence of a potential $V=V(z)$ corresponding to the Hamiltonian 
\begin{align}\label{e.def-H_1}
 \hat{H}_1 \equiv \frac{\hat{p}^2}{2m} + V(\hat{z})\, .
\end{align}
Here we have included the subscript $1$ in order to indicate that we consider the motion in the gravitational field of an atom in the internal state $\ket{g_1}$.

Moreover, we emphasize that the mass $m$ appearing in $\hat{H}_1$ is the inertial mass $m_\i$. For the sake of simplicity in notation we have dropped for the time being the subscript $\i$. However, we will return to the question of the appearance of inertial and gravitational mass when we discuss the phase shift in the Kasevich-Chu interferometer.

The dynamics of the center-of-mass motion between two laser pulses separated in time by $T$ reads
\begin{align}
 \ket{\psi(T)}&=\e^{-\i \hat{H}_1 T/\hbar} \ket{\psi_0}\equiv \hat{U}_1 \ket{\psi_0} \, ,
\end{align}
where $\ket{\psi_0}$ is the initial state.

\subsection{The interferometer as an operator sequence}
%
We are now in the position to provide the representation-free approach~\cite{Kajari10,Schleich13,SchleichNJP13} towards the Kasevich-Chu interferometer representing the beam splitters, mirrors and the time evolutions in the gravitational field in terms of unitary operators. The phase shift in the interferometer is then a consequence of the non-commutativity of the time evolutions in the gravitational field for an atom in the excited and in the ground state.

\subsubsection{Order of events in the two paths}
%

\setlength{\tabcolsep}{0pt}
\renewcommand{\arraystretch}{1.5} 

 \begin{table}
 \begin{center}
   \caption{Kasevich-Chu interferometer for a three-level atom interacting with three short laser pulses at $t=0$, $t=T$, and $t=2T$ represented in a space-time diagram. At the bottom we show the unitary operators as well as the normalization factors associated with the propagation of the atom in the two different states $\ket{g_1}$ and $\ket{g_2}$ through the interferometer. Since unitary operators always act on the states to their right, time flows in this diagram to the left.}
   \label{tab.KC}
   
\begin{tabular}{R{1.8cm}L{0.5cm}C{1.4cm}C{0.35cm}C{1.4cm}C{0.35cm}C{1.4cm}p{0.9cm}p{1.9cm}p{1.0cm}}
    \noalign{\hrule height .04cm}\\[-0.3cm] 
    \multicolumn{10}{c}{
    \includegraphics[width=11cm]{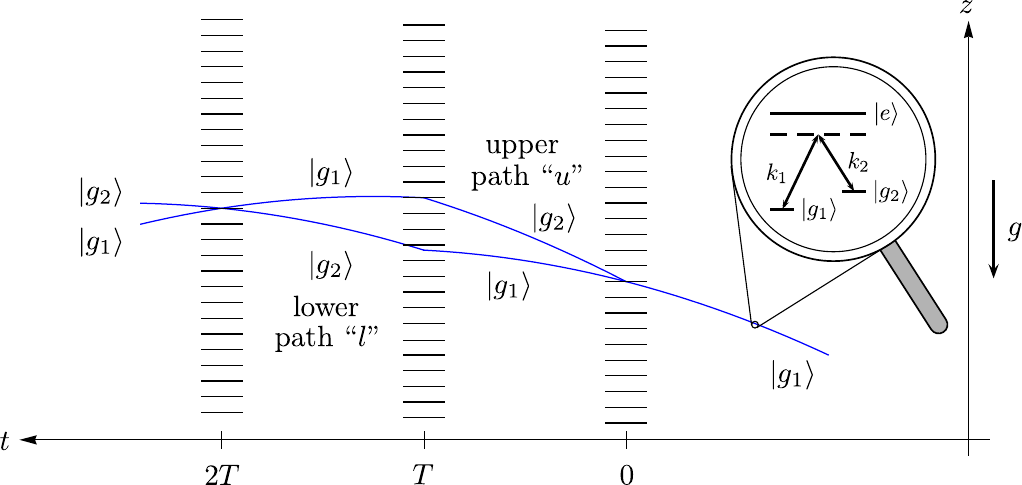}}\\[0.0cm]
     \hline
      $\ket{\psi_u}=$ & $\mathcal{N}_u$ & $\mathbbm{1}$  &  $\hat{U}_1$ & $\hat{U}^{(-)}(T)$ & $\hat{U}_1$ & $\hat{U}^{(+)}(0)$ & $\ket{\psi_0}$ & ``upper path''&\\
      $\ket{\psi_l}=$ & $\mathcal{N}_l$ & $\hat{U}^{(-)}(2T)$  &  $\hat{U}_1$ & $\hat{U}^{(+)}(T)$ & $\hat{U}_1$ & $\mathbbm{1}$ & $\ket{\psi_0}$ & ``lower path''&\\
     \hline
      $\mathcal{N}_u=-\frac12=$&  & $\frac{1}{\sqrt{2}}$  &  $\times$ & $(-\i)$ & $\times$ & $\frac{-\i}{\sqrt{2}}$ &  & ``upper path''&\\
      $\mathcal{N}_l=-\frac12=$&  & $\frac{-\i}{\sqrt{2}}$  &  $\times$ & $(-\i)$ & $\times$ & $\frac{1}{\sqrt{2}}$ &  & ``lower path''&\\[0.2cm]
     \noalign{\hrule height .04cm}
   \end{tabular}
 \end{center}
 \end{table}

\setlength{\tabcolsep}{6pt}
\renewcommand{\arraystretch}{1.0}

The initial state 
\begin{align}
\ket{\Psi_0}\equiv \ket{\psi_0}\ket{g_1}
\end{align}
of the complete system consisting of the center-of-mass motion $\ket{\psi_0}$ and the internal state $\ket{g_1}$ has to be propagated through the interferometer. For this purpose we follow the upper and the lower path shown in table~\ref{tab.KC} and focus for the time being only on the atoms exiting in the state $\ket{g_1}$. We first concentrate on the unitary operations and then take into account the numerical factors $1/\sqrt{2}$ and $(-\i)$ due to the beam splitters and the mirrors.

We start our discussion with the upper path where the atom gets excited by the first laser pulse at $t=0$  into $\ket{g_2}$. This process is associated with the unitary operator $\hat{U}^{(+)}(0)$. Then the atom propagates in the gravitational field corresponding to the unitary operator $\hat{U}_1$ followed by the operator $\hat{U}^{(-)}(T)$ representing the de-excitation to $\ket{g_1}$ due to the mirror pulse at $t=T$. The subsequent propagation in the state $\ket{g_1}$ through the gravitational field is again described by the unitary operator $\hat{U}_1$. Since we concentrate on atoms exiting the interferometer in $\ket{g_1}$ and the atom is already in this state, the action of the third laser pulse at $t=2T$ is given by the identity operator $\mathbbm{1}$. As a result, we find in complete accordance with the bottom part of table~\ref{tab.KC}, on the upper path the unitary operator
\begin{align}
 \hat{U}_u=\mathbbm{1}\,\hat{U}_1 \hat{U}^{(-)}(T)\hat{U}_1\hat{U}^{(+)}(0) \equiv \e^{-\i[\phi(T)-\phi(0)]} \hat{U}_1 \hat{U}_2 \, , \label{eq:U_u}
\end{align}
where we have introduced the abbreviation
\begin{align}
\hat{U}_2&\equiv \e^{-\i k\hat{z}}\e^{-\i\hat{H}_1 T/\hbar} \e^{ik\hat{z}}= \e^{-\i\hat{H}_2 T/\hbar} 
\end{align}
where 
\begin{align}
 \hat{H}_2 \equiv \frac{(\hat{p}+\hbar k)^2}{2m} + V(\hat{z})\, . \label{e.def-H_2}
\end{align}
In the last step we have combined the displacement operators with the unitary time evolution in the gravitational field which has introduced in the kinetic energy the shift of the momentum by $\hbar k$. Hence, the propagation in the excited state $\ket{g_2}$ is governed by the Hamiltonian $\hat{H}_2$, whereas in the ground state $\ket{g_1}$ it is given by $\hat{H}_1$ as expressed by the unitary operator $\hat{U}_1$.

On the lower path the atom first remains in its ground state $\ket{g_1}$ while it interacts with the first laser pulse at $t=0$ and then propagates in this internal state through the gravitational field, until it gets excited by the second laser pulse at $t=T$. The last propagation in the gravitational field in $\ket{g_2}$ is followed by a de-excitation into the exit state $\ket{g_1}$ at $t=2T$. As a consequence, we have in complete accordance with table~\ref{tab.KC} the sequence of unitary operations
\begin{align}
 \hat{U}_l= \hat{U}^{(-)}(2T)\hat{U}_1\hat{U}^{(+)}(T)\hat{U}_1\, \mathbbm{1}\equiv \e^{-\i[\phi(2T)-\phi(T)]} \hat{U}_2 \hat{U}_1 \, . \label{eq:U_l}
\end{align}
Apart from phase factors $\hat{U}_u$ and $\hat{U}_l$ differ in the order of $\hat{U}_1$ and $\hat{U}_2$ reflecting the fact that the order of events is different in the two paths. Indeed, in the upper path the atom first propagates in $\ket{g_2}$  and then in $\ket{g_1}$, while in the lower path it is just the opposite.

Next, we briefly address the question of the numerical factors. For every beam splitter we have to include a factor $1/\sqrt{2}$ and every time the internal state is changed we multiply by $(-\i)$. At a mirror, we have to include a factor of $(-\i)$. As shown in table~\ref{tab.KC}, we obtain for the upper path the numerical factor
\begin{align}
     \mathcal{N}_u&=\frac{1}{\sqrt{2}}\cdot(-\i)\cdot\frac{-\i}{\sqrt{2}}=-\frac12 \, .
\end{align}
The fact that the order of events is different on the lower path is of no consequence for the corresponding numerical factor $\mathcal{N}_l$ and leads to an identical result, that is
\begin{align}
     \mathcal{N}_l&=\frac{-\i}{\sqrt{2}}\cdot(-\i)\cdot\frac{1}{\sqrt{2}}=-\frac12 =\mathcal{N}_u \, .
\end{align}
We conclude by emphasizing that the quantum states of the center-of-mass motion and the atomic ones are entangled. In table~\ref{tab.complete-state-KC} we summarize the complete state at the various stages of the interferometer.

\setlength{\tabcolsep}{0pt}
\renewcommand{\arraystretch}{1.5} 

 \begin{table}
 \begin{center}
   \caption{Quantum state $\ket{\Psi}$ of the complete system consisting of the center-of-mass motion and internal states $\ket{g_1}$ and $\ket{g_2}$ at the different stages of the Kasevich-Chu interferometer in terms of unitary operators.}
   \label{tab.complete-state-KC}
  \begin{tabular}{p{0cm}L{2.0cm}L{1.7cm}C{0.5cm}L{9.3cm}p{0cm}}
     \noalign{\hrule height .04cm}\\[-0.3cm]
      & before \newline first pulse & $\ket{\Psi(0-\epsilon)}$ & $=$ & $\ket{\psi_0} \ket{g_1}$ &  \\
      & after \newline first pulse & $\ket{\Psi(0+\epsilon)}$ & $=$ & $\frac{1}{\sqrt{2}}\left[\ket{\psi_0} \ket{g_1} - \i\, \hat{U}^{(+)}(0) \ket{\psi_0} \ket{g_2} \right]$ &  \\
      & before \newline second pulse & $\ket{\Psi(T-\epsilon)}$ & $=$ & $\frac{1}{\sqrt{2}}\left[\hat{U}_1 \ket{\psi_0} \ket{g_1} - \i\, \hat{U}_1 \hat{U}^{(+)}(0) \ket{\psi_0} \ket{g_2} \right]$ &  \\
      & after \newline second pulse & $\ket{\Psi(T+\epsilon)}$ & $=$ & $\frac{-\i}{\sqrt{2}}\left[\hat{U}^{(+)}(T) \hat{U}_1 \ket{\psi_0} \ket{g_2} - \i\, \hat{U}^{(-)}(T) \hat{U}_1 \hat{U}^{(+)}(0) \ket{\psi_0} \ket{g_1} \right]$ & \\
      & before \newline third pulse & $\ket{\Psi(2T-\epsilon)}$ & $=$ & $\frac{-\i}{\sqrt{2}}\left[\hat{U}_1 \hat{U}^{(+)}(T) \hat{U}_1 \ket{\psi_0} \ket{g_2} - \i\, \hat{U}_1 \hat{U}^{(-)}(T) \hat{U}_1 \hat{U}^{(+)}(0) \ket{\psi_0} \ket{g_1} \right]$ &  \\       
      & after \newline third pulse & $\ket{\Psi(2T+\epsilon)}$ & $=$ & $\frac{-\i}{2} \left\{-\i\left[\hat{U}^{(-)}(2T) \hat{U}_1 \hat{U}^{(+)}(T) \hat{U}_1 + \hat{U}_1 \hat{U}^{(-)}(T) \hat{U}_1 \hat{U}^{(+)}(0)\right]  \ket{\psi_0} \ket{g_1} \right.$ & \\
     & & & &  $ \left. + \left[\hat{U}_1 \hat{U}^{(+)}(T) \hat{U}_1 - \hat{U}^{(+)}(2T) \hat{U}_1 \hat{U}^{(-)}(T) \hat{U}_1 \hat{U}^{(+)}(0)  \right] \ket{\psi_0} \ket{g_2} \right\}$ &  \\    
     [0.2cm]
     \noalign{\hrule height .04cm}
   \end{tabular}
	\end{center}
 \end{table}

\setlength{\tabcolsep}{6pt}
\renewcommand{\arraystretch}{1.0} 

\subsubsection{Exit port probability and phase-shift operator}
In the next step, we connect the unitary operators with the number of atoms in the exit ports. When we again restrict ourselves to the atoms in the state $\ket{g_1}$ the state $\ket{\psi_1}$ corresponding to the center-of-mass motion in the exit port of the interferometer is a coherent superposition
\begin{align}
 \ket{\psi_1}\equiv\ket{\psi_u} + \ket{\psi_l}
\end{align}
of the states
\begin{align}
 \ket{\psi_u}\equiv\mathcal{N}_u\, \e^{-\i [\phi(T)- \phi(0)]} \hat{U}_1 \hat{U}_2 \ket{\psi_0}
\label{eq:KC:psiu}
\end{align}
and
\begin{align}
 \ket{\psi_l}\equiv\mathcal{N}_l\, \e^{-\i [\phi(2T)- \phi(T)]} \hat{U}_2 \hat{U}_1 \ket{\psi_0}
\label{eq:KC:psil}
\end{align}
created by unitary time evolution along the upper and the lower path, respectively.

Hence, the probability $P_{g_1}$ to be at the exit of the interferometer in the internal state $\ket{g_1}$ reads
\begin{align}
P_{g_1} \equiv \left\langle \psi_1| \psi_1\right\rangle = \left\langle \psi_u|\psi_u\right\rangle + \left\langle \psi_l|\psi_l\right\rangle + \left\langle \psi_l|\psi_u\right\rangle +\left\langle \psi_u|\psi_l\right\rangle\, ,
\end{align}
and translates with the definitions eqs.~\eqref{eq:KC:psiu} and~\eqref{eq:KC:psil} of the states $\ket{\psi_u}$ and $\ket{\psi_l}$ into
\begin{align}
P_{g_1} = \frac12 \left[ 1 + \frac12 \left(\e^{\i\delta\phi} \sandwich{\psi_0}{\hat{U}_\varphi}{\psi_0} +\text{c.c.} \right) \right]\, .
\label{eq:probability}
\end{align}
Here we have introduced the abbreviations 

\begin{align}
 \delta\phi \equiv \phi(2T)-2\phi(T)+\phi(0)
 \label{eq:KC:laserphase}
\end{align}
for the combination of the phases of the three laser pulses at the times $t=0$, $T$ and $2T$ and the phase-shift operator 
\begin{align}
\hat{U}_\varphi\equiv\hat{U}_1^\dagger \hat{U}_2^\dagger\hat{U}_1 \hat{U}_2\, .
\label{eq:KC:phaseshiftoperator}
\end{align}
The interference term in the exit probability $P_{g_1}$ is determined by the real part of the product of the phase factor $\exp(\i\delta \phi)$ governed by the difference of the laser phases and the expectation value $\langle \hat{U}_\varphi \rangle$ of the phase-shift operator. Here, the initial state $\ket{\psi_0}$ of the center-of-mass motion is used in the average.

The phases of the three laser pulses enter into the interference term in equation \eqref{eq:probability} as a discrete second derivative in time. This fact stands out most clearly when we use the Taylor expansion
\begin{align}
 \phi(t) &\cong \phi(0) + \dot{\phi}(0)\, t + \frac12 \ddot{\phi}(0)\, t^2+\dots
\end{align}
of $\phi=\phi(t)$ which with the definition~\eqref{eq:KC:laserphase} of $\delta \phi$ leads us to the expression
\begin{align}
 \delta \phi \cong \ddot{\phi}(0)\, T^2\,.
\end{align}
Indeed, $\delta\phi$ is the discrete acceleration of the laser phase as expressed by the second derivative of the phase at $t=0$. Here dots denote derivatives with respect to time.

Moreover, the phase-shift operator $\hat{U}_\varphi$ enjoys a very intuitive interpretation: We propagate along a closed path, first forward in time in  the state $\ket{g_2}$, and then in $\ket{g_1}$ corresponding to the upper path. Moreover, we go backwards in time, first in $\ket{g_2}$ and then in $\ket{g_1}$, representing the time-reversed lower path.

%
%
%
\subsubsection{Acceleration in phase-shift operator}
%
An important ingredient of the phase-shift operator is the operator product
\begin{align}
 \hat{U}_1\hat{U}_2 \equiv \e^{-\i \hat{H}_1 T/\hbar} \e^{-\i\hat{H}_2 T/\hbar}
\label{eq:upper-opseq}
\end{align}
where the two Hamiltonians $\hat{H}_1$ and $\hat{H}_2$ do not commute with each other. Indeed, we find from the definitions equation \eqref{e.def-H_1} and equation \eqref{e.def-H_2} the representation
\begin{align}
 \hat{H}_2 \equiv \hat{H}_1 + \frac{\hbar k}{m} \hat{p} + \frac{(\hbar k)^2}{2m}
\label{eq:KC:H2}
\end{align}
which yields the commutator
\begin{align}
 [\hat{H}_1,\hat{H}_2]= \frac{\hbar k}{m}[\hat{H}_1 ,\hat{p}]\,.
\end{align}

With the Heisenberg equation of motion
\begin{align}
 \dot{\hat{p}}=\frac{\i}{\hbar}[\hat{H}_1 ,\hat{p}]
\end{align}
the commutator reads
\begin{align}
 [\hat{H}_1,\hat{H}_2] = -\i \frac{\hbar^2 k}{m} \dot{\hat{p}} = -\i \hbar^2 k \ddot{\hat{z}}
\end{align}
and is proportional to the acceleration $\ddot{\hat{z}}$ of the particle. We emphasize that this fact is independent of the form of the potential.

Since the two Hamiltonians $\hat{H}_1$ and $\hat{H}_2$ do not commute we cannot combine the two unitary operators $\hat{U}_1$ and $\hat{U}_2$ into a single one by simply adding $\hat{H}_1$ and $\hat{H}_2$. As expressed by the Baker-Campbell-Hausdorff theorem there will be corrections involving nested commutators made out of the commutator of $\hat{H}_1$ and $\hat{H}_2$. Hence, it is the acceleration of the atom that determines the phase-shift operator.

This fact stands out most clearly when we cast the phase-shift operator defined by eq.~\eqref{eq:KC:phaseshiftoperator} into the form
\begin{align}\label{eq:UphiAB}
\hat{U}_\varphi= \e^{\hat{A}} \e^{\hat{B}}\e^{-\hat{A}} \e^{-\hat{B}}\, ,
\end{align}
where we have introduced the abbreviations
\begin{align}
\hat{A}\equiv \i \frac{T}{\hbar} \hat{H}_1
\end{align}
and
\begin{align}
\hat{B}\equiv \i \frac{T}{\hbar} \hat{H}_2 =\i \frac{T}{\hbar} \left[\hat{H}_1 + \frac{\hbar k}{m} \hat{p}(0) + \frac{(\hbar k)^2}{2m} \right] \, .
\label{eq:KC:B}
\end{align}
In the last step we have used the representation eq.~\eqref{eq:KC:H2} of $\hat{H}_2$ and $\hat{p}(0)$ is the momentum operator in the Heisenberg picture at $t=0$.

Indeed, with the help of the operator identity~\cite{Louisell}
\begin{align}
\e^{\hat{A}} \e^{\hat{B}}\e^{-\hat{A}} = \exp\left[ \e^{\hat{A}}\, \hat{B}\, \e^{-\hat{A}} \right] \equiv \exp\left[{\hat{B}(T)}\right]
\end{align}
and the Baker-Campbell-Hausdorff relation the phase-shift operator in the form equation \eqref{eq:UphiAB} reads
\begin{align}
\hat{U}_\varphi =\e^{\hat{B}(T)} \e^{-\hat{B}}=\exp\left(\hat{B}(T)- \hat{B} -\frac12 [\hat{B}(T),\hat{B}]+\ldots\right)\, .
\end{align}
When we recall the definition~\eqref{eq:KC:B} of the operator $\hat{B}$ we find with
\begin{align}
\hat{B}(T)=\i \frac{T}{\hbar}  \exp\left[\i \frac{T}{\hbar} \hat{H}_1\right]\hat{H}_2 \exp\left[-\i\frac{T}{\hbar}  \hat{H}_1\right]= \i\frac{T}{\hbar} \left[\hat{H}_1 + \frac{\hbar k}{m} \hat{p}(T) + \frac{(\hbar k)^2}{2m} \right]\, ,
\end{align}
where
\begin{align}
\hat{p}(T) \equiv \exp\left[\i \frac{T}{\hbar} \hat{H}_1\right]\hat{p}(0) \exp\left[-\i\frac{T}{\hbar}  \hat{H}_1\right]
\end{align}
is the momentum operator at time $T$ in the Heisenberg picture, the relation
\begin{align}
 \hat{B}(T) = \hat{B} + \i \frac{k}{m} \left[\hat{p}(T)- \hat{p}(0) \right]T\,.
\label{eq:KC:BT}
\end{align}

Hence, we arrive at the expression
\begin{align}
\hat{U}_\varphi \equiv \exp\left\{  \i \frac{k}{m} \left[ \hat{p}(T)-\hat{p}(0)\right] T - \frac{1}{2} [\hat{B}(T),\hat{B}]+\ldots \right\}
\label{eq:KC:Uphi}
\end{align}
for the phase-shift operator, which involves apart from commutators between $\hat{B}(T)$ and $\hat{B}$ the difference $\hat{p}(T)-\hat{p}(0)$ between the momentum operators at $T$ and $0$, that is the acceleration of the atom. We emphasize that this result is valid for an arbitrary potential $V=V(z)$.
%
\subsection{Application to special potentials}
%
So far our description of the Kasevich-Chu interferometer has not made any use of the form of the potential $V$. We now illustrate the results obtained in the preceding section and, in particular, analyze the phase-shift operator $\hat{U}_\varphi$ defined by eq.~\eqref{eq:KC:Uphi} for two specific examples for $V$: (\textit{i}) a linear gravitational potential, and (\textit{ii}) a linear gravitational potential including gravity gradients.

We show that in the first case the phase-shift operator reduces to a c-number phase factor. In contrast, the presence of gravity gradients leads, apart from additional c-number phase factors, to the emergence of the displacement operator containing the position and the momentum operator. As a result, the interference term in the exit probability $P_{g_1}$ depends on the initial state $\ket{\psi_0}$ of the center-of-mass motion, determines the contrast, and introduces a state-dependent phase shift.
%
\subsubsection{Linear potential}
%
In order to analyze the expression~\eqref{eq:KC:Uphi} for the phase-shift operator we now consider the linear potential
\begin{align}
 V_{\g}(z) \equiv m_\g gz\, ,
\end{align}
where we have again included the subscript $\g$ to denote the gravitational mass $m_{\g}$. 

The equation of motion resulting from this potential gives rise to the momentum operator
\begin{align}
\hat{p}(T)=\hat{p}(0)-m_\g  g T
\end{align}
at time $T$ and eq.~\eqref{eq:KC:BT} reduces to
\begin{align}
\hat{B}(T)= \hat{B}-\i \delta \varphi_\g\, ,
\end{align}
where we have introduced the phase shift
\begin{align}
\delta \varphi_\g \equiv \frac{m_\g}{m_\i}k g T^2 
\end{align}
in the Kasevich-Chu interferometer in the presence of a linear gravitational potential.
Here we have also recalled that in eq.~\eqref{eq:KC:BT} the quantity $m$ denotes the inertial mass $m_\i$. As a result, only the ratio of the two masses enters into the phase shift.

Since $\hat{B}(T)$ and $\hat{B}$ only differ by a c-number, the higher order corrections in the Baker-Campbell-Hausdorff theorem vanish and the phase-shift operator 
\begin{align}
\hat{U}_\varphi = \e^{-\i \delta \varphi_{\g}}
\label{eq:KC:UphiLin}
\end{align}
is a c-number phase factor. 

As a consequence, the exit port probability
\begin{align}\label{eq:P_g_1-linear}
P_{g_1} &= \frac12 \left[ 1+ \frac12 \left(\e^{\i(\delta\phi-\delta\varphi_{\g})}+ \text{c.c.}\right)\right] = \frac12 \left[ 1+  \cos\left(\delta\phi-\delta\varphi_{\g}\right)\right]
\end{align}
following from eq.~\eqref{eq:probability} is independent of the initial state $\ket{\psi_0}$ of the center-of-mass motion. Here we have assumed that $\ket{\psi_0}$ is normalized.

Hence, in the case of a constant acceleration the total phase shift $\delta\phi- \delta\varphi_{\g}$ in the Kasevich-Chu interferometer is given by the difference of two phases: (\textit{i}) The phase $\delta \phi$ is the discrete second derivative in time, that is the acceleration of the laser phase, and (\textit{ii}) the phase $\delta \varphi_{\g}$, which is due to the gravitational acceleration of the atom, involves the ratio of the gravitational and inertial mass and reflects the non-commutativity of the two time-evolutions on the two paths. We emphasize that in this representation-free description of the Kasevich-Chu interferometer no trajectories and no actions enter.

%
\subsubsection{Gravity gradients}
%
\label{subsubsec:Gravity-gradients}
So far we have discussed the interferometer in the presence of a linear gravitational potential. We now extend our consideration to include gravity gradients as expressed by the potential 
\begin{align}
 V_\Gamma(z)\equiv m_\text{g}gz+\frac{1}{2} m_\text{g}\Gamma z^2\,.
\end{align}
Although we now focus on the limit of weak gravity gradients, that is $\Gamma T^2\ll1$, we emphasize that an exact treatment is possible~\cite{Kleinert,Audretsch1994,Antoine2003,Zeller14,Roura14} and is in fact pursued in the phase-space approach of section~\ref{subsec:analysis_in_wigner_space}.

In appendix \ref{app:Gradients} we obtain in this limit the phase-shift operator
\begin{align}
\hat{U}_\varphi \cong \exp\left[ -\i \left( \delta \tilde{\varphi}_\g+ \frac{m_\g}{m_\i}\Gamma T^2  \frac{\hbar k^2}{2 m_\i}T\right)\right] \hat{D}(\Delta \tilde{z},\Delta \tilde{p}) \label{eq:U_phi}
\end{align}
where we have introduced the modified phase shift
\begin{align}
\delta \tilde{\varphi}_\g \equiv \delta \varphi_\g \left(1- \frac{7}{12} \frac{m_\g}{m_\i}\Gamma T^2 \right)
\end{align}
arising from gravity gradients and have recalled the definition~\eqref{eq:Doperator} of the displacement operator $\hat{D}$ with the shifts
\begin{align}
 \Delta \tilde{z} \equiv -\frac{m_\g}{m_\i}\frac{\hbar k}{m_\i} \Gamma T^3
\end{align}
and
\begin{align}
 \Delta \tilde{p} \equiv -\frac{m_\g}{m_\i}{\hbar k} \Gamma T^2
\end{align}
in position and momentum.

When we compare the expression~\eqref{eq:U_phi} for $\hat{U}_\varphi$ to the corresponding one for the linear potential, eq.~\eqref{eq:KC:UphiLin}, we find three modifications: (\textit{i}) The phase shift $\delta\varphi_\g$ due to the linear acceleration of the atom is reduced by $7(m_\g/m_\i)\Gamma T^2/12$, (\textit{ii}) an additional phase factor appears which is proportional to the recoil frequency $\hbar k^2/(2m_\i)$, and (\textit{iii}) the displacement operator $\hat{D}$ contains the operators of the coordinate and the momentum. 

Throughout this section we have kept track of the gravitational and inertial mass. For small gravity gradients $\Gamma T^2 \ll 1$, the masses enter always as a ratio. However, there are two exceptions: In the phase due to the recoil frequency, as well as in the displacement $\Delta \tilde{z}$. Here in addition the inertial mass appears. The resulting unusual combination $m_\g/m_\i^2$ of masses is a consequence of the fact that the momentum of the photon is $\hbar k$ and the inertial mass is needed for dimensional reasons. In section~\ref{subsec:analysis_in_wigner_space} we analyze the Kasevich-Chu interferometer using a Wigner phase-space approach and go beyond the limit of weak gravity gradients. In particular, we show that this unfamiliar appearance of the inertial and gravitational mass is still present.

We now calculate the probability $P_{g_1}$ to find the atom in the state $\ket{g_1}$ in the exit of the interferometer. With the help of the expression~\eqref{eq:U_phi} for $\hat{U}_\varphi$ we find from eq.~\eqref{eq:probability}

\begin{align}
P_{g_1} &= \frac12 \left[ 1+ \frac12 \left(\exp \left[ \i\left( \delta\phi-\delta\tilde{\varphi}_{\g} -\frac{m_\g}{m_\i} \frac{\hbar k^2}{2m_\i} \Gamma T^3 \right) \right] \bra{\psi_0} \hat{D}(\Delta \tilde{z},\Delta \tilde{p}) \ket{\psi_0} + \text{c.c.}\right)\right] \,.
\end{align}
Due to the presence of $\hat{D}$ the probability $P_{g_1}$ now crucially depends on the initial state $\ket{\psi_0}$ through the expectation value $\langle\hat{D}\rangle\equiv \bra{\psi_0}\hat{D}\ket{\psi_0}$ which we can interpret as the overlap of the initial state $\ket{\psi_0}$ displaced by $\hat{D}$ in phase space by the amount $\Delta \tilde{z}$ and $\Delta \tilde{p}$, and the original state $\ket{\psi_0}$. This overlap decreases starting from unity as we increase the shifts $\Delta \tilde{z}$ and $\Delta \tilde{p}$. As a result, the amplitude of the interference term, that is the contrast or the visibility of the fringes, decreases~\cite{Zeller14,Roura14}.

Moreover, since the expectation value of a unitary operator such as $\hat{D}$ is not necessarily real, that is
\begin{align}\label{eq:Derwzerlegt}
 \langle\hat{D}\rangle = |\langle\hat{D}\rangle| \e^{\i\beta}
\end{align}
we obtain an additional phase $\beta=\beta(\Delta \tilde{z},\Delta\tilde{p})$ which depends not only on the shifts $\Delta  \tilde{z}$ and $\Delta \tilde{p}$ but also on the initial state $\ket{\psi_0}$ of the center-of-mass motion.

Hence, we arrive at the expression
\begin{align}\label{eq:Pg1gradient}
P_{g_1} &= \frac12 \left[ 1+ |\langle\hat{D}\rangle| (\Delta  \tilde{z},\Delta \tilde{p})\cos\left( \delta\phi-\delta \tilde{\varphi}_{\g} - \frac{m_\g}{m_\i}\frac{\hbar k^2}{2m_\i} \Gamma T^3  +\beta(\Delta  \tilde{z},\Delta \tilde{p}) \right) \right]
\end{align}
for the probability $P_{g_1}$ to find the atom at the exit of the interferometer in $\ket{g_1}$.

%
\section{Wigner phase-space description of the Kasevich-Chu interferometer}
%
\label{subsec:analysis_in_wigner_space}
In the preceding section we have used the representation-free description of the Kasevich-Chu interferometer to derive an approximate expression for the resulting exit probability. However, the origin of the individual terms, and in particular the shifts $\Delta \tilde{z}$ and $\Delta \tilde{p}$ is hidden behind an opaque curtain of operator algebra. They come to light in the description of the Kasevich-Chu interferometer in terms of the Wigner function which constitutes the topic of the present section. 

We have already shown that the phase shift results, apart from the difference $\delta \phi$ of the laser phases, from the non-commutativity of the time evolutions in the two different momentum states associated with the internal states $\ket{g_1}$ and $\ket{g_2}$. In a corresponding analysis from quantum phase space we are tempted to propagate the initial Wigner function according to the two different Hamiltonians using the quantum Liouville equation~\eqref{eq:quant_Liouville}.

However, such an approach is incapable of explaining the phase shift in the interferometer. Indeed, we now show that the Wigner function of the center-of-mass motion at the exit of the interferometer does not only consist of the original Wigner function propagated along the upper and the lower path but involves an interference term. Thus the final state is very similar to a Schr\"odinger cat. The main goal of this section is to derive this interference term from first principles.

For this purpose we first express the Wigner function of the center-of-mass motion in terms of state vectors and unitary time evolutions familiar from the Schr\"odinger formulation of quantum mechanics. We then translate this formalism into phase space and define a Wigner matrix which reflects the different time evolutions along the different paths and, in particular, their interference. For these propagations we derive quantum Liouville equations which are fundamentally different from the one discussed in section~\ref{subsec:Q-Liouville-eq}. For the special example of a linear gravitational potential including gravity gradients we calculate the end points of the classical trajectories along the two paths and the interference path and find the relevant phase shift. We also identify the origin of the loss of contrast which is intimately connected to the presence of the interference term and outline a technique to avoid it.

\subsection{Wigner function from state vector}
In the Kasevich-Chu interferometer the center-of-mass motion and the internal states are entangled. According to table~\ref{tab.complete-state-KC} the state $\ket{\Psi}$ of the complete system at any time during the sequence of laser pulses is of the form
\begin{align}
\ket{\Psi}=\ket{\psi_1}\ket{g_1}+\ket{\psi_2}\ket{g_2} \label{eq:state-psi_1+psi_2}
\end{align}
with the normalization condition
\begin{align}
1 = \left\langle \Psi | \Psi \right\rangle = \left\langle \psi_1 | \psi_1 \right\rangle+\left\langle \psi_2 | \psi_2 \right\rangle\,.
\end{align}

We now analyze the quantum state of motion $\ket{\psi_1}$ at the exit of the interferometer when the atom is in the internal state $\ket{g_1}$. From table~\ref{tab.complete-state-KC} we find the representation
\begin{align}
\ket{\psi_1}= -\frac{1}{2}\left[\hat{U}_l+\hat{U}_u\right] \ket{\psi_0}\,
\end{align}
which gives rise to the operator
\begin{align}\label{eq:psi_1-psi_1}
\ket{\psi_1} \bra{\psi_1}=& \frac{1}{4} \left[\hat{U}_u \ket{\psi_0}\bra{\psi_0}\hat{U}_u^\dagger + \hat{U}_l  \ket{\psi_0}\bra{\psi_0}\hat{U}_l^\dagger + \hat{U}_u \ket{\psi_0}\bra{\psi_0}\hat{U}_l^\dagger + \hat{U}_l \ket{\psi_0}\bra{\psi_0}\hat{U}_u^\dagger \right]\,.
\end{align}

The first two contributions correspond to the propagation of the initial state $\ket{\psi_0}$ of the center-of-mass motion along the upper and lower path, respectively. The last two terms represent the interference of the two paths and therefore have a more complicated time evolution. Indeed, the bra and the ket vectors evolve along different paths as expressed by different unitary operators.

This fact stands out most clearly in the corresponding Wigner function $W_{g_1}$ corresponding to the state $\ket{\psi_1}$ of the center-of-mass motion of the atom in the state $\ket{g_1}$ at the exit of the interferometer. Indeed, when we recall the decompositions eqs.~\eqref{eq:U_u} and \eqref{eq:U_l} of the unitary operators $\hat{U}_u$ and $\hat{U}_l$ we obtain upon substitution of eq.~\eqref{eq:psi_1-psi_1} into the definition~\eqref{equ:Wigner_2}  of the Wigner function the expression
\begin{align}\label{eq:W_g_1}
W_{g_1}=\frac{1}{4}\left[W_u +  W_l+ \e^{\i \delta \phi}W_{i}+\e^{-\i \delta \phi} W_{i}^*\right]
\end{align}
for $W_{g_1}$. Here we have introduced the Wigner functions
\begin{align}
W_{u} \equiv W_{\underarc[.9]{{\scriptsize $1$\underarc[.9]{$22$}$1$}}}\left[ \ket{\psi_0}\bra{\psi_0}\right] \label{eq:W_u}
\end{align}
and
\begin{align}
W_{l} \equiv W_{\underarc[.9]{{\scriptsize $2$\underarc[.9]{$11$}$2$}}}\left[ \ket{\psi_0}\bra{\psi_0}\right]\label{eq:W_l}
\end{align}
together with
\begin{align}\label{eq:W_i}
W_{i} \equiv W_{\underarc[.9]{\scriptsize{$1$\underarc[.9]{$21$}$2$}}}\left[ \ket{\psi_0}\bra{\psi_0}\right]
\end{align}
corresponding to the upper, lower, and interference path in terms of the Wigner matrix
\begin{align}
W_{\underarc[.9]{\scriptsize{$i$\underarc[.9]{$jk$}$l$}}}\left[\hat{\mathcal{O}} \right]  \equiv	\frac{1}{2\pi\hbar} \int\limits\limits_{-\infty}^{\infty} \!\! \drm \zeta \
				      \e^{-\i p \zeta /\hbar} \ \braket{z + \frac{1}{2}\, \zeta |\hat{U}_i \hat{U}_j\hat{\mathcal{O}}\hat{U}_k^\dagger \hat{U}_l^\dagger| z- \frac{1}{2}\, \zeta } \label{eq:W_ijkl}
\end{align}
of forth rank.

Here the first pair of indices $i$ and $j$ denote the two unitary operators $\hat{U}_i \hat{U}_j$ corresponding to the two time evolutions, and the second one represents the hermitian conjugate part $\hat{U}_k^\dagger \hat{U}_l^\dagger$ of these evolutions. The first time evolution is given by the inner pair of indices, whereas the second one is expressed by the outer ones, as indicated in eqs.~\eqref{eq:W_u},~\eqref{eq:W_l}, and~\eqref{eq:W_i} by the arcs.

When we compare the expressions~\eqref{eq:W_u} and~\eqref{eq:W_l} for $W_u$ and $W_l$ we note that they are symmetric in their indices. Indeed, in each time evolution the two operators $\hat{U}_m$ and $\hat{U}_n^\dagger$ have identical indices, that is $m=n$. However, $W_u$ and $W_l$ differ in their order of time evolutions. Whereas in $W_u$ we first propagate with $\hat{U}_2$ and then with $\hat{U}_1$, we see in $W_l$ first the action of $\hat{U}_1$ and then of $\hat{U}_2$.

The symmetry in indices, that is in time evolution, is broken in the interference term $W_i$ defined by eq.~\eqref{eq:W_i} where we propagate first $\ket{\psi_0}$ by $\hat{U}_2$ and $\bra{\psi_0}$ by $\hat{U}_1^\dagger$, and then the resulting state, that is $(\hat{U}_2\ket{\psi_0})$ by $\hat{U}_1$ and $(\bra{\psi_0}\hat{U}_1^\dagger)$ by $\hat{U}_2^\dagger$. It is this asymmetric propagation which will provide us with the phase observed in the Kasevich-Chu interferometer.

\subsection{Translation into phase space}
We emphasize that our description of the Kasevich-Chu interferometer outlined in the preceding section still rests on the Schr{\"o}dinger formulation of quantum mechanics since it is based on the initial density operator $\hat{\rho}_0 \equiv \ket{\psi_0}\bra{\psi_0}$ and unitary time-evolution operators. Therefore, we now need to translate this formalism into phase space and derive quantum Liouville equations for $W_u$, $W_l$, and $W_i$ which allow us to determine those quantities directly from phase space without going through Hilbert space.

For this purpose we derive in appendix~\ref{app:subsec:applications} equations of motion for the Wigner matrix
\begin{align}
W_{il} \equiv	\frac{1}{2\pi\hbar} \int\limits\limits_{-\infty}^{\infty} \!\! \drm \zeta \
				      \e^{-\i p \zeta /\hbar} \ \braket{z + \frac{1}{2}\, \zeta |\hat{U}_i \hat{\rho}_0\hat{U}_l^\dagger| z- \frac{1}{2}\, \zeta }
\end{align}
of the density operator $\hat{\rho}_0$, the solution of which is the propagation of the initial Wigner function
\begin{align}\label{eq:W_0}
W_0 \equiv	\frac{1}{2\pi\hbar} \int\limits\limits_{-\infty}^{\infty} \!\! \drm \zeta \
				      \e^{-\i p \zeta /\hbar} \ \braket{z + \frac{1}{2}\, \zeta | \hat{\rho}_0| z- \frac{1}{2}\, \zeta }
\end{align}
corresponding to $\hat{\rho}_0\equiv \ket{\psi_0}\bra{\psi_0}$by an appropriate integral operator $\mathcal{T}_{il}$ such that
\begin{align}
W_{il} =	\mathcal{T}_{il}\left[ W_0 \right]\,.
\end{align}

Indeed, we show that the Wigner function $W_{22}$ of an atom being excited at time $t_0$ into 
the state $\ket{g_2}$, propagated for a time $t$ in that state and then returning to $\ket{g_1}$ reads
\begin{align}
  W_{22}(z,p;t_0+t) =&\int\limits_{-\infty}^{\infty} \!\!\drm z_0 \int\limits_{-\infty}^{\infty} \!\!\drm p_0 \,
			 \ \delta[z-\bar z(z_0,p_0+\hbar k; t)] \\
			  &\times \delta[p+ \hbar k - \bar p(z_0,p_0+\hbar k; t)] 
			  \ W_{22}(z_0,p_0;t_0)\,.\label{eq:Wigner_excited}
\end{align}
Hence, the time-evolution operator $\mathcal{T}_{22}$ is given by a phase space integral containing a delta function kernel ensuring propagation along classical trajectories. Here we have assumed that the potential giving rise to the motion is at most quadratic in the coordinate $z$.

Likewise, we find according to appendix~\ref{app:sec:Quantum-Liouville-equation} the expression 
\begin{align}
W_{11}(z,p;t_0+t) 	=& 	\int\limits_{-\infty}^{\infty} \!\!\drm z_0 \int\limits_{-\infty}^{\infty} \!\!\drm p_0  \, 
			\ \delta[z-\overline{z}(z_0,p_0;t)] \\
			&\times  \delta[p-\overline{p}(z_0,p_0;t)]
			\ W_{11}(z_0,p_0;t_0) \label{eq:Wigner_ground}
\end{align}
for the Wigner function $W_{11}$ of an atom propagated in $\ket{g_1}$ which defines in this way $\mathcal{T}_{11}$. 

When we compare the expression for the Wigner function $W_{11}$ corresponding to propagation in $\ket{g_1}$, eq.~\eqref{eq:Wigner_ground}, to $W_{22}$, eq.~\eqref{eq:Wigner_excited}, that is evolution in $\ket{g_2}$ we note two major differences: (\textit{i}) The initial momentum $p_0$ is replaced by $p_0+\hbar k$ corresponding 
to the interaction with the laser pulse, and (\textit{ii}) the final momentum $p=\overline p - \hbar k$ is the momentum $\overline p$ due to motion in the 
force field reduced by the momentum transfer $\hbar k$ due to the transition to $\ket{g_1}$.

Hence, the Wigner functions $W_{11}$ and $W_{22}$ of the motion propagate differently in phase space. This difference is a consequence of the interaction of the atom with the laser promoting 
the atom from $\ket{g_1}$ to $\ket{g_2}$ and back.

In addition we have to keep track of the motion of the Wigner functions corresponding to the interference terms. 
According to appendix~\ref{app.subsubsec:Excited-state-QLiouville-eq} they are given 
by the expressions
\begin{align}
W_{21}(z,p;t_0+t) 	=& 	\int\limits_{-\infty}^{\infty} \!\!\drm z_0 \int\limits_{-\infty}^{\infty} \!\!\drm p_0 \ \e^{-\i k(z-z_0)} \, 
			\ \delta[z-\overline{z}(z_0,p_0 + \frac{1}{2} \hbar k;t)] \\
			& \times \delta[p + \frac{1}{2}\hbar k-\overline{p}(z_0,p_0 + \frac{1}{2}\hbar k;t)]
			\ W_{21}(z_0,p_0;t_0)\,,\label{eq:Wigner_21}
\end{align}
and
\begin{align}
W_{12}(z,p;t_0+t) 	=& 	\int\limits_{-\infty}^{\infty} \!\!\drm z_0 \int\limits_{-\infty}^{\infty} \!\!\drm p_0 \ \e^{\i k(z-z_0)} \, 
			\ \delta[z-\overline{z}(z_0,p_0 + \frac{1}{2} \hbar k;t)] \\
			& \times \delta[p + \frac{1}{2}\hbar k-\overline{p}(z_0,p_0 + \frac{1}{2}\hbar k;t)]
			\ W_{12}(z_0,p_0;t_0)\, .\label{eq:Wigner_12}
\end{align}
In contrast to the time-evolution operators $\mathcal{T}_{22}$ and $\mathcal{T}_{11}$ defined by the phase space integrations eqs.~\eqref{eq:Wigner_excited} and ~\eqref{eq:Wigner_ground} the operators $\mathcal{T}_{21}$ and $\mathcal{T}_{12}$ corresponding to the propagation $W_{21}$ and $W_{12}$ given by eqs.~\eqref{eq:Wigner_21} and~\eqref{eq:Wigner_12} contain not only a momentum change of $\hbar k / 2$ rather than $\hbar k $ but also have the phase factors 
$\exp[-\i k(z-z_0)]$ for $W_{21}$, and $\exp[\i k(z-z_0)]$ for $W_{12}$.
These contributions will give rise to the phase shift in the Kasevich-Chu interferometer.
%
%
%
%
\subsection{Dynamics through the interferometer}
%
We are now in the position to analyze the Kasevich-Chu interferometer in Wigner phase space. For this purpose we first discuss 
the time evolution of the Wigner function along the upper and the lower path, and then turn to the dynamics of the interference terms. Here we take advantage of the fact that we can express the Wigner matrix $W_{\underarc[.9]{{\scriptsize$i$\underarc[.9]{$jk$}$l$}}}$ of forth order by a sequence
\begin{align}
W_{\underarc[.9]{\scriptsize{$i$\underarc[.9]{$jk$}$l$}}} =	\mathcal{T}_{il}\left[\mathcal{T}_{jk}\left[ W_0 \right] \right]\,.
\end{align}
of two propagations $\mathcal{T}_{jk}$ and $\mathcal{T}_{il}$.

In terms of these propagators we find the expressions
\begin{align} \label{eq:W_u-with-T}
W_{u} =	\mathcal{T}_{11}\left[\mathcal{T}_{22}\left[ W_0 \right] \right]=\mathcal{T}_{11}\left[ W_{22} \right]
\end{align}
and
\begin{align}
W_{l} =	\mathcal{T}_{22}\left[\mathcal{T}_{11}\left[ W_0 \right] \right]=\mathcal{T}_{22}\left[ W_{11} \right]
\end{align}
together with
\begin{align}
W_{i} =	\mathcal{T}_{12}\left[\mathcal{T}_{21}\left[ W_0 \right] \right]=\mathcal{T}_{12}\left[ W_{21} \right]
\end{align}
for the Wigner functions $W_u$, $W_l$, and $W_i$ contributing to the Wigner function $W_{g_1}$ given by eq.~\eqref{eq:W_g_1} and describing the center-of-mass motion of the atom exiting the Kasevich-Chu interferometer in the internal state $\ket{g_1}$.

%
%
\subsubsection{Upper path}
%
We start our analysis with the upper path where according to eq.~\eqref{eq:W_u-with-T}, we have to combine $\mathcal{T}_{11}$ and $\mathcal{T}_{22}$ defined by eqs.~\eqref{eq:Wigner_excited} and~\eqref{eq:Wigner_ground}.

Indeed, from eq.~\eqref{eq:Wigner_ground} we find with $W_u = \mathcal{T}_{11}[W_{22}]$ the expression
\begin{align}
W_u = W_{11}(z,p;2T) 	=&	\int\limits_{-\infty}^{\infty} \!\!\drm z_T \int\limits_{-\infty}^{\infty} \!\!\drm p_T  \, 
			   \delta[z-\overline{z}(z_T,p_T;T)]  \\
			 & \times\delta[p-\overline{p}(z_T,p_T;T)]
			   \ W_{22} (z_T,p_T;T)\label{eq:Wigner_u_(2T)}
\end{align}
for the Wigner function at time $2T$, that is at the end of the pulse sequence. Here $\overline{z}(z_T,p_T;T)$ and 
$\overline{p}(z_T,p_T;T)$ denote the coordinate and momentum, respectively, obtained by propagating $(z_T,p_T)$ over 
the time $T$. Moreover, we have used the fact that the initial state of the propagation $\mathcal{T}_{11}$ is $W_{22}$.

With the help of the relation $W_{22}=\mathcal{T}_{22}[W_0]$ and the explicit form of $\mathcal{T}_{22}$ given by eq.~\eqref{eq:Wigner_excited} we obtain the Wigner function
\begin{align}
W_{22} (z_T,p_T;T) 	=& 	\int\limits_{-\infty}^{\infty} \!\!\drm z_0 \int\limits_{-\infty}^{\infty} \!\!\drm p_0  \, 
			\ \delta[z_T-\overline{z}(z_0,p_0+\hbar k;T)] \\
			& \times \delta[p_T+\hbar k - \overline{p}(z_0,p_0+\hbar k;T)]
			\ W_{22} (z_0,p_0;0) \, .\label{eq:Wigner_u_(T)}
\end{align}
Indeed, $W_{22}$ emerges from the initial Wigner function 
\begin{align}
  W_{22} (z,p;0) &\equiv  W_0(z,p)
\end{align}
by first increasing the initial momentum $p_0$ by the photon recoil $\hbar k$, then propagating it along the classical trajectory, and finally reducing the so-obtained 
momentum by $\hbar k$ due to the transition from $\ket{g_2}$ to $\ket{g_1}$.

When we now substitute eq.~\eqref{eq:Wigner_u_(T)} into eq.~\eqref{eq:Wigner_u_(2T)}, interchange the order of the integrations, 
and perform the integration over the phase space variables $z_T$ and $p_T$ with the help of the two delta functions, we realize the sequential propagation $\mathcal{T}_{11}[\mathcal{T}_{22}[W_0]]$ and arrive at 
\begin{align}
W_u (z,p;2T) 	=& 		\int\limits_{-\infty}^{\infty} \!\!\drm z_0 \int\limits_{-\infty}^{\infty} \!\!\drm p_0  \, 
				\ \delta\bigg(z-\overline{z}\Big[\overline{z}(z_0,p_0+\hbar k;T),\overline{p}(z_0,p_0+\hbar k;T)-\hbar k;T\Big]\bigg) \\
				& \times\delta\bigg(p-\overline{p}\Big[\overline{z}(z_0,p_0+\hbar k;T),\overline{p}(z_0,p_0+\hbar k;T)-\hbar k;T\Big]\bigg)
				\ W_0(z_0,p_0)\,.\label{eq:W_u(2T)}
\end{align}
Hence, the final point $(z_u,p_u)$ in phase space is obtained in four steps: (\textit{i}) We first replace in the point $(z_0,p_0)$ the initial momentum $p_0$ by $p_0 + \hbar k$, (\textit{ii}) propagate $(z_0,p_0+\hbar k)$ according to Newtonian mechanics, 
(\textit{iii}) subtract from the so-obtained momentum the photon momentum $\hbar k$, and ($iv$) propagate this point again according to the 
Newton law. 

\begin{figure}[ht]
\centering
\def\svgwidth{0.8\columnwidth}
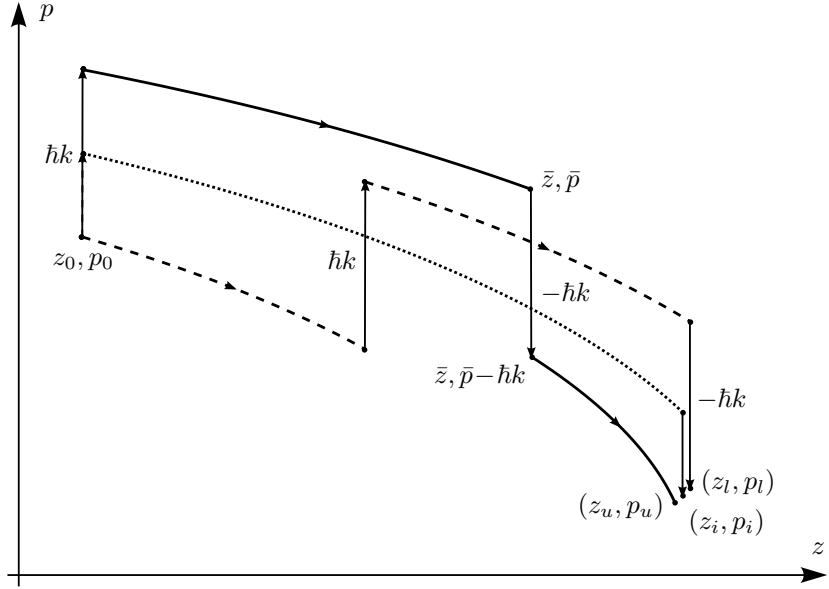
\caption{Motion of a point $(z_0,p_0)$ in phase space along the upper ($u$) and lower path ($l$) of the Kasevich-Chu interferometer indicated by solid and dashed lines, respectively.
	  The motion of the interference term represented by the dotted curve always runs between the solid and the dashed lines since the initial and all the following momentum displacements are $\hbar k/2$ rather than $\hbar k$. We display a situation in which the phase space trajectories corresponding to the upper and lower paths do not end at 
identical points, that is $(z_u,p_u) \neq (z_l,p_l)$. In this case the Kasevich-Chu interferometer does not close in phase space. This phenomenon is either due to gravity gradients, or an asymmetric laser pulse configuration.}
\label{fig.phase_space_trajectory}
\end{figure}

As a result, the motion of the initial point $(z_0,p_0)$ in phase space is an alternating sequence 
\begin{align}
\label{sequenceu}
\begin{pmatrix}   z_0 \\ p_0 \end{pmatrix} \!\!
 	&\xrightarrow[\text{of photon}]{\text{absorption}}\!\!
\begin{pmatrix}   z_0 \\ p_0+\hbar k \end{pmatrix} \!\!
	 \xrightarrow[\text{motion}]{\text{Newtonian}}\!\!
\begin{pmatrix}   \overline{z} \\ \overline{p} \end{pmatrix} \!\!
	 \xrightarrow[\text{of photon}]{\text{emission}}\!\!
\begin{pmatrix}   \overline{z} \\ \overline{p}-\hbar k \end{pmatrix} \!\!
	 \xrightarrow[\text{motion}]{\text{Newtonian}}\!\!
\begin{pmatrix}   z_u \\ p_u \end{pmatrix}
\end{align}
of shifts along the momentum axis and Newtonian mechanics as depicted in fig.~\ref{fig.phase_space_trajectory}. The script on the arrows indicate the physical processes causing the evolution in phase space, such as the instantaneous absorption or emission of a photon, or the Newtonian motion due to the gravitational field for a time $T$.

%

\subsubsection{Lower path}
Next we analyze the motion along the lower path corresponding to the relation $W_l = \mathcal{T}_{22}[W_{11}]$. Hence, according to 
eq.~\eqref{eq:Wigner_excited} the Wigner function $W_{22}$ at $2T$ reads
\begin{align}
W_l =  W_{22}(z,p;2T)	=& 	\int\limits_{-\infty}^{\infty} \!\!\drm z_T \int\limits_{-\infty}^{\infty} \!\!\drm p_T  \, 
				  \ \delta[z-\overline{z}(z_T,p_T+\hbar k;T)] \\
				  & \times \delta[p+\hbar k-\overline{p}(z_T,p_T+\hbar k;T)]
				  \ W_{11}(z_T,p_T;T)\, . \label{eq:W_l(2T)first}
\end{align}
Here we have already used the fact that the initial Wigner function for the propagation $\mathcal{T}_{22}$ is the final 
Wigner function $W_{11}$ of $\mathcal{T}_{11}$.

Indeed, we find with the help of eq.~\eqref{eq:Wigner_ground} representing the propagation $\mathcal{T}_{11}$ the result
\begin{align}
\label{eq:W_l(T)}
W_{11}(z_T,p_T;T) 	=& 	\!\! \int\limits_{-\infty}^{\infty} \!\!\drm z_0 \!\!\int\limits_{-\infty}^{\infty} \!\!\drm p_0  \!
			\ \delta[z_T-\overline{z}(z_0,p_0;T)]  \delta[p_T-\overline{p}(z_0,p_0;T)]
			\ W_0(z_0,p_0)\,.
\end{align}
Here we have taken advantage of $W_{11}(z,p;0) \equiv W_0(z_0,p_0)$.

When we substitute eq.~\eqref{eq:W_l(T)} into eq.~\eqref{eq:W_l(2T)first} and perform the integration over the intermediate phase space point $(z_T,p_T)$ we arrive at the expression
\begin{align}
W_l(z,p;2T) 	=& 	\int\limits_{-\infty}^{\infty} \!\!\drm z_0 \int\limits_{-\infty}^{\infty} \!\!\drm p_0  \, 
			\ \delta\bigg( z-\overline{z}\Big[\overline{z}(z_0,p_0;T),\overline{p}(z_0,p_0;T)+\hbar k ; T\Big]\bigg) \\
			& \times \delta\bigg( p+\hbar k -\overline{p}\Big[\overline{z}(z_0,p_0;T),\overline{p}(z_0,p_0;T)+\hbar k ; T\Big]\bigg)\
			 W_0(z_0,p_0)\,, \label{eq:W_l(2T)}
\end{align}
for the final Wigner function propagated along the lower path.

In comparison to the Wigner function $W_u$ given by eq.~\eqref{eq:W_u(2T)} and corresponding to the upper path now the order 
of events in the sequence of shifts and Newtonian motion is interchanged. Again we obtain the final point $(z_l,p_l)$ in phase 
space in four steps: (\textit{i}) we first propagate the initial point $(z_0,p_0)$ according to the Newtonian mechanics, (\textit{ii}) then replace in 
the so-obtained phase-space point $(\overline{z},\overline{p})$ the momentum $\overline{p}$ by $\overline{p}+\hbar k$ taking 
into account the absorption of the photon, (\textit{iii}) propagate $(\overline{z}, \overline{p}+\hbar k)$ according to Newtonian 
mechanics, and ($iv$) finally subtract from the so-obtained momentum the photon momentum $\hbar k$ due to the transition 
into the atomic ground state. 

Hence, we deal with the alternating sequence
\begin{align}
\label{eq:sequence_1}
\begin{pmatrix} z_0\\ p_0\end{pmatrix}\!\!
 \xrightarrow[\text{motion}]{\text{Newtonian}}\!\!
\begin{pmatrix}\overline{z}\\ \overline{p}\end{pmatrix}\!\!
 \xrightarrow[\text{of photon}]{\text{absorption}}\!\!
\begin{pmatrix} \overline{z}\\ \overline{p}+\hbar k\end{pmatrix} \!\!
 \xrightarrow[\text{motion}]{\text{Newtonian}}\!\!
\begin{pmatrix} z_l\\ p_l+\hbar k\end{pmatrix}\!\!
 \xrightarrow[\text{of photon}]{\text{emission}}\!\!
\begin{pmatrix} z_l\\ p_l\end{pmatrix}
\end{align}
of shifts in momentum and Newtonian motion as indicated by the dashed trajectory in figure~\ref{fig.phase_space_trajectory}.
\subsubsection{Interference term}
Finally we turn to the discussion of the interference term $W_i=\mathcal{T}_{12}[W_{21}]$. 
According to eq.~\eqref{eq:Wigner_12} we find for the propagation $\mathcal{T}_{12}$ the expression
\begin{align}
W_i = W_{12}(z,p;2T) 	=& 	\int\limits_{-\infty}^{\infty} \!\!\drm z_T \int\limits_{-\infty}^{\infty} \!\!\drm p_T  \, \e^{\i k(z-z_T)}
			\ \delta[z-\overline{z}(z_T,p_T+\frac{1}{2}\hbar k;T)] \\
			& \times \delta[p+\frac{1}{2}\hbar k-\overline{p}(z_T,p_T+\frac{1}{2}\hbar k;T)]
			\ W_{21}(z_T,p_T;T) \,.\label{eq:Wigner_21(2T)}
\end{align}
Here we have used already the fact that the initial Wigner function of $\mathcal{T}_{12}$ is the final Wigner function of the propagation
$\mathcal{T}_{21}$ which according to eq.~\eqref{eq:Wigner_21} reads
\begin{align}
W_{21}(z_T,p_T;T) 	=& 	\int\limits_{-\infty}^{\infty} \!\!\drm z_0 \int\limits_{-\infty}^{\infty} \!\!\drm p_0  \, \e^{-\i k(z_T-z_0)}
			\ \delta[z_T-\overline{z}(z_0,p_0+\frac{1}{2}\hbar k;T)] \\
			& \times \delta[p_T+\frac{1}{2}\hbar k-\overline{p}(z_0,p_0+\frac{1}{2}\hbar k;T)]
			\ W_0(z_0,p_0)\,.\label{eq:Wigner_21(T)}
\end{align}
The Wigner function $W_0$ defined by eq.~\eqref{eq:W_0} serves as the starting point of the propagation $\mathcal{T}_{21}$.

Substitution of eq.~\eqref{eq:Wigner_21(T)} into eq.~\eqref{eq:Wigner_21(2T)} and integration over the intermediate 
point $(z_T,p_T)$ yields the final result
\begin{align}
W_i(z,p;2T) 	=& \!\!	\int\limits_{-\infty}^{\infty}\!\!\drm z_0 \!\!\int\limits_{-\infty}^{\infty} \!\!\drm p_0  \, \e^{\i \delta \varphi(z_0,p_0)}
			\ \delta\bigg( z-\overline{z}\Big[\overline{z}(z_0,p_0+\frac{1}{2}\hbar k;T),\overline{p}(z_0,p_0+ \frac{1}{2}\hbar k;T) ; T\Big]\bigg) \\
			& \times \delta\bigg( p+\frac{1}{2}\hbar k -\overline{p}\Big[\overline{z}(z_0,p_0+\frac{1}{2}\hbar k ;T),\overline{p}(z_0,p_0+\frac{1}{2}\hbar k ;T) ; T\Big]\bigg) W_0(z_0,p_0)\,, \label{eq:W_i(2T)}
\end{align}
where the phase shift 
\begin{align}
\delta \varphi\equiv k \bigg( \overline{z}\Big[\overline{z}(z_0,p_0+\frac{1}{2}\hbar k;T),\overline{p}(z_0,p_0+ \frac{1}{2}\hbar k;T) ; T\Big]- 2 \overline{z}(z_0,p_0+ \frac{1}{2}\hbar k;T)+ z_0  \bigg)
\end{align}
is now expressed solely in terms of the initial phase-space point $(z_0,p_0)$. 

The first term in the expression for $\delta\varphi$ represents the final coordinate after the propagation of the initial point in phase space $(z_0,p_0+\hbar k/2)$ over a time $2T$ according to Newtonian mechanics: Here we first propagate this point for a time $T$, and then the resulting phase-space point for another time $T$. Hence, we obtain the relation  
\begin{align}
\overline{z}\Big[\overline{z}(z_0,p_0+\frac{1}{2}\hbar k;T),\overline{p}(z_0,p_0+ \frac{1}{2}\hbar k;T) ; T\Big]= \overline{z}(z_0,p_0+ \frac{1}{2}\hbar k;2T)
\end{align}
and the phase shift 
\begin{align}
\delta \varphi\equiv k \delta z \equiv k \bigg( \overline{z}(z_0,p_0+ \frac{1}{2}\hbar k;2T)- 2 \overline{z}(z_0,p_0+ \frac{1}{2}\hbar k;T)+ z_0  \bigg) \label{eq:varphi-with-trajectories}
\end{align}
takes the form of a discrete second derivative which is reminiscent of the difference $\delta\phi$ defined by eq.~\eqref{eq:KC:laserphase} of the laser phases.

According to the expression~\eqref{eq:W_i(2T)} for the Wigner function $W_i$ we have the following sequence
\begin{align}
\label{eq:sequence_2}
\begin{pmatrix} z_0\\ p_0 \end{pmatrix}\!\!&
 \xrightarrow[\text{of photon}]{\text{absorption}}\!\!
\begin{pmatrix} z_0\\ p_0+\frac{1}{2}\hbar k \end{pmatrix}\!\!
 \xrightarrow[\text{motion}]{\text{Newtonian}}\!\!
\begin{pmatrix} \overline{z}\\ \overline{p} \end{pmatrix}\!\!
 \xrightarrow[\text{of photon}]{\text{emission}}\!\!
\begin{pmatrix} \overline{z}\\ \overline{p}- \frac{1}{2}\hbar k \end{pmatrix}\\
 &\xrightarrow[\text{of photon}]{\text{absorption}}\!\!
\begin{pmatrix}  \overline{z}\\ \overline{p}- \frac{1}{2}\hbar k + \frac{1}{2}\hbar k \end{pmatrix}
=
\begin{pmatrix} \overline{z}\\ \overline{p} \end{pmatrix}\!\!
 \xrightarrow[\text{motion}]{\text{Newtonian}}\!\!
\begin{pmatrix} z_i\\ p_i+\frac{1}{2}\hbar k \end{pmatrix}\!\!
 \xrightarrow[\text{of photon}]{\text{emission}}\!\!
\begin{pmatrix} z_i\\ p_i \end{pmatrix}
\end{align}
of alternating momentum shifts and Newtonian motion shown in Fig.~\ref{fig.phase_space_trajectory} by the dotted curve.

When we compare this sequence to the ones given by eqs.~\eqref{sequenceu} and~\eqref{eq:sequence_1}, and corresponding to the upper and lower paths 
we recognize two major differences: (\textit{i}) the shift in momentum is now not in units of $\hbar k$ but $\hbar k/2$, and (\textit{ii}) there is no shift half-way through the interferometer. In contrast to the 
upper and lower paths which both jump down or up, the interference path continues without a jump.
%
%
%
%
\subsubsection{Specific example}
So far we have not specified the potential $V$ except that it is at most quadratic in the coordinate $z$, which is the 
underlying assumption of the quantum Liouville equations derived in appendix~\ref{app:subsec:applications} and used throughout. We now illustrate the representation of the Kasevich-Chu interferometer in terms of the Wigner functions obtained in the preceding sections by considering the specific gravitational potential
\begin{align}
V_\Gamma(z)\equiv m g z + \frac{1}{2} m \Gamma z^2 \label{eq:V_Gamma}
\end{align}
including gravity gradients.

In appendix~\ref{app:phase-space-dynamics} we calculate the end points $(z_u,p_u)$ and $(z_l,p_l)$ of the phase space trajectories corresponding to the atom moving along the upper and lower path of the interferometer starting from the initial point $(z_0,p_0)$. Here we establish the connections
\begin{align}
\begin{pmatrix} z_u\\ p_u \end{pmatrix}
=
\begin{pmatrix} z_i\\ p_i \end{pmatrix}
+
\frac{1}{2}\begin{pmatrix} \Delta z\\ \Delta p \end{pmatrix}
\end{align}
and
\begin{align}
\begin{pmatrix} z_l\\ p_l \end{pmatrix}
=
\begin{pmatrix} z_i\\ p_i \end{pmatrix}
-
 \frac{1}{2}\begin{pmatrix} \Delta z\\ \Delta p \end{pmatrix} \, ,
\end{align}
where 
\begin{align}
z_i \equiv& - 2 \frac{g}{\Gamma} \sin^2\left( \sqrt{\Gamma}T\right)+ \frac{\hbar k}{2 m \sqrt{\Gamma}} \sin\left(\sqrt{\Gamma}2T\right)\\
&+z_0 \cos\left( \sqrt{\Gamma}2T\right) + \frac{p_0}{m \sqrt{\Gamma}} \sin\left(  \sqrt{\Gamma}2T\right) \label{eq:z_i-exact}
\end{align}
and
\begin{align}
p_i \equiv& - m \frac{g}{\sqrt{\Gamma}} \sin\left(  \sqrt{\Gamma}2T\right)-\hbar k \sin^2\left(\sqrt{\Gamma}T\right)\\
&- m \sqrt{\Gamma} z_0 \sin\left( \sqrt{\Gamma}2T\right) + p_0 \cos\left(  \sqrt{\Gamma}2T\right)\label{eq:p_i-exact}
\end{align}
denote the final coordinate and momentum of the interference term.

Hence, the Wigner functions $W_u$ and $W_l$ corresponding to the upper and lower path are symmetrically located with respect to the interference term $W_i$ and shifted by the amount
\begin{align}
\begin{pmatrix} \Delta z\\ \Delta p \end{pmatrix}
\equiv
\begin{pmatrix}
\frac{\hbar k}{ m} \frac{\delta \mathscr{s}}{\sqrt{\Gamma}}\\
\hbar k \delta \mathscr{c} 
\end{pmatrix}\,.\label{eq:Delta_z-Delta_p}
\end{align} 
 Here we have introduced the abbreviations
\begin{align}
\delta \mathscr{s} \equiv \sin\left( \sqrt{\Gamma}2T \right)-2 \sin \left( \sqrt{\Gamma}T\right) \label{eq:delta_s}
\end{align}
and
\begin{align}
\delta \mathscr{c} \equiv \cos\left( \sqrt{\Gamma}2T \right)-2 \cos \left( \sqrt{\Gamma}T\right)+1 \,. \label{eq:delta_c}
\end{align}
The shifts in phase space defined by eq.~\eqref{eq:Delta_z-Delta_p} contain two ingredients: (\textit{i}) the momentum transfer $\hbar k$ by the laser, and (\textit{ii}) the discrete second derivatives $\delta \mathscr{s}$ and $\delta\mathscr{c}$ of the time evolution in the harmonic oscillator potential representing the gravity gradients. 

Indeed, the shifts arise from the difference of the displacements along the two paths triggered by the momentum kicks $\pm \hbar k$ due to the interaction of the atom with the laser pulses at $t=0$, $T$, and $2T$. We start our discussion with $\Delta p$ which has a rather elementary explanation.

On the upper path the displacement in momentum at $t=0$ by $+\hbar k$ propagates for $2T$ in the harmonic oscillator potential, whereas the reverse shift by $-\hbar k$ at $T$ experiences the oscillator only for $T$ giving rise to the final momentum change $\Delta p_u \equiv \hbar k \cos \left( \sqrt{\Gamma}2T \right)- \hbar k \cos \left(\sqrt{\Gamma}T \right)$. Likewise, on the lower path the increase in $p$ by $+\hbar k$ at $T$ propagates for $T$, whereas the decrease at $2T$ by $-\hbar k$ does not have any time left to evolve resulting in $\Delta p_l \equiv \hbar k \cos \left( \sqrt{\Gamma}T \right)- \hbar k \cos \left(\sqrt{\Gamma}0 \right)$. As a result, we obtain the expression $\Delta p_u-\Delta p_l = \hbar k \delta \mathscr{c}= \Delta p$ for the total separation $\Delta p \equiv p_u-p_l$ between the final momenta $p_u$ and $p_l$ of the upper and the lower path.

In the course of time the momentum changes $\pm \hbar k$ introduce position changes consisting of the product of $\zeta_0 \equiv \hbar k/(m\sqrt{\Gamma})$ and sine functions. Indeed, along the upper path the initial momentum increase by $+\hbar k$ at $t=0$ leads during the time period $2T$ to the displacement $\zeta_0 \sin\left(\sqrt{\Gamma}2T\right)$ in position, whereas the decrease by $-\hbar k$ at $T$ yields during the remaining time $T$ a shift of $-\zeta_0 \sin\left(\sqrt{ \Gamma}T \right)$ resulting in the displacement $\Delta z_u \equiv \zeta_0 \sin \left( \sqrt{\Gamma}2T \right)- \zeta_0 \sin \left(\sqrt{\Gamma}T \right)$ of the coordinate along the upper path. Likewise, on the lower path we obtain from the increase of the momentum by $+\hbar k$ at $T$ and propagation for $T$ the shift $\zeta_0 \sin \left(\sqrt{\Gamma}T \right)$. However, the decrease at $2T$ does not have any time to accumulate and leads us to the shift in position $\Delta z_l \equiv \zeta_0 \sin \left( \sqrt{\Gamma}T \right)- \zeta_0 \sin \left(\sqrt{\Gamma}0 \right)$. As a result, we obtain the total separation $\Delta z_u - \Delta z_l = \zeta_0 \delta \mathscr{s}=\Delta z$ between the final positions $z_u$ and $z_l$ of the two trajectories corresponding to the upper and the lower path.

\subsection{Phase shift and loss of contrast}
In the preceding section we have analyzed the final locations of the Wigner functions $W_u$, $W_l$, and $W_i$ reached by the propagation of the original Wigner function $W_0$ along the respective paths in phase space. In particular, we have shown that in the presence of gravity gradients they are located at different points in phase space. We now analyze the consequences resulting from this separation of the three Wigner functions.

For this purpose we first express the probability $P_{g_1}$ to find the atom at the exit of the interferometer in the state $\ket{g_1}$ in terms of phase-space integrals of $W_u$, $W_l$, and $W_i$. We then make the connection to the representation-free description of the Kasevich-Chu interferometer discussed in section~\ref{sec:KC_interferometer}, which allows us to identify the origin of the loss of contrast due to gravity gradients already mentioned in section~\ref{subsubsec:Gravity-gradients}.

\subsubsection{Integration over phase space}
The expressions eqs.~\eqref{eq:W_u(2T)},~\eqref{eq:W_l(2T)}, and~\eqref{eq:W_i(2T)} for the Wigner functions 
$W_u$, $W_l$, and $W_i$ corresponding to the upper, the lower, and the interference path allow us to obtain the 
probability
\begin{align}\label{eq:Pg_g_1-from-overline-W}
P_{g_1}\equiv \overline{W}_{g_1}\equiv \int\limits_{-\infty}^{\infty} \!\!\drm z \int\limits_{-\infty}^{\infty} \!\!\drm p  \, W_{g_1}=\frac{1}{4}\left[ \overline{W}_u+\overline{W}_l + \e^{\i \delta \phi} \overline{W}_i+\e^{-\i \delta \phi} \overline{W}_i^*\right]
\end{align}
for an atom to be at the exit port of the interferometer in the state $\ket{g_1}$. Here we have used eq.~\eqref{eq:W_g_1} and introduced the abbreviation
\begin{align}
\overline{W}_{j}\equiv \int\limits_{-\infty}^{\infty} \!\!\drm z \int\limits_{-\infty}^{\infty} \!\!\drm p  \, W_{j}
\end{align}
for the integration of the Wigner function $\overline{W}_j$ over all phase space.

With the help of the delta functions in eqs.~\eqref{eq:W_u(2T)},~\eqref{eq:W_l(2T)}, and~\eqref{eq:W_i(2T)} we can easily perform this integration and find in the case of $\overline{W}_u$ and $\overline{W}_l$ the result
\begin{align}\label{eq:overline-W_u}
\overline{W}_u=\overline{W}_l= 	\int\limits_{-\infty}^{\infty} \!\!\drm z_0 \int\limits_{-\infty}^{\infty} \!\!\drm p_0  \, W_0(z_0,p_0) = 1 \,,
\end{align}
which is identical to the phase-space integral of the initial Wigner function $W_0$. This identity is a direct consequence of the conservation of particles, which constitutes the basic assumption of the Liouville theorem.

However, in the case of the interference term $W_i$ we obtain from eq.~\eqref{eq:W_i(2T)} the expression
\begin{align}
\overline{W}_i=  	\int\limits_{-\infty}^{\infty} \!\!\drm z_0 \int\limits_{-\infty}^{\infty} \!\!\drm p_0  \, \e^{\i \delta\varphi(z_0,p_0)}W_0(z_0,p_0) \label{eq:N_i}
\end{align}
and the initial Wigner function $W_0$ is modulated by a phase factor with argument $\delta\varphi$.

In appendix~\ref{app:subsec:phase_shift} we derive the explicit expression
\begin{align} \label{eq:phi-phase-space}
\delta \varphi = g \frac{\Delta p}{\hbar \Gamma}  + \frac{1}{2}k\Delta z+ \frac{1}{\hbar}\left(\Delta p \, z_0 +\Delta z\, p_0 \right)
\end{align}
for $\delta\varphi$ due to the motion in the potential $V_\Gamma$, eq.~\eqref{eq:V_Gamma}, including gravity gradients.

When we substitute eq.~\eqref{eq:phi-phase-space} into eq.~\eqref{eq:N_i} we arrive at
\begin{align}
\overline{W}_i = \exp\left[\i \left(g \frac{\Delta p}{\hbar \Gamma}  + \frac{1}{2}k\Delta z\right)\right] \widetilde{W}_0\left( \Delta z , \Delta p \right)\,, \label{eq:final-exact-overline-W_i}
\end{align}
where
\begin{align}\label{eq:Fourier-transform}
\widetilde{W}_0(\zeta, q)\equiv\int\limits_{-\infty}^{\infty} \!\!\drm z \int\limits_{-\infty}^{\infty} \!\!\drm p  \, \e^{\i(\zeta p + qz)/\hbar} \,W_0(z,p)
\end{align}
is the two-dimensional Fourier transform of the initial Wigner function.

It is interesting to note that this quantity is sometimes~\cite{Stenholm80} referred to as the Shirley function~\cite{Shirley77,Shirley80}.
However, for the present discussion deeper insight springs from the fact that $\widetilde{W}_0$ is identical to the expectation value $\braket{\hat D}$ 
of the displacement operator $\hat D$ defined by eq.~\eqref{eq:Doperator} and performed with the help of the Wigner function. Indeed, with eqs.~\eqref{eq:Derwartungswert} and~\eqref{eq:Fourier-transform} we find
\begin{align}\label{}
  \bra{\psi_0} \hat D \ket{\psi_0} &= \int\limits_{-\infty}^{\infty} \!\!\drm z \int\limits_{-\infty}^{\infty} \!\!\drm p \, D_W \,W_0 = \widetilde{W}_0\,,
\end{align}
and when we recall the representation eq.~\eqref{eq:Derwzerlegt} of $\braket{\hat D}$ in amplitude $|\braket{\hat D}|$ and phase $\beta$,
we derive from eq.~\eqref{eq:final-exact-overline-W_i} the expression
\begin{align}\label{eq.W_i-overline-with-D}
  \overline{W}_i = |\braket{\hat D}| (\Delta z,\Delta p) \exp\left\lbrace\i \left[g \frac{\Delta p}{\hbar \Gamma}  + \frac{1}{2}k\Delta z + \beta(\Delta z,\Delta p)\right]\right\rbrace 
\end{align}
for the phase-space integral $\overline{W}_i$ of the Wigner function $W_i$ corresponding to the interference path.
%
%
%
\subsubsection{Exit probability from phase space}
%
%
Now we have all ingredients for the probability  $P_{g_1}$ to exit the interferometer in the internal state $\ket{g_1}$ given 
by eq.~\eqref{eq:Pg_g_1-from-overline-W}, which together with eqs.~\eqref{eq:overline-W_u} and~\eqref{eq.W_i-overline-with-D} reads
\begin{align}\label{eq:general-P_g_1}
P_{g_1}= \frac{1}{2}\left[ 1+ |\braket{\hat D}|(\Delta z,\Delta p) \cos \left(\delta \phi +g \frac{\Delta p}{\hbar \Gamma}  + \frac{1}{2}k\Delta z+ \beta\left(\Delta z, \Delta p \right) \right) \right]\,. 
\end{align}
When we compare this expression for $P_{g_1}$ valid for arbitrary values of $\sqrt{\Gamma}T$ to the corresponding one for the limit of weak gravity gradients given by eq.~\eqref{eq:Pg1gradient} we recognize the same overall structure but three important differences:
(\textit{i}) Whereas in eq.~\eqref{eq:Pg1gradient} the arguments of $\braket{\hat D}$ are the shifts $\Delta \tilde{z}$ and $\Delta \tilde{p}$ the corresponding 
arguments in eq.~\eqref{eq:general-P_g_1} are $\Delta z$ and $\Delta p$, (\textit{ii}) the phase of the interference term is slightly 
different, and (\textit{iii}) the role of the inertial and gravitational mass is not obvious since we have not distinguished them in this calculation.

We now address each of these points and start with the last one. According to appendix~\ref{app:Gradients} we can include the two types of masses when we replace the linear acceleration $g$ by $ (m_\g/m_\i) g$, and the gravity gradient $\Gamma$ by $(m_\g/m_\i)\Gamma$. All other masses appearing are the inertial mass $m_\i$. Hence, for the separation $\Delta z$ given by eq.~\eqref{eq:Delta_z-Delta_p} we find
\begin{align}
\Delta z = \frac{\hbar k}{\sqrt{m_\i m_\g}} \frac{\delta \mathscr{s} }{\sqrt{\Gamma}}
\end{align}
and the two masses appear in the form $\sqrt{m_\i m_\g}$. Moreover, in the arguments of the discrete derivatives $\delta \mathscr{s}$ and $\delta \mathscr{c}$ defined by eqs.~\eqref{eq:delta_s} and\eqref{eq:delta_c} we get an additional factor of $\sqrt{m_\g/m_\i}$.

Next we address the distinctions (\textit{i}) and (\textit{ii}) which disappear in the limit of weak gravity gradients, that is for $\sqrt{\Gamma}T \ll 1$. Indeed, in this limit the expressions eqs.~\eqref{eq:delta_s} and~\eqref{eq:delta_c} for the discrete second derivatives $\delta \mathscr{s}$ and $\delta \mathscr{c}$ reduce to
\begin{align}
\delta \mathscr{s} \cong - \left( \sqrt{\Gamma}T \right)^3 
\end{align}
and
\begin{align}
\delta \mathscr{c} \cong - \Gamma T^2 \left(1-\frac{7}{12}\Gamma T^2\right)\,,
\end{align}
and the displacement in phase space defined by eqs.~\eqref{eq:Delta_z-Delta_p} reads
\begin{align}\label{eq:displ_phase_space}
\begin{pmatrix} \Delta z\\ \Delta p \end{pmatrix}
\cong -
\begin{pmatrix}
\frac{\hbar k}{ m }T \,(\Gamma T^2) \\
\hbar k  \Gamma T^2  \left[1-\frac{7}{12}\Gamma T^2\right]
\end{pmatrix} \ .
\end{align}
Consequently, the expression~\eqref{eq:general-P_g_1} for $P_{g_1}$ leads to the approximate one given by eq.~\eqref{eq:Pg1gradient}
in the appropriate limit.
%
%
\subsubsection{Schr\"odinger cat}
%
%
We conclude our discussion of the Kasevich-Chu interferometer seen from quantum phase space by noting that the connection between the limit of $P_{g_1}$ for weak gravity gradients  based on the representation-free approach 
and the exact treatment using the Wigner function technique established in this section also brings to light the origin of the 
shifts $\Delta \tilde{z}$ and $\Delta \tilde{p}$. According to eq.~\eqref{eq:displ_phase_space} they are the corresponding limits of the separations 
$\Delta z$ and $\Delta p$ between the Wigner functions $W_u$ and $W_l$ obtained by propagating along the upper and the lower paths.

Moreover, the loss of contrast discussed already in section~\ref{subsubsec:Gravity-gradients} is a consequence of this separation. Indeed, the interference term 
$W_i$ located half-way between $W_u$ and $W_l$ displays interference fringes in phase space similar to a Schr\"odinger cat. Their 
period is determined by the separation in phase space between $W_u$ and $W_l$. Since the probability $P_{g_1}$ is determined by an integration over all phase space, the contribution from 
$W_i$ is reduced due to the fringes. In addition, the further $W_u$ and $W_l$ are apart, the faster the modulation of the original Wigner 
function, and the lower the contrast. Thus, only when this separation vanishes, that is $W_u$ and $W_l$, and consequently also $W_i$ 
are centered at the same point in phase space, do we find perfect contrast.

For the present example of the potential $V_\Gamma$ given by eq.~\eqref{eq:V_Gamma} it is the presence of the gravity gradient, that 
is  a non-zero value $ \Gamma$ which leads to a non-vanishing separation. Hence, only for $\Gamma=0$ do we find $\Delta z=\Delta p=0$ 
and thus perfect contrast as reflected by the expression~\eqref{eq:P_g_1-linear} for $P_{g_1}$.

We refer to interferometers with either $\Delta z \neq 0$ \textit{or} $\Delta p \neq 0$, or $\Delta z \neq 0$ \textit{and} $\Delta p \neq 0$ as \textit{open 
interferometers}. Mechanisms that lead to such interferometers are, besides gravity gradients, rotations or an asymmetric 
timing of the light pulses. Open atom interferometer geometries viewed from phase space are also discussed in~\cite{Zeller14}. 
There, as well as in~\cite{Roura14}, strategies to overcome the loss of visibility are demonstrated. We do emphasize 
that our treatment can also be generalized to open atom interferometer geometries beyond the Kasevich-Chu interferometer.
%
%
\section{Summary and outlook}
%
\label{sec:Summary}
The Wigner function is the perfect tool to study the propagation of an ensemble of atoms in a gravitational field, which is central to tests of the equivalence principle. Indeed, due to the equivalence of the classical and the quantum Liouville equation for potentials which are at most quadratic the weak equivalence principle also holds true in the \textit{dynamics} of quantum particles. Nevertheless, unusual combinations of inertial and gravitational mass may arise due to the mass-dependence of the initial distribution, and the observed distributions being the marginals of the phase-space distribution.

A new insight into the Kasevich-Chu interferometer is offered by the Wigner function. Since the internal states of the atom 
change during the propagation through the interferometer the familiar quantum Liouville equation does not suffice to describe 
the process. Indeed, the internal states and the associated time evolution of the center-of-mass motion promote the Wigner function to a Wigner matrix. Here, it is the off-diagonal element which gives rise to the phase shift. Even for realizations of interferometers by momentum transfer without the change of an internal state, such a Wigner matrix formalism can be developed to describe the occurrence of the interference.

Throughout these lectures we have concentrated on the most elementary aspects but emphasize that many extensions are possible. Here we list only a few.

It is remarkable that the proper-times accumulated along the two paths in the Kasevich-Chu interferometer are identical~\cite{Mueller10}. This fact is intimately connected~\cite{Greenberger12} to the type of mirror acting on the atom. Indeed, in the Kasevich-Chu interferometer the mirror is realized by a constant momentum transfer $ \hbar k$. In contrast to such a soft mirror a hard wall reflects \textit{every} momentum. Indeed, for atoms hard mirrors exist and rely on evanescent waves~\cite{Cohen93,Grimm97} rather than on standing or running waves. When we now replace the laser pulse of the Kasevich-Chu interferometer at $T$ representing the soft mirror by a hard mirror the two paths display~\cite{Giese13} a difference in proper-time.

There exist many similarities and differences between neutron~\cite{Rauch+Werner} and atom optics. It has recently been argued~\cite{Greenberger12} that the mirrors in neutron optics provide a complete inversion of the momentum. However, this claim has been disputed~\cite{Lemmel13} and has opened up a new direction of research. 

The role of proper-time in quantum mechanics~\cite{Greenberger01}, and in particular, in atom optics has many facets. Wave-particle duality and complementarity as expressed~\cite{Scully91} by which-path information versus interference dictates that interference fringes must disappear when we gain information about the path of the interfering particle. This idea has recently been applied~\cite{Zych11,Zych12} to the concept of interference of proper times. A proper-time difference between two clocks propagating along the two arms of an interferometer leads to a loss of contrast. Since the difference between the Kasevich-Chu interferometer and the hard-mirror analogue lies in the proper time, it is interesting to consider the complementarity aspect~\cite{Zych11,Zych12} in this context. In this way new aspects in the debate on the gravitational redshift will open up~\cite{GRS-redshift}.

These examples demonstrate in a striking way the excitement, the opportunities, and the scientific challenges offered by the 
interface of gravity and quantum mechanics. We are well aware of the fact that we are only at the beginning and still have a long way 
to go before we can unite the two great theories of the twentieth century. Nevertheless, we are encouraged in our efforts by the 
observation of the American  philosopher John Dewey (1859-1952) in \textit{The Quest for Certainty}: 
\textit{"Every great advance in science has issued from a new audacity of imagination"}.

\appendix
\section{Weyl-Wigner correspondence of the displacement operator}
\label{app:Weyl-Wigner-displacement-operator}
In this appendix we evaluate the Weyl-Wigner correspondence
\begin{align}\label{WeylWignerCorresp}
  D_W	&\equiv		 \int\limits_{-\infty}^{\infty} \!\! \drm \zeta \
				      \e^{-\i p \zeta /\hbar} \ \bra{z + \frac{1}{2}\, \zeta} \hat{D} \ket{z- \frac{1}{2}\, \zeta }
\end{align}
of the displacement operator
\begin{align}\label{DisplacementOperator}
 \hat{D}	&\equiv	e^{\i (\xi \hat{p} + q \hat{z})/\hbar} = e^{\i \xi \hat{p}/\hbar}\,e^{\i q \hat{z}/\hbar}\,e^{-\i \xi q /(2 \hbar)}
\end{align}
where in the last step we have used the Baker-Campbell-Hausdorff theorem.

When we substitute $\hat{D}$ in the form of the product given by eq.~\eqref{DisplacementOperator} into the definition~\eqref{WeylWignerCorresp} and apply the exponential of the position operator onto $ \ket{z- \frac{1}{2}\, \zeta }$ we arrive at  
\begin{align}  D_W	=	e^{-\i \xi q /(2 \hbar)}  \int\limits_{-\infty}^{\infty} \!\! \drm \zeta \
				      \e^{-\i p \zeta /\hbar}\,  \e^{\i q (z-\zeta/2)/\hbar} \ \bra{z + \frac{1}{2}\, \zeta}  e^{\i \xi \hat{p}/\hbar}  \ket{z- \frac{1}{2}\, \zeta }\,.
\end{align}

With the help of the relation
\begin{align} 
e^{\i \xi \hat{p}/\hbar}\ket{z}=\ket{z - \xi}
\end{align}
we finally find
\begin{align}  D_W	=	e^{-\i \xi q /(2 \hbar)}  \int\limits_{-\infty}^{\infty} \!\! \drm \zeta \
				      \e^{-\i p \zeta /\hbar}\,  \e^{\i q (z-\zeta/2)/\hbar} \ \bra{z + \frac{1}{2}\, \zeta}  z- \frac{1}{2}\, \zeta -\xi \rangle \,,
 \end{align}
which due to the ortogonality of the position eigenstates, that is 
\begin{align}
\bra{z'}\, z'' \rangle = \delta (z' - z'') 
\end{align}
reduces to
\begin{align}
D_W = e^{\i (\xi p + q z)/\hbar} \,.
 \end{align}
Hence, the Weyl-Wigner correspondence $D_W$ of the operator $\hat{D}$ is identical to $\hat{D}$ where the operators  $\hat{z}$ and  $\hat{p}$ are replaced by the c-numbers $z$ and $p$, respectively.
%
\section{Quantum mechanics in phase space}
\label{app.quantum_liouville}
\label{app.weyl-wigner}
In section~\ref{sec:Weyl-Wigner-correspondence} we have introduced the Weyl-Wigner correspondence 
\begin{align}
  \mathcal{O}_W (z,p)	&\equiv	\int\limits_{-\infty}^{\infty} \!\! \drm \zeta \
				      \e^{-\i p \zeta/\hbar} \ \bra{z + \frac{1}{2}\, \zeta} \hat{\mathcal{O}} \ket{z- \frac{1}{2}\, \zeta} \label{eq:app_Weyl-Wigner}
\end{align}
of an arbitrary operator $\hat{\mathcal{O}}$. We now derive expressions for the Weyl-Wigner correspondence of products $\hat{\mathcal{V}}\hat{\mathcal{O}}$ and $\hat{\mathcal{O}}\hat{\mathcal{V}}$, or $\hat{\mathcal{K}}\hat{\mathcal{O}}$ and $\hat{\mathcal{O}}\hat{\mathcal{K}}$, consisting of $\hat{\mathcal{O}}$ and operators
\begin{align}
\mathcal{V}(\hat{z})\equiv \sum\limits_j \mathcal{V}_j \hat{z}^j
\end{align}
and 
\begin{align}
\mathcal{K}(\hat{p})\equiv \sum\limits_j \mathcal{K}_j \hat{p}^j
\end{align}
expressed as a power series of the position and the momentum operators $\hat{z}$ and $\hat{p}$ with expansion c-number coefficients $\mathcal{V}_j$ and $\mathcal{K}_j$. This analysis leads us straight to the Bopp operators~\cite{Bopp61} and allows us to derive not only the quantum Liouville equation and the Schr\"odinger equation in phase space, but also the equations of motion for the dynamics of the atoms in the Kasevich-Chu interferometer viewed from Wigner phase space.

\subsection{Bopp operators}
We begin by introducing the Bopp operators for functions that consist solely of the position operator and then turn to the corresponding problem of the momentum operator. Here, we translate the Weyl-Wigner correspondence~\eqref{eq:app_Weyl-Wigner} into momentum space which allows us to immediately read-off the relevant expressions for the Bopp operators for functions of the momentum operator.

\subsubsection{Position operator}
We start our discussion by considering the Weyl-Wigner correspondence 
\begin{align}
  \left( z\mathcal{O}\right)_W (z,p)	&\equiv	\int\limits_{-\infty}^{\infty} \!\! \drm \zeta \
				      \e^{-\i p \zeta/\hbar} \ \bra{z + \frac{1}{2}\, \zeta}\hat{z} \hat{\mathcal{O}} \ket{z- \frac{1}{2}\, \zeta}
\end{align}
of the operator product $\hat{z} \hat{\mathcal{O}}$, which with the help of the eigenvalue equation $\hat{z}\ket{z}=z\ket{z}$ for the hermitian operator $\hat{z}$ yields
\begin{align}
  \left( z\mathcal{O}\right)_W &=	\int\limits_{-\infty}^{\infty} \!\! \drm \zeta \
				      \e^{-\i p \zeta/\hbar} \left(z+\frac{1}{2} \zeta \right)\ \bra{z + \frac{1}{2}\, \zeta} \hat{\mathcal{O}} \ket{z- \frac{1}{2}\, \zeta}\,,
\end{align}
or
\begin{align}
  \left( z\mathcal{O}\right)_W &\equiv\left(z-\frac{1}{2} \frac{\hbar}{\i}\frac{\partial}{\partial p} \right)	\int\limits_{-\infty}^{\infty} \!\! \drm \zeta \
				      \e^{-\i p \zeta/\hbar} \ \bra{z + \frac{1}{2}\, \zeta} \hat{\mathcal{O}} \ket{z- \frac{1}{2}\, \zeta}\,.
\end{align}
As a consequence, we obtain the relation
\begin{align}
  \left( z\mathcal{O}\right)_W &=\hat{z}^{(l)}_B\mathcal{O}_W\,, \label{eq:EW-z_B^l}
\end{align}
where we have introduced the Bopp operator
\begin{align}\label{eq:app.z_B^l}
\hat{z}^{(l)}_B\equiv z-\frac{1}{2}\frac{\hbar}{\i} \frac{\partial}{\partial p} \,. 
\end{align}
Here we have added a superscript $l$ in order to reflect the fact that in the operator product $\hat{z}\hat{\mathcal{O}}$ the position operator $\hat{z}$ is located to the left of $\hat{\mathcal{O}}$.

The role of the Bopp operator $\hat{z}_B^{(l)}$ is analogous to that of the operators in a specific representation of quantum mechanics. Indeed, in the Schr\"odinger formulation we can express a state either in position or momentum representation, and the corresponding momentum operator is determined either by differentiation with respect to the coordinate, or by multiplication by the corresponding c-number variable. The operators in a specific representation act on the corresponding wave functions. Likewise, the Bopp operators act on the Weyl-Wigner correspondence $\mathcal{O}_W$ of the operator $\hat{\mathcal{O}}$. Therefore, $\mathcal{O}_W$ plays the role of the wave function of the Schr\"odinger formulation.

Indeed, the Wigner approach is a phase-space representation of quantum mechanics and the corresponding operators of position and momentum take a form which is different from the familiar one. Due to the fact that we now work in phase space they involve both position \textit{and} momentum in multiplication and differentiation. The factor $1/2$ in front of the derivative is a consequence of the definition~\eqref{eq:app_Weyl-Wigner} of the Weyl-Wigner correspondence and in particular, of the coordinate $z$ being located half-way between the jump of length $\zeta$.

Next we consider the Weyl-Wigner correspondence
\begin{align}
  \left( \mathcal{O}z\right)_W &\equiv	\int\limits_{-\infty}^{\infty} \!\! \drm \zeta \
				      \e^{-\i p \zeta/\hbar} \ \bra{z + \frac{1}{2}\, \zeta} \hat{\mathcal{O}} \hat{z}\ket{z- \frac{1}{2}\, \zeta}
\end{align}
of the product $\mathcal{\hat{O}}\hat{z}$ and find by an analogous calculation the expression
\begin{align}\label{eq.app.(Oz)_W}
  \left( \mathcal{O}z\right)_W &=\hat{z}^{(r)}_B\mathcal{O}_W\,,
\end{align}
where the corresponding Bopp operator
\begin{align}\label{eq:app.z_B^r}
\hat{z}^{(r)}_B\equiv z+\frac{1}{2}\frac{\hbar}{\i} \frac{\partial}{\partial p}
\end{align}
with the superscript $r$ reflects the fact that in the original operator product $\hat{\mathcal{O}}\hat{z}$ the operator $\hat{z}$ is to the right of $\hat{\mathcal{O}}$. However, we emphasize that in eq.~\eqref{eq.app.(Oz)_W} the Bopp operator $\hat{z}_B^{(r)}$ is still to the left of $\mathcal{O}_W$ since it is a phase-space operator acting on $\mathcal{O}_W$.

We now turn to the case of an operator $\hat{z}^j \mathcal{\hat{O}}$ consisting of the product of an arbitrary power $j$ of $\hat{z}$ and $\hat{\mathcal{O}}$. When we recall the relation
\begin{align}
\hat{z}^j \hat{\mathcal{O}}= \hat{z}\left( \hat{z}^{j-1}\hat{\mathcal{O}}\right)\equiv \hat{z}\hat{\mathcal{O}}^\prime\,,
\end{align}
where $\hat{\mathcal{O}}^\prime$ is another arbitrary operator, we find with the identity eq.~\eqref{eq:EW-z_B^l} the result
\begin{align}
  \left( z^j\mathcal{O}\right)_W &=\hat{z}^{(l)}_B\mathcal{O}_W^\prime\,.
\end{align}
Iterating this formula we arrive at the product rule
\begin{align}
  \left( z^j\mathcal{O}\right)_W &=\left(\hat{z}^{(l)}_B\right)^j\mathcal{O}_W\,,
\end{align}
that is powers of $\hat{z}$ translate into powers of the Bopp operators $\hat{z}_B^{(l)}$.

Similarly, we can derive the substitution law
\begin{align}
  \left(\mathcal{O}z^j\right)_W &=\left(\hat{z}^{(r)}_B\right)^j\mathcal{O}_W\,.
\end{align}
We are now in the position to state the final results
\begin{align}\label{eq:app.V_l}
  \left(\mathcal{V}(z)\mathcal{O}\right)_W &=\mathcal{V}(\hat{z}^{(l)}_B)\mathcal{O}_W
\end{align}
and
\begin{align}\label{eq:app.V_r}
  \left(\mathcal{O}\mathcal{V}(z)\right)_W &=\mathcal{V}(\hat{z}^{(r)}_B)\mathcal{O}_W
\end{align}
for a function $\mathcal{V}$ of the position operator.

\subsubsection{Momentum operator}
Finally we turn to the Weyl-Wigner correspondence of the product of $\hat{\mathcal{O}}$ and powers of the momentum operator $\hat{p}$. Here it is advantageous to first translate the Weyl-Wigner correspondence eq.~\eqref{eq:app_Weyl-Wigner} into momentum space.

For this purpose we introduce two complete sets of momentum states $\ket{p-\frac{1}{2}\tilde{p}}$ and $\ket{p+\frac{1}{2}\tilde{\tilde{p}}}$ into eq.~\eqref{eq:app_Weyl-Wigner} and find with the relation 
\begin{align}
 \braket{z|p} &=\frac{1}{\sqrt{2\pi \hbar}}\e^{\i p z /\hbar}
\end{align}
from
\begin{align}
 \mathcal{O}_W &= \frac{1}{4}\int\limits_{-\infty}^{\infty} \!\! \drm \tilde{p} \!\! \int\limits_{-\infty}^{\infty} \!\! \drm \tilde{\tilde{p}} \!\!	\int\limits_{-\infty}^{\infty} \!\! \drm \zeta \
				      \e^{-\i p \zeta/\hbar} \ \braket{z + \frac{1}{2}\, \zeta|p-\frac{1}{2}\tilde{p}}\bra{p-\frac{1}{2}\tilde{p}} \hat{\mathcal{O}}\ket{p+\frac{1}{2}\tilde{\tilde{p}}} \braket{p+\frac{1}{2}\tilde{\tilde{p}}|z- \frac{1}{2}\, \zeta}
\end{align}
the formula
\begin{align}
 \mathcal{O}_W &= \int\limits_{-\infty}^{\infty} \!\! \drm \tilde{p} \ \int\limits_{-\infty}^{\infty} \!\! \drm \tilde{\tilde{p}}\ \e^{-\i (\tilde{p}+\tilde{\tilde{p}})z/(2\hbar)}\bra{p-\frac{1}{2}\tilde{p}} \hat{\mathcal{O}}\ket{p+\frac{1}{2}\tilde{\tilde{p}}} \frac{1}{2\pi \hbar}	\int\limits_{-\infty}^{\infty} \!\! \drm \left(\frac{\zeta}{4} \right) \
				      \e^{-\i (\tilde{p}-\tilde{\tilde{p}}) \zeta/(4\hbar)}\,.
\end{align}

The integration over $\zeta$ provides us with a Dirac delta function with the argument $\tilde{p}-\tilde{\tilde{p}}$, which allows us to perform the integration over $\tilde{\tilde{p}}$, and we arrive at
\begin{align}\label{eq:app_Weyl-Wigner_momentum}
  \mathcal{O}_W (z,p)	&\equiv	\int\limits_{-\infty}^{\infty} \!\! \drm \pi \
				      \e^{-\i z \pi /\hbar} \ \bra{p- \frac{1}{2}\, \pi} \hat{\mathcal{O}} \ket{p+ \frac{1}{2}\, \pi}\,.
\end{align}

When we compare this expression for the Weyl-Wigner correspondence in momentum space to the corresponding one in position space given by expression~\eqref{eq:app_Weyl-Wigner} we note that they are similar, but: (\textit{i}) the bra- and ket-vectors $\bra{z+\frac{1}{2}\zeta}$ and $\ket{z-\frac{1}{2}\zeta}$ are replaced by $\bra{p-\frac{1}{2}\pi}$ and $\ket{p+\frac{1}{2}\pi}$, respectively, and (\textit{ii}) the integration over the separation $\zeta$ in position is replaced by the separation $\pi$ in momentum.

Since the signs in the phase factors $\exp \left(-\i p \zeta/\hbar \right)$ and $\exp \left(-\i z \pi/\hbar \right)$ are identical we can immediately deduce the relations
\begin{align}
  \left( p\mathcal{O}\right)_W &=\hat{p}^{(l)}_B\mathcal{O}_W
\end{align}
and
\begin{align}
  \left( \mathcal{O}p\right)_W &=\hat{p}^{(r)}_B\mathcal{O}_W\,,
\end{align}
where the Bopp operators
\begin{align}\label{eq:app.p_B^l}
\hat{p}^{(l)}_B\equiv p+\frac{1}{2}\frac{\hbar}{\i} \frac{\partial}{\partial z}
\end{align}
and
\begin{align}\label{eq:app.p_B^r}
\hat{p}^{(r)}_B\equiv p-\frac{1}{2}\frac{\hbar}{\i} \frac{\partial}{\partial z}
\end{align}
arise from the corresponding counterparts $\hat{z}_B^{(l)}$ and $\hat{z}_B^{(r)}$, defined by eqs.~\eqref{eq:app.z_B^l} and~\eqref{eq:app.z_B^r}, by exchanging $z$ and $p$, and by replacing the signs in front of the derivatives. This change is a consequence of the interchange of the sign in the bra- and ket vectors in the two representations eqs.~\eqref{eq:app_Weyl-Wigner} and~\eqref{eq:app_Weyl-Wigner_momentum} of the Weyl-Wigner correspondence.

We conclude that by induction we can verify the relations
\begin{align}
  \left(p^j\mathcal{O}\right)_W &=\left(\hat{p}^{(l)}_B\right)^j\mathcal{O}_W
\end{align}
and
\begin{align}
  \left(\mathcal{O}p^j\right)_W &=\left(\hat{p}^{(r)}_B\right)^j\mathcal{O}_W\,,
\end{align}
valid for arbitrary powers $j$, which finally yield the identities 
\begin{align}\label{eq:app.K_l}
  \left(\mathcal{K}(p)\mathcal{O}\right)_W &=\mathcal{K}(\hat{p}^{(l)}_B)\mathcal{O}_W
\end{align}
and
\begin{align}\label{eq:app.K_r}
  \left(\mathcal{O}\mathcal{K}(p)\right)_W &=\mathcal{K}(\hat{p}^{(r)}_B)\mathcal{O}_W
\end{align}
for the expansion of the operator $\mathcal{K}$ into powers of the momentum operator.

\subsubsection{Hamiltonian}
We are now in the position to determine the Weyl-Wigner correspondence of the operators $\hat{\mathcal{H}}\hat{\mathcal{O}}$ and $\hat{\mathcal{O}}\hat{\mathcal{H}}$ consisting of the product of the Hamiltonian
\begin{align}
 \hat{\mathcal{H}}(\hat{z},\hat{p})\equiv \mathcal{K}(\hat{p}) + \mathcal{V}(\hat{z})
\end{align}
and an arbitrary operator $\hat{\mathcal{O}}$.

At this point it is important to note that $\hat{\mathcal{H}}$ does not involve products of $\hat{z}$ and $\hat{p}$. For this reason, and due to the linearity of the Weyl-Wigner correspondence, we can immediately apply the relations eqs.~\eqref{eq:app.V_l} and~\eqref{eq:app.K_l} to obtain
\begin{align}
(\mathcal{H}\mathcal{O})_W=(\mathcal{K}\mathcal{O})_W +(\mathcal{V}\mathcal{O})_W = \mathcal{K}(\hat{p}_B^{(l)})\mathcal{O}_W + \mathcal{V}(\hat{z}_B^{(l)})\mathcal{O}_W \, ,
\end{align}
or
\begin{align}\label{eq:app.HOw}
(\mathcal{H}\mathcal{O})_W=\mathcal{H}(\hat{z}_B^{(l)},\hat{p}_B^{(l)})\mathcal{O}_W,
\end{align}
where the Bopp operators $\hat{z}_B^{(l)}$ and $\hat{p}_B^{(l)}$ are defined by eqs.~\eqref{eq:app.z_B^l} and~\eqref{eq:app.p_B^l}.

Similarly, we can establish with the help of eqs.~\eqref{eq:app.V_r} and~\eqref{eq:app.K_r} the identity 
\begin{align}\label{eq:app.OHw}
 (\mathcal{O}\mathcal{H})_W=\mathcal{H}(\hat{z}_B^{(r)},\hat{p}_B^{(r)})\mathcal{O}_W
\end{align}
for the operator product $\hat{\mathcal{O}}\hat{\mathcal{H}}$ where the Bopp operators $\hat{z}_B^{(r)}$ and $\hat{p}_B^{(r)}$ are defined by eqs.~\eqref{eq:app.z_B^r} and~\eqref{eq:app.p_B^r}.

\subsection{Quantum Liouville equations}
\label{app:subsec:applications}

In the preceding section we have derived the Weyl-Wigner correspondence of the product of an arbitrary operator $\hat{\mathcal{O}}$ and an operator $\hat{\mathcal{H}}$ consisting of the sum of functions of position and momentum operators. These results allow us now to translate the Schr\"odinger equation, or the von Neumann equation into phase space creating a quantum Liouville equation.

\subsubsection{Time evolution of Wigner matrix}
For this purpose we first consider a slightly more general problem which is central to understanding the Kasevich-Chu interferometer in Wigner phase space. We consider the time evolution of the Wigner matrix
\begin{align}
 W_{jk}\equiv \frac{1}{2\pi \hbar} \int\limits_{-\infty}^{\infty} \!\! \drm \zeta \
				      \e^{-\i p \zeta/\hbar} \ \bra{z + \frac{1}{2}\, \zeta} \hat{U}_j \hat{\rho} \hat{U}^\dagger_k \ket{z- \frac{1}{2}\, \zeta} \label{eq:app.Wignermatrix}
\end{align}
of the density operator $\hat{\rho}$ determined by the unitary time-evolution operator
\begin{align}
 \hat{U}_j\equiv \e^{-\i \hat{H}_jt/\hbar}
\end{align}
for $j,k=1,2$ in terms of the Hamiltonians
\begin{align}\label{eq:app.H1}
 \hat{H}_1\equiv\frac{\hat{p}^2}{2m}+ V(\hat{z})
\end{align}
and
\begin{align}\label{eq:app.H2}
 \hat{H}_2\equiv\frac{(\hat{p}+\hbar k)^2}{2m}+ V(\hat{z}) = \hat{H}_1+ \frac{\hbar k}{m} \hat{p} + \frac{(\hbar k)^2}{2m}.
\end{align}
Here $\hat{H}_1$ denotes the Hamiltonian for the motion of the atom through the potential $V$ in the ground state $\ket{g_1}$. In contrast, the Hamiltonian $\hat{H}_2$ corresponds to the excitation to $\ket{g_2}$ associated with an momentum increase by $\hbar k$, motion through $V$ in $\ket{g_2}$, and finally a de-excitation to $\ket{g_1}$ connected with a reduction of the momentum by $\hbar k$.

The \textit{quantum Liouville equation} is then a special case of these results. Indeed, it is the equation of motion for the matrix element $W_{11}$. Since the time evolution $W_{22}$ is determined by $\hat{H}_2$ corresponding to excitation into $\ket{g_2}$, motion in $\ket{g_2}$, and de-excitation to $\ket{g_1}$ we refer to the resulting equation of motion in phase space as the \textit{excited state quantum Liouville equation}. Moreover, the off-diagonal elements $W_{12}$ and $W_{21}$ governed in their time evolution by two different Hamiltonians $\hat{H}_1$ and $\hat{H}_2$ are essential in understanding the interference between the two paths in the Kasevich-Chu interferometer. For this reason we call the corresponding evolution equations in phase space \textit{interference quantum Liouville equations}.

\subsubsection{General equations of motion}
We first obtain from the definition 
\begin{align}\label{eq:app.O_mit_H_1,2}
\hat{\rho}_{jk}(t)\equiv \hat{U}_j\hat{\rho}\hat{U}_k^\dagger \equiv\e^{-\i \hat{H}_{j}t/\hbar}\ \hat{\rho} \ \e^{\i \hat{H}_{k}t/\hbar}
\end{align}
of the time-dependent operator by direct differentiation with respect to time
the equation of motion
\begin{align}\label{eq:app_motion-of-hat-rho}
\frac{\partial\hat{\rho}_{jk}}{\partial t}= -\frac{\i}{\hbar}\left[ \hat{H}_{j}\hat{\rho}_{jk}(t)-\hat{\rho}_{jk}(t)\hat{H}_{k}\right]\,,
\end{align}
which for $j=k$ constitutes the von Neumann equation.

With the help of eqs.~\eqref{eq:app.HOw} and \eqref{eq:app.OHw} we immediately find the phase-space differential equation
\begin{align}\label{eq:app_motion-of-rho_W}
\frac{\partial}{\partial t}(\rho_{jk})_W= -\frac{\i}{\hbar}\left[ H_{j} (\hat{z}_B^{(l)},\hat{p}_B^{(l)}) -H_{k}(\hat{z}_B^{(r)},\hat{p}_B^{(r)})\right](\rho_{jk})_W
\end{align}
for the Weyl-Wigner correspondence $(\rho_{jk})_W$ of the density operator $\hat{\rho}_{jk}$, which after multiplication of both sides by $1/(2\pi\hbar)$ leads us to the equation of motion
\begin{align}\label{eq:app.Wignermatrixeom}
   \frac{\partial}{\partial t} W_{jk} &=  - \frac{\i}{\hbar} \left[H_j(\hat{z}_B^{(l)},\hat{p}_B^{(l)}) - H_k (\hat{z}_B^{(r)},\hat{p}_B^{(r)})  \right] W_{jk}\, 
\end{align}
for the Wigner matrix $W_{jk}$.

\subsubsection{Quantum Liouville equation}
\label{app:sec:Quantum-Liouville-equation}
In the case of $j=k=1$ eq.\eqref{eq:app_motion-of-hat-rho} reduces to the familiar von Neumann equation of a non-relativistic particle moving in a potential $V$, and the corresponding equation of motion, eq.~\eqref{eq:app.Wignermatrixeom}, for the element $W_{11}$ of the Wigner matrix turns into
\begin{align}\label{eq:app.weyl-wigner-equation-motion}
   \frac{\partial}{\partial t} W_{11} &=  - \frac{\i}{\hbar} \left[H_1(\hat{z}_B^{(l)},\hat{p}_B^{(l)}) - H_1 (\hat{z}_B^{(r)},\hat{p}_B^{(r)})  \right] W_{11}\,.
\end{align}

From the explicit form eq.~\eqref{eq:app.H1} of the Hamiltonian $\hat{H}_1$ together with the definitions eqs.~\eqref{eq:app.z_B^l},~\eqref{eq:app.p_B^l} and~\eqref{eq:app.z_B^r},~\eqref{eq:app.p_B^r} of the Bopp operators $(\hat{z}^{(l)},\hat{p}^{(l)})$ and $(\hat{z}^{(r)},\hat{p}^{(r)})$ we find
\begin{align}
   \frac{\partial}{\partial t} W_{11} =&- \frac{\i}{\hbar} \frac{1}{2m}\left[\left(p+ \frac{1}{2}\frac{\hbar}{\i}\frac{\partial}{\partial z} \right)^2 -\left(p- \frac{1}{2}\frac{\hbar}{\i}\frac{\partial}{\partial z} \right)^2 \right]W_{11}\\
   &- \frac{i}{\hbar} \left[ V(z- \frac{1}{2} \frac{\hbar}{\i} \frac{\partial}{\partial p})-V(z+\frac{1}{2} \frac{\hbar}{\i}\frac{\partial}{\partial p})\right]W_{11}\,.
\end{align}

When we expand the potential $V$ into a Taylor series and move the first derivative with respect to momentum, which is independent of $\hbar$, to the left-hand side, we arrive at the quantum Liouville equation
\begin{align}\label{eq:app.full-quantum-Liouville}
\left(\frac{\partial}{\partial t} + \frac{p}{m} \frac{\partial}{\partial z} -  \frac{\partial V}{\partial z} \frac{\partial}{\partial p} \right)  W_{11}	
	&=	\sum_{l=1}^\infty \frac{(-1)^l (\hbar/2)^{2l}}{(2l+1)!} \frac{\partial^{2l+1} V}{\partial z^{2l+1}} 
			    \frac{\partial^{2l+1}}{\partial p^{2l+1}}\,  W_{11} \, ,
\end{align}
or
\begin{align}\label{eq:app_quant_Liouville}
  \mathcal{\hat{L}}\, W(z,p;t) 	&=	\hat{\mathcal{L}}_\text{odd}\, W(z,p;t)\, .
\end{align}
Here we have recalled the time-evolution operator 
\begin{align}\label{eq:app.liouvilleOp}
\hat{\mathcal{L}}\equiv \left(\frac{\partial}{\partial t} + \frac{p}{m} \frac{\partial}{\partial z} -  \frac{\partial V}{\partial z} \frac{\partial}{\partial p} \right)
\end{align}
of the classical Liouville equation defined in eq.~\eqref{equ:liouville_equ} and the operator
\begin{align}
\hat{\mathcal{L}}_\text{odd}\equiv \sum_{l=1}^\infty \frac{(-1)^l (\hbar/2)^{2l}}{(2l+1)!} \frac{\partial^{2l+1} V}{\partial z^{2l+1}} \frac{\partial^{2l+1}}{\partial p^{2l+1}}
\end{align}
involves only the odd derivatives of the potential, and the odd derivatives of the Wigner function with respect to the momentum.

For the special case of a potential that is at most quadratic in the coordinate, $\hat{\mathcal{L}}_\text{odd}$ vanishes and the quantum Liouville equation~\eqref{eq:app_quant_Liouville} reduces to the classical one, that is
\begin{align}
  \left(\frac{\partial}{\partial t} + \frac{p}{m} \frac{\partial}{\partial z} - 
			  \frac{\partial V}{\partial z} \frac{\partial}{\partial p} \right) W_{11}(z,p;t) 	
	&=	0\, .
\end{align}

Therefore, the Wigner function $W_{11}(z,p;t_0+t)$ at time $t_0+t$ arises from $W_{11}(z_0,p_0;t_0)$ at $t=t_0$ by propagating each point $(z_0,p_0)$ along the classical trajectories $(\overline{z},\overline{p})$, which leads us to the expression
\begin{align}\label{eq:app_Wigner+trajectories}
W_{11}(z,p;t_0+t) 	&= 	\int\limits_{-\infty}^{\infty} \!\!\drm z_0 \int\limits_{-\infty}^{\infty} \!\!\drm p_0  \, 
			\ \delta[z-\overline{z}(z_0,p_0;t)] \, \delta[p-\overline{p}(z_0,p_0;t)]
			\ W_{11}(z_0,p_0;t_0)\,.
\end{align}
%

\subsubsection{Excited state quantum Liouville equation}
\label{app.subsubsec:Excited-state-QLiouville-eq}
In the next example we consider the time evolution of the Wigner matrix element $W_{22}$ corresponding to the propagation with the Hamiltonian $\hat{H}_2$.

With $j=k=2$ and the explicit form, eq.~\eqref{eq:app.H2} of the Hamiltonian $\hat{H}_2$ we obtain from eq.~\eqref{eq:app.Wignermatrixeom} the equation of motion
\begin{align}\label{eq:app.excited_state_motion}
   \frac{\partial}{\partial t} W_{22} &=  - \frac{\i}{\hbar} \left[H_1(\hat{z}_B^{(l)},\hat{p}_B^{(l)}) - H_1 (\hat{z}_B^{(r)},\hat{p}_B^{(r)})  \right] W_{22}- \frac{\i}{\hbar} \frac{\hbar k}{m}\left[\hat{p}_B^{(l)}-\hat{p}_B^{(r)}  \right] W_{22}\, .
\end{align}

With the definitions eqs.~\eqref{eq:app.p_B^l} and~\eqref{eq:app.p_B^r} of $\hat{p}_B^{(l)}$ and $\hat{p}_B^{(r)}$, we find
\begin{align}
\hat{p}_B^{(l)}-\hat{p}_B^{(r)}=-\i \hbar \frac{\partial}{\partial z}
\end{align}
and using eq.~\eqref{eq:app.full-quantum-Liouville} for the term in the first brackets on the right-hand side of eq.~\eqref{eq:app.excited_state_motion}, we arrive at the partial differential equation
\begin{align}
  \left(\frac{\partial}{\partial t} + \frac{p+\hbar k}{m} \frac{\partial}{\partial z} - 
			  \frac{\partial V}{\partial z} \frac{\partial}{\partial p} \right) W_{22}(z,p;t) 	
	&=	\hat{\mathcal{L}}_\text{odd}\,  W_{22}(z,p;t)\, .
\end{align}
The only difference to eq.~\eqref{eq:app.full-quantum-Liouville} is that the factor $p$ in front of the first derivative with respect to the coordinate $z$ in the classical Liouville operator $\hat{\mathcal{L}}$ has now been replaced by $p+\hbar k$. 

Hence, for potentials up to quadratic in the position $z$ where $\hat{\mathcal{L}}_\text{odd}$ vanishes, we find the solution
\begin{align}
W_{22}(z,p;t_0+t) 	&= 	\int\limits_{-\infty}^{\infty} \!\!\drm z_0 \int\limits_{-\infty}^{\infty} \!\!\drm p_0  \, 
			\ \delta[z-\overline{z}(z_0,p_0+\hbar k;t)] \\
			& \times \delta[p+\hbar k-\overline{p}(z_0,p_0+\hbar k;t)]
			\ W_{22}(z_0,p_0;t_0)
\end{align}
with the initial distribution $W_{22}(z_0,p_0;t_0)$.
\subsubsection{Interference quantum Liouville equations}
Finally we address the problem of finding the equations of motion of $W_{21}$ or $W_{12}$ where $j=2$ and $k=1$ or $j=1$ and $k=2$.

With the explicit forms, eqs.~\eqref{eq:app.H1} and~\eqref{eq:app.H2} of the Hamiltonians $\hat{H}_1$ and $\hat{H}_2$, eq.~\eqref{eq:app.Wignermatrixeom} reduces to
\begin{align}
\frac{\partial}{\partial t}W_{21} =- \frac{\i}{\hbar} \left[ H_2(\hat{z}_B^{(l)},\hat{p}_B^{(l)})-H_1(\hat{z}_B^{(r)},\hat{p}_B^{(r)})\right] W_{21}- \frac{\i}{\hbar} \left[ 
\frac{\hbar k}{m} \hat{p}_B^{(l)} + \frac{(\hbar k)^2}{2m}\right] W_{21}\,. 
\end{align}
The left hand-side and the first term on the right hand-side correspond to the quantum Liouville equation~\eqref{eq:app_quant_Liouville}, and with the definition~\eqref{eq:app.p_B^l} of the Bopp operator $\hat{p}_B^{(l)}$ we find 
\begin{align}
\hat{\mathcal{L}}\,W_{21} =\hat{\mathcal{L}}_\text{odd} \,W_{21}- \frac{\i}{\hbar} \left[ 
\frac{\hbar k}{m} \left(p+\frac{1}{2}\frac{\hbar}{\i} \frac{\partial}{\partial z} \right) + \frac{(\hbar k)^2}{2m}\right] W_{21}\,. 
\end{align}

When we recall the explicit form eq.~\eqref{eq:app.liouvilleOp} of the classical Liouville operator $\hat{\mathcal{L}}$ we arrive at the equation of motion
\begin{align}
\left[ \frac{\partial}{\partial t}+ \frac{1}{m}\left(p+\frac{1}{2}\hbar k\right)\frac{\partial}{\partial z}- \frac{\partial V}{\partial z} \frac{\partial}{\partial p}\right]W_{21} =\hat{\mathcal{L}}_\text{odd}\, W_{21}- \i \frac{k}{m} \left(p+ \frac{1}{2}\hbar k \right) W_{21}\,
\end{align}
for the Wigner matrix element $W_{21}$ determining the interference term. 

In the next step we restrict ourselves to potentials at most quadratic in the position variable $z$ such that $\hat{\mathcal{L}}_\text{odd}$ vanishes which leads us to the expression
\begin{align} \label{eq:app_liouville_interference}
\left[ \frac{\partial}{\partial t}+ \frac{1}{m}\left(p+\frac{1}{2}\hbar k\right)\frac{\partial}{\partial z}- \frac{\partial V}{\partial z} \frac{\partial}{\partial p}\right]W_{21} =- \i \frac{k}{m} \left(p+ \frac{1}{2} \hbar k\right) W_{21}\,. 
\end{align}
Thus the time evolution of the Wigner function $W_{21}$ corresponding to the part of the interference term contains the momentum variable in the form $p+\hbar k /2$ rather than $p$ or $p+\hbar k$. The factor $1/2$ in the momentum shift is typical for interference effects. Moreover, we note the inhomogeneity in the evolution equation which involves the complex unit $\i$ explicitly and creates a complex phase factor in the solution.

Indeed, it is straightforward to verify that
\begin{align}
W_{21}(z,p;t_0+t) 	=& 	\int\limits_{-\infty}^{\infty} \!\!\drm z_0 \int\limits_{-\infty}^{\infty} \!\!\drm p_0  \, \e^{-\i k(z-z_0)}
			\ \delta[z-\overline{z}(z_0,p_0+\frac{1}{2}\hbar k;t)] \\
			& \times \delta[p+\frac{1}{2}\hbar k-\overline{p}(z_0,p_0+\frac{1}{2}\hbar k;t)]
			\ W_{21}(z_0,p_0;t_0)
\end{align}
satisfies the Liouville equation~\eqref{eq:app_liouville_interference} with the initial distribution $W_{21}(z_0,p_0;t_0)$.

We conclude by briefly discussing the time evolution of the Wigner matrix element $W_{12}$ which follows from eq.~\eqref{eq:app.Wignermatrixeom} for $j=1$ and $k=2$. Indeed, we find from eq.~\eqref{eq:app.Wignermatrixeom} with the explicit forms of the Hamiltonians $\hat{H}_1$ and $\hat{H}_2$, and the appropriate Bopp operators the equation of motion
\begin{align}
\left[ \frac{\partial}{\partial t}+ \frac{1}{m}\left(p+\frac{1}{2}\hbar k\right)\frac{\partial}{\partial z}- \frac{\partial V}{\partial z} \frac{\partial}{\partial p}\right]W_{12} =\hat{\mathcal{L}}_\text{odd}\, W_{12}+ \i \frac{k}{m} \left(p+ \frac{1}{2}\hbar k \right) W_{12}\,
\end{align}
for $W_{12}$.

Hence, the only difference compared to the equation of motion eq.~\eqref{eq:app_liouville_interference} is the sign of the inhomogeneity which for vanishing $\hat{\mathcal{L}}_\text{odd}$ gives rise to the expression
\begin{align}
W_{12}(z,p;t_0+t) 	= &	\int\limits_{-\infty}^{\infty} \!\!\drm z_0 \int\limits_{-\infty}^{\infty} \!\!\drm p_0  \, \e^{\i k(z-z_0)}
			\ \delta[z-\overline{z}(z_0,p_0+\frac{1}{2}\hbar k;t)] \\
			& \times \delta[p+\frac{1}{2}\hbar k-\overline{p}(z_0,p_0+\frac{1}{2}\hbar k;t)]
			\ W_{12}(z_0,p_0;t_0)
\end{align}
for $W_{12}$.

\subsection{Schr\"odinger equation in phase space}
We conclude by translating the eigenvalue equation
\begin{align}
\hat{H} \ket{E} = E \ket{E}
\end{align}
for the energy eigenstate $\ket{E}$ into Wigner phase space.

For this purpose we first multiply both sides of this equation from the right by 
$\bra{E}$ which yields the operator equation
\begin{align}
\hat H \ket{E}\bra{E} = E \ket{E}\bra{E}\,.
\end{align}
Next we consider its Weyl-Wigner correspondence which leads us to the partial differential equation
\begin{align}
  H\left(\hat{z}_B^{(l)} ,\hat{p}_B^{(l)} \right) W_{E} 	&= E\  W_{E} 
\end{align}
in phase space for the Wigner function $W_{E}$ of $\ket{E}$.

When we take the real and imaginary parts of this equation we arrive at
\begin{align}
 \frac{1}{2}\left[ H\left(\hat{z}_B^{(l)} ,\hat{p}_B^{(l)} \right)+H\left(\hat{z}_B^{(r)} ,\hat{p}_B^{(r)} \right)\right] W_{E} 	&= E\  W_{E}
\label{eq:app:WignerEreal}
\end{align}
and
\begin{align}
 \frac{1}{2 \i}\left[ H\left(\hat{z}_B^{(l)} ,\hat{p}_B^{(l)} \right)-H\left(\hat{z}_B^{(r)} ,\hat{p}_B^{(r)} \right)\right] W_{E} 	&= 0\, .
 \label{eq:app:WignerEimaginary}
\end{align}
Here we have used $\left[H\left(\hat{z}_B^{(l)} ,\hat{p}_B^{(l)} \right)\right]^*=H\left(\hat{z}_B^{(r)} ,\hat{p}_B^{(r)} \right)$ and the fact that $E$ and $W_E$ are real.

Equation~\eqref{eq:app:WignerEimaginary} is the remnant of the quantum Liouville equation~\eqref{eq:app.weyl-wigner-equation-motion} for an energy eigenstate where the partial derivative with respect to time vanishes, and eq.~\eqref{eq:app:WignerEreal} is the phase-space analogue of the time-independent Schr\"odinger equation. It represents the energy eigenvalue equation in phase space.

When we recall the definitions eqs.~\eqref{eq:app.z_B^l},~\eqref{eq:app.p_B^l}, and~\eqref{eq:app.z_B^r}, \eqref{eq:app.p_B^r} of the Bopp operators together with the explicit form of
\begin{align}
\hat{H}\equiv\frac{\hat{p}^2}{2m}+V(\hat{z})\,,
\end{align}
the time-independent Schr{\"o}dinger equation in phase space reads
\begin{align}
  \left[\frac{p^2}{2m} 	+ V - \frac{\hbar^2}{8m} \frac{\partial^2}{\partial z^2} 
			+\sum_{l=1}^\infty \frac{(-1)^l (\hbar/2)^{2l}}{(2l)!} \frac{\partial^{2l} V}{\partial z^{2l}} 
			    \frac{\partial^{2l}}{\partial p^{2l}} \right] W_{E}
		  &= E\  W_{E}\,.
\end{align}
Due to the appearance of the sum of the Weyl-Wigner correspondences of $(H\rho)_W$ and $ (\rho H)_W$ in eq.~\eqref{eq:app:WignerEreal} only the even derivatives of the potential enter. Moreover, $\hbar$ appears explicitly, even for a constant potential.

\section{Phase-shift operator in the presence of weak gradients}
\label{app:Gradients}
In this appendix, we evaluate the phase-shift operator 
\begin{align}
 \hat{U}_\varphi\equiv \e^{\hat{B}(T)}\e^{-\hat{B}}
\label{eq:app:Uphi}
\end{align}
defined by the operators
\begin{align}
\hat{B}(T)\equiv \i \frac{T}{\hbar}\left[\hat{H}_1+\frac{\hbar k}{m_\i} \hat{p}(T)+ \frac{(\hbar k)^2}{2m_\i}\right]
\label{eq:app:BT}
\end{align}
and
\begin{align}
\hat{B}\equiv \i \frac{T}{\hbar}\left[\hat{H}_1+\frac{\hbar k}{m_\i} \hat{p}(0)+ \frac{(\hbar k)^2}{2m_\i}\right]
\label{eq:app:B}
\end{align}
in the presence of gravity gradients, that is for the potential 
\begin{align}
 V_\Gamma(z)\equiv m_\g gz+\frac{1}{2}m_\g \Gamma z^2\, ,
\end{align}
in the limit of weak gravity gradients given by $\Gamma T^2 \ll1$.

In order to highlight the role of the inertial and gravitational mass $m_\i$ and $m_\g$ in the phase-shift operator $\hat{U}_\varphi$ we throughout this appendix distinguish between the two masses. Indeed, the mass appearing explicitly in the operators $\hat{B}(T)$ and $\hat{B}$ defined by eqs.~\eqref{eq:app:BT} and~\eqref{eq:app:B} is the inertial mass $m_\i$. In addition, the gravitational mass $m_\g$ enters into $\hat{B}(T)$ and $\hat{B}$ through the potential $ V_\Gamma$ in the Hamiltonian 
\begin{align}
\hat{H}_1 \equiv \frac{\hat{p}^2}{2m_\i}+V_\Gamma(\hat{z})\,.
\end{align}

Our calculation proceeds in three steps: (\textit{i}) We first solve the Heisenberg equation of motion for the position operator $\hat{z}$ of a harmonic oscillator in the presence of a constant acceleration, and derive the momentum operator $\hat{p}(T)$ at the time $T$ in the limit of weak gravity gradients. (\textit{ii}) We then recall a generalized Baker-Campbell-Hausdorff formula appropriate for this case, and (\textit{iii}) finally evaluate the relevant commutation relations.
%
\subsection{Time-dependent momentum operator}
%
The Heisenberg equation of motion
\begin{align}
m_\i \ddot{\hat{z}}+m_\g  \Gamma \hat{z}= -m_\g g
\end{align}
of the coordinate $\hat{z}$ due to the potential $V_\Gamma$ takes the form
\begin{align} \label{app.eq.of-motion-prime}
 \ddot{\hat{z}}+ \Gamma' \hat{z}= - g' \,
\end{align}
when we introduce the mass-dependent gradient and acceleration, that is
\begin{align}\label{app.eq.Gamma-prime}
\Gamma'\equiv\frac{m_\g}{m_\i} \Gamma 
\end{align}
and
\begin{align}\label{app.eq.g-prime}
g'\equiv \frac{m_\g}{m_\i} g\,.
\end{align}

From the solution
\begin{align}
\hat{z}(t)=- \frac{g'}{\Gamma'}+\left(\hat{z}(0)+ \frac{g'}{\Gamma'} \right)  \cos\left( \sqrt{\Gamma'}t \right) + \frac{\hat{p}(0)}{m_\i \sqrt{\Gamma'}} \sin\left( \sqrt{\Gamma'}t \right)
\end{align}
of eq.~\eqref{app.eq.of-motion-prime} we find the momentum operator
\begin{align}
\hat{p}(T)=-m_\i g' \frac{\sin \left(\sqrt{\Gamma'}T \right)}{\sqrt{\Gamma'}}- m_\i \sqrt{\Gamma'} \sin \left(\sqrt{\Gamma'}T \right) \hat{z}(0)+\cos  \left(\sqrt{\Gamma'}T \right) \hat{p}(0)\, ,
\end{align}
which in the limit of weak gravity gradients
\begin{align}
\epsilon \equiv \Gamma'T^2 \ll 1
\end{align}
reduces to
\begin{align}
\hat{p}(T)\cong \hat{p}(0) -m_\i g' T \left( 1-  \frac{1}{6}\epsilon \right)- m_\i \Gamma' T \left[ \hat{z}(0)+ \frac{1}{2} \frac{\hat{p}(0)}{m_\i} T \right]\, .
\label{eq:app:pT}
\end{align}
%
\subsection{Generalized Baker-Campbell-Hausdorff formula}
%
In the next step we substitute the so-calculated momentum operator eq.~\eqref{eq:app:pT} into the definition~\eqref{eq:app:BT} of the operator $\hat{B}(T)$ and find with the help of eq.~\eqref{eq:app:B} determining $\hat{B}$ the explicit form
\begin{align}
\hat{B}(T) &\cong\hat{B} - \i \delta \varphi_\g \left( 1 - \frac{1}{6}\epsilon\right)- \epsilon \hat{C}\, , 
\label{e.B(T)_C}
\end{align}
where we have recalled the phase shift 
\begin{align}
 \delta \varphi_{\g}=k g' T^2 = \frac{m_\g}{m_\i} kgT^2
\label{eq:app:deltaphig}
\end{align}
in the Kasevich-Chu interferometer in a constant gravitational field. In the last step we have made use of the definition~\eqref{app.eq.g-prime} of $g'$. Moreover, we have introduced the abbreviation
\begin{align}
\hat{C}\equiv \i k \left[ \hat{z}(0)+ \frac{1}{2} \frac{\hat{p}(0)}{m_\i} T \right] \, .
\end{align}

In order to evaluate the phase-shift operator $\hat{U}_\varphi$ given by eq.~\eqref{eq:app:Uphi} we recall the slightly more complicated Baker-Campbell-Hausdorff formula~\cite{Wilcox67}
\begin{align}
 \e^{\hat{B}(T)}\e^{-\hat{B}}=&\exp\left(\hat{B}(T)- \hat{B}- \frac12 [\hat{B}(T),\hat{B}]-\frac{1}{12}[\hat{B}(T)+\hat{B},[\hat{B}(T),\hat{B}]]\right. \nonumber \\
 & \left.-\frac{1}{24}[\hat{B},[\hat{B}(T),[\hat{B}(T),\hat{B}]]] + \ldots\right)\, ,
\label{eq:app:fullBCH}
\end{align}
which with the explicit form eq.~\eqref{e.B(T)_C} of $\hat{B}(T)$ resulting in $[\hat{B}(T),\hat{B}]=- \epsilon [\hat{C},\hat{B}]$ reduces in lowest order in $\epsilon$ to 
\begin{align}
\hat{U}_\varphi \cong&\,\e^{-\i \delta\varphi_\g\left( 1 -\epsilon/6\right)} \\
&\times \exp\left\lbrace \epsilon\left[- \hat{C}+\frac{1}{2} [\hat{C},\hat{B}]+ \frac{1}{6}[\hat{B},[\hat{C},\hat{B}]] +\frac{1}{24}[\hat{B},[\hat{B},[\hat{C},\hat{B}]]]\right]+\ldots\right\rbrace\, .
\label{eq:app:uphizwischen}
\end{align}
%
\subsection{Commutation relations and final expression}
%
In the last step we evaluate the commutators determining $\hat{U}_\varphi$. Since  the argument of the exponential in eq.~\eqref{eq:app:uphizwischen} defining $\hat{U}_\varphi$ is already linear in $\epsilon$, it suffices to calculate the commutators neglecting terms proportional to $\epsilon$. In this way we find after minor algebra from the expression~\eqref{eq:app:B} for $\hat{B}$ the relation
\begin{align}
[\hat{C},\hat{B}]\cong -\i k T  \frac{\hat{p}(0)}{m_\i} -\i \frac{\hbar k^2}{m_\i} T + \frac{\i}{2} \delta \varphi_\g 
\label{eq:app:KomCB}
\end{align}
and thus
\begin{align}
[\hat{B},[\hat{C},\hat{B}]]\cong \i \delta \varphi_\g \, .
\label{eq:app:KomBCB}
\end{align}
Here we have recalled the definition~\eqref{eq:app:deltaphig} of $\delta\varphi_{\g}$.

Since the double-commutator eq.~\eqref{eq:app:KomBCB} is a c-number the higher order corrections in the Baker-Campbell-Hausdorff formula eq.~\eqref{eq:app:uphizwischen} vanish and the expansion truncates.

When we substitute the commutators eqs.~\eqref{eq:app:KomCB} and~\eqref{eq:app:KomBCB} into the expression~\eqref{eq:app:uphizwischen} for $\hat{U}_\varphi$ we arrive at the approximation
\begin{align}
\hat{U}_\varphi\cong \exp\left[ -\i \delta \varphi_\g \left(1- \frac{7}{12}\Gamma' T^2 \right)\right]\exp\left[ -\i \Gamma' T^2  \frac{\hbar k^2}{2 m_\i}T\right]\exp\left[ -\i \Gamma' T^2  k \left( \hat{z}(0)+ \frac{\hat{p}(0)}{m_\i}T \right)\right]\,,
\end{align}
or
\begin{align}
\hat{U}_\varphi\cong& \exp\left[ -\i \delta \varphi_\g \left(1- \frac{7}{12} \frac{m_\g}{m_\i}\Gamma T^2 \right)\right]\exp\left[ -\i \frac{m_\g}{m_\i} \Gamma T^2  \frac{\hbar k^2}{2 m_\i}T\right] \\
&\times \exp\left[ -\i \frac{m_\g}{m_\i}\Gamma T^2  k \left( \hat{z}(0)+ \frac{\hat{p}(0)}{m_\i}T \right)\right]
\end{align}
of the phase-shift operator $\hat{U}_\varphi$ in the presence of weak gravity gradients. In the last step we have recalled the definition~\eqref{app.eq.Gamma-prime} of the mass-dependent gradient $\Gamma'$ and have obtained an expression which distinguishes between the inertial and gravitational mass.

\section{Phase-space dynamics and phase shift}
\label{app:phase-space-dynamics}
In this appendix we derive explicit expressions for the endpoints of the phase space trajectories in the Kasevich-Chu interferometer corresponding to the upper and the lower paths of the interferometer, as well as to the one representing the interference term. Moreover, we obtain an expression for the phase shift.

Throughout this appendix we consider the potential
\begin{align}
 V_\Gamma \equiv mgz+\frac12 m \Gamma z^2
\end{align}
giving rise after the time $T$ to the coordinate
\begin{align}
 \bar{z}(z_0,p_0;T)=-\frac{g}{\Gamma} + \left(z_0+\frac{g}{\Gamma} \right)\mathscr{c} + \frac{p_0}{m\sqrt{\Gamma}} \mathscr{s}
\label{eq:appWig:zbar}
\end{align}
and the momentum
\begin{align}
 \bar{p}(z_0,p_0;T)=-m\sqrt{\Gamma} \left(z_0+\frac{g}{\Gamma} \right)\mathscr{s} + p_0 \mathscr{c}\,,
\label{eq:appWig:pbar}
\end{align}
provided we have started at $z_0$ and $p_0$. Here we have introduced the abbreviations $\mathscr{c}\equiv\cos\left(\sqrt{\Gamma}T\right)$ and $\mathscr{s}\equiv\sin\left(\sqrt{\Gamma}T\right)$.

\subsection{Endpoints}

In this section we propagate the point $(z_0,p_0)$ in phase space through the interferometer corresponding to the upper, the lower and the interference path. In this way we show that the endpoints of the upper and the lower path are symmetrically located around the one of the interference path, that is, the latter is always half-way between the former.

\subsubsection{Upper path}

We start our analysis by considering the upper path of the interferometer. From the expression~\eqref{eq:W_u(2T)} for the Wigner function $W_u$ at the exit of the interferometer we find with 
\begin{align}
 \bar{z}(z_0,p_0+j\hbar k;T)=-\frac{g}{\Gamma} + \left(z_0+\frac{g}{\Gamma} \right)\mathscr{c} + \frac{p_0+j\hbar k}{m\sqrt{\Gamma}}\mathscr{s}
\label{eq:appWig:zbarj}
\end{align}
and
\begin{align}
 \bar{p}(z_0,p_0+j\hbar k;T)=-m\sqrt{\Gamma} \left(z_0+\frac{g}{\Gamma} \right)\mathscr{s} + (p_0+j\hbar k)\mathscr{c}
\label{eq:appWig:pbarj}
\end{align}
where $j=0$, $j=1/2$, and $j=1$, for the end coordinate
\begin{align}
 z_u\equiv \bar{z}\left[\bar{z}(z_0,p_0+\hbar k;T),\bar{p}(z_0,p_0+\hbar k;T) -\hbar k;T \right]
\end{align}
the formula
\begin{align}
 z_u=-\frac{g}{\Gamma}+ \left[\left(z_0+\frac{g}{\Gamma} \right)\mathscr{c} +  \frac{p_0+\hbar k}{m\sqrt{\Gamma}}\mathscr{s} \right]\mathscr{c}+\left[-\left(z_0+\frac{g}{\Gamma} \right)\mathscr{s}+ \frac{p_0+\hbar k}{m\sqrt{\Gamma}}\mathscr{c}-\frac{\hbar k}{m\sqrt{\Gamma}}\right]\mathscr{s}\,,
\end{align}
that is
\begin{align}
 z_u=-\frac{g}{\Gamma}\left[1-(\mathscr{c}^2-\mathscr{s}^2) \right] + z_0 (\mathscr{c}^2-\mathscr{s}^2)+ \frac{p_0}{m\sqrt{\Gamma}}2\mathscr{s}\mathscr{c}+\frac{\hbar k}{m\sqrt{\Gamma}}\mathscr{s}\mathscr{c} +\frac{\hbar k}{m\sqrt{\Gamma}}(\mathscr{s}\mathscr{c}-\mathscr{s})\,.
\label{eq:appWig:zupper}
\end{align}
Here we have used $j=1$.

Likewise, we obtain from eq.~\eqref{eq:W_u(2T)} the final momentum
\begin{align}
 p_u\equiv \bar{p}\left[\bar{z}(z_0,p_0+\hbar k;T),\bar{p}(z_0,p_0+\hbar k;T) -\hbar k;T \right]\, ,
\end{align}
which with eq.~\eqref{eq:appWig:pbar}, as well as eqs.~\eqref{eq:appWig:zbarj} and~\eqref{eq:appWig:pbarj} for $j=1$ takes the form
\begin{align}
 p_u=-m\sqrt{\Gamma}\left[\left(z_0+\frac{g}{\Gamma} \right)\mathscr{c} +  \frac{p_0+\hbar k}{m\sqrt{\Gamma}}\mathscr{s} \right]\mathscr{s} +\left[-m\sqrt{\Gamma}\left(z_0+\frac{g}{\Gamma} \right)\mathscr{s}+ (p_0+\hbar k)\mathscr{c}-\hbar k \right]\mathscr{c}\,,
\end{align}
or
\begin{align}
 p_u=-m\frac{g}{\sqrt{\Gamma}}2\mathscr{s}\mathscr{c} - m\sqrt{\Gamma} z_0 2\mathscr{s}\mathscr{c} + p_0(\mathscr{c}^2-\mathscr{s}^2)-\hbar k \mathscr{s}^2 +\hbar k (\mathscr{c}^2-\mathscr{c})\,.
\label{eq:appWig:pupper}
\end{align}

\subsubsection{Lower path}
Next we turn to the lower path and derive from the corresponding Wigner function $W_l$ given by eq.~\eqref{eq:W_l(2T)} the final position
\begin{align}
 z_l\equiv \bar{z}\left[\bar{z}(z_0,p_0;T),\bar{p}(z_0,p_0;T) +\hbar k;T \right]\,,
\end{align}
which with eqs.~\eqref{eq:appWig:zbarj} and~\eqref{eq:appWig:pbarj} for $j=0$, and eq.~\eqref{eq:appWig:zbar} yields
\begin{align}
 z_l=-\frac{g}{\Gamma}+ \left[\left(z_0+\frac{g}{\Gamma} \right)\mathscr{c} +  \frac{p_0\mathscr{s}}{m\sqrt{\Gamma}} \right]\mathscr{c}+\left[-\left(z_0+\frac{g}{\Gamma} \right)\mathscr{s}+ \frac{p_0c+\hbar k}{m\sqrt{\Gamma}}\right]\mathscr{s}\,,
\end{align}
or
\begin{align}
 z_l=-\frac{g}{\Gamma}\left[1-(\mathscr{c}^2-\mathscr{s}^2) \right] + z_0 (\mathscr{c}^2-\mathscr{s}^2)+ \frac{p_0}{m\sqrt{\Gamma}}2\mathscr{s}\mathscr{c}+\frac{\hbar k}{m\sqrt{\Gamma}}\mathscr{s}\mathscr{c} -\frac{\hbar k}{m\sqrt{\Gamma}}(\mathscr{s}\mathscr{c}-\mathscr{s})\,.
\label{eq:appWig:zlower}
\end{align}

Similarly, we find the final momentum
\begin{align}
 p_l\equiv \bar{p}\left[\bar{z}(z_0,p_0;T),\bar{p}(z_0,p_0;T) +\hbar k;T \right] -\hbar k
\end{align}
determined by the Wigner function $W_l$ given by eq.~\eqref{eq:W_l(2T)} when we substitute eqs.~\eqref{eq:appWig:zbarj} and~\eqref{eq:appWig:pbarj} for $j=0$ into eq.~\eqref{eq:appWig:pbar} and arrive at
\begin{align}
 p_l=-m\sqrt{\Gamma}\left[\left(z_0+\frac{g}{\Gamma} \right)\mathscr{c} +  \frac{p_0}{m\sqrt{\Gamma}}\mathscr{s} \right]\mathscr{s} +\left[-m\sqrt{\Gamma}\left(z_0+\frac{g}{\Gamma} \right)\mathscr{s}+ p_0\mathscr{c}+\hbar k \right]\mathscr{c}-\hbar k\,,
\end{align}
or
\begin{align}
 p_l=-m\frac{g}{\sqrt{\Gamma}}2\mathscr{s}\mathscr{c} - m\sqrt{\Gamma} z_0 2\mathscr{s}\mathscr{c} + p_0(\mathscr{c}^2-\mathscr{s}^2)-\hbar k \mathscr{s}^2 -\hbar k (\mathscr{c}^2-\mathscr{c})\,.
\label{eq:appWig:plower}
\end{align}
Here we have made use of the trigonometric relation
\begin{align}
\mathscr{c}^2+\mathscr{s}^2=1\,. \label{app.eq:cos2+sin2=1}
\end{align}
\subsubsection{Interference term}

From the expression~\eqref{eq:W_i(2T)} for the Wigner function $W_i$ corresponding to the interference term we can read-off the formula
\begin{align}
z_i\equiv \bar{z}\left[\bar{z}(z_0,p_0+\frac12\hbar k;T),\bar{p}(z_0,p_0+\frac12\hbar k;T) ;T \right]
\end{align}
for the end point $z_i$ which with eq.~\eqref{eq:appWig:zbar}, and eqs.~\eqref{eq:appWig:zbarj} and~\eqref{eq:appWig:pbarj} for $j=1/2$, leads us to
\begin{align}
 z_i=-\frac{g}{\Gamma}+ \left[\left(z_0+\frac{g}{\Gamma} \right)\mathscr{c} +  \frac{p_0+\hbar k/2}{m\sqrt{\Gamma}}\mathscr{s} \right]\mathscr{c}+\left[-\left(z_0+\frac{g}{\Gamma} \right)\mathscr{s}+ \frac{p_0+\hbar k/2}{m\sqrt{\Gamma}}\mathscr{c}\right]\mathscr{s}\,,
\end{align}
or
\begin{align}
 z_i=-\frac{g}{\Gamma}\left[1-(\mathscr{c}^2-\mathscr{s}^2) \right] + z_0 (\mathscr{c}^2-\mathscr{s}^2)+ \frac{p_0}{m\sqrt{\Gamma}}2\mathscr{s}\mathscr{c}+\frac{\hbar k}{m\sqrt{\Gamma}}\mathscr{s}\mathscr{c}\,.
\label{eq:appWig:zi}
\end{align}

Likewise, we obtain from eq.~\eqref{eq:W_i(2T)} the final momentum
\begin{align}
 p_i\equiv \bar{p}\left[\bar{z}(z_0,p_0+\frac12\hbar k;T),\bar{p}(z_0,p_0+\frac12 \hbar k;T) ;T \right]-\frac12 \hbar k
\end{align}
by substituting  eqs.~\eqref{eq:appWig:zbarj} and~\eqref{eq:appWig:pbarj} for $j=1/2$ into eq.~\eqref{eq:appWig:pbar}, that is
\begin{align}
 p_i=&-m\sqrt{\Gamma}\left[\left(z_0+\frac{g}{\Gamma} \right)\mathscr{c} +  \frac{p_0+\hbar k/2}{m\sqrt{\Gamma}}\mathscr{s} \right]\mathscr{s}\\
 & +\left[-m\sqrt{\Gamma}\left(z_0+\frac{g}{\Gamma} \right)\mathscr{s}+ (p_0+\frac{1}{2}\hbar k)\mathscr{c} \right]\mathscr{c} -\frac{1}{2} \hbar k \, ,
\end{align}
or
\begin{align}
 p_i=-m\frac{g}{\sqrt{\Gamma}}2\mathscr{s}\mathscr{c} - m\sqrt{\Gamma} z_0 2\mathscr{s}\mathscr{c} + p_0(\mathscr{c}^2-\mathscr{s}^2)-\hbar k \mathscr{s}^2\,.
\label{eq:appWig:pi}
\end{align}
Here we have made use again of the trigonometric relation eq.~\eqref{app.eq:cos2+sin2=1}.

\subsubsection{Compact representation}

When we compare the expressions for $z_u$ and $p_u$ given by eqs.~\eqref{eq:appWig:zupper} and~\eqref{eq:appWig:pupper}, as well as the ones for $z_l$ and $p_l$, eqs.~\eqref{eq:appWig:zlower} and~\eqref{eq:appWig:plower}, to the ones for $z_i$ and $p_i$, determined by eqs.~\eqref{eq:appWig:zi} and~\eqref{eq:appWig:pi}, we find the component representation
\begin{align}
\begin{pmatrix} z_u\\p_u\end{pmatrix}  =\begin{pmatrix} z_i\\ p_i\end{pmatrix}  +  \frac{1}{2}\begin{pmatrix} \Delta z\\ \Delta p\end{pmatrix}
\end{align}
as well as
\begin{align}
\begin{pmatrix} z_l\\p_l\end{pmatrix}  = \begin{pmatrix} z_i\\ p_i\end{pmatrix}  - \frac{1}{2}\begin{pmatrix} \Delta z\\ \Delta p\end{pmatrix}\,,
\end{align}
where 
\begin{align}
 \Delta z\equiv \frac{\hbar k}{m \sqrt{\Gamma}}2 (\mathscr{s}\mathscr{c}-\mathscr{s}) \equiv \frac{\hbar k}{m} \frac{\delta \mathscr{ s}}{ \sqrt{\Gamma}} \label{eq:app.Delta-z}
\end{align}
and
\begin{align}
 \Delta p \equiv \hbar k 2(\mathscr{c}^2-\mathscr{c})\equiv \hbar k \delta \mathscr{c}\,.  \label{eq:app.Delta-p}
\end{align}
Here we have introduced the abbreviations
\begin{align}
 \delta \mathscr{s} \equiv 2\mathscr{s}\mathscr{c} -2\mathscr{s} = \sin\left(\sqrt{\Gamma}2T\right)-2\sin\left(\sqrt{\Gamma}T\right) \label{eq:app.delta-s}
\end{align}
and
\begin{align}
 \delta \mathscr{c} \equiv2 \mathscr{c}^2- 2\mathscr{c}= \cos\left(\sqrt{\Gamma}2T\right)-2\cos\left(\sqrt{\Gamma}T\right)+1\,. \label{eq:app.delta-c}
\end{align}
In the last step we have recalled the definitions of $\mathscr{c}$ and $\mathscr{s}$ together with familiar trigonometric relations.

Hence, the end points of the phase space trajectories of the upper and lower paths are shifted symmetrically with respect to the one of the interference term.
\subsection{Phase shift}
\label{app:subsec:phase_shift}
Finally, we evaluate the phase shift
\begin{align}
\delta \varphi\equiv k \lbrace \overline{z}(z_0,p_0+ \frac{1}{2}\hbar k;2T)- 2 \overline{z}(z_0,p_0+ \frac{1}{2}\hbar k;T)+ z_0  \rbrace
\end{align}
given by eq.~\eqref{eq:varphi-with-trajectories}.
For the first term, that is, the coordinate at time $2T$, we use eq.~\eqref{eq:appWig:zbarj} with $j=1/2$ and replace $T$ by $2T$ in the arguments of the trigonometric functions $\mathscr{c}$ and $\mathscr{s}$. From eq.~\eqref{eq:appWig:zbarj} we find with $j=1/2$ the position $\overline{z}$ at the intermediate time $T$. Hence, we arrive with the definitions eqs.~\eqref{eq:app.delta-s} and~\eqref{eq:app.delta-c} of the discrete derivatives of the trigonometric functions  at the expression
\begin{align}
\delta \varphi =  g \frac{k\delta \mathscr{c}}{\Gamma}  + \frac{1}{2} k \frac{ \hbar k}{m } \frac{\delta \mathscr{s}}{\sqrt{\Gamma}}+ k \delta \mathscr{c}\, z_0 + \frac{k }{m} \frac{\delta \mathscr{s}}{\sqrt{\Gamma}} p_0\,.
\end{align}

When we recall the definitions eqs.~\eqref{eq:app.Delta-z} and~\eqref{eq:app.Delta-p} of the shifts $\Delta z$ and $\Delta p$ we obtain the final expression
\begin{align}
\delta \varphi = g \frac{\Delta p}{\hbar \Gamma}  + \frac{1}{2}k\Delta z+ \frac{1}{\hbar} \left( \Delta p z_0 +\Delta z p_0\right)
\end{align}
for the phase shift.

\acknowledgments
We would like to thank H.\ Abele, M.\ Aspelmeyer, Ch.\ J.\ Bord\'{e}, \v{C}.\ Brukner, C.\ Cohen-Tannoudji, D.\ Giulini, D.\ M.\ Greenberger, N.\ L.\ Harshman, D.\ Heim, E.\ Kajari, H.\ Lemmel, H.\ M\"uller, R.\ F.\ O'Connell, A.\ Peters, I.\ Pikovski, E.\ M.\ Rasel, H. \ Rauch, S.\ Reynaud, S.\ Stenholm, G.\ S\"u{\ss}mann, K.\ Vogel, S. \ Werner, P.\ Wolf, A.\ Zeilinger, and M.\ Zych for many fruitful discussions. In particular, we are most grateful to the organizers of the school, M.\ Kasevich, G.\ Tino, and F.\ Sorrentino for a most stimulating workshop. As part of the QUANTUS collaboration, this project was supported by the German Space Agency DLR with funds provided by the Federal Ministry of Economics and Technology (BMWi) under grant no. DLR 50 WM 0837. Moreover, W.\ P.\ S. thanks Texas A\&M University for a Texas A\&M Institute for Advanced Study (TIAS) Faculty Fellowship. E.\ G. and W.\ P.\ S. are grateful to Texas A\&M for the gracious hospitality which was essential for the completion of these lecture notes.

\bibliographystyle{varenna}
\bibliography{bibliography_varenna13}

\end{document}